\documentclass[11pt]{article}
\usepackage{enumerate}
\usepackage{amsfonts}
\usepackage{graphicx}
\begin{document}
\def\be{\begin{equation}}
\def\bea{\begin{eqnarray}}
\def\ee{\end{equation}}
\def\eea{\end{eqnarray}}
\def\d{\partial}
\def\eps{\varepsilon}
\def\la{\lambda}
\def\b{\bigskip}
\def\nn{\nonumber \\}
\def\p{\partial}
\def\t{\tilde}
\def\h{{1\over 2}}
\def\be{\begin{equation}}
\def\bea{\begin{eqnarray}}
\def\ee{\end{equation}}
\def\eea{\end{eqnarray}}
\def\b{\bigskip}
\def\u{\uparrow}
\newcommand{\comment}[2]{#2}

\makeatletter
\def\blfootnote{\xdef\@thefnmark{}\@footnotetext}  
\makeatother

\begin{center}
{\LARGE (Non)--Integrability of Geodesics in D--brane Backgrounds}
\\
\vspace{18mm}
{\bf   Yuri Chervonyi  and   Oleg Lunin}
\vspace{14mm}

Department of Physics,\\ University at Albany (SUNY),\\ Albany, NY 12222, USA\\ 

\vskip 10 mm

\blfootnote{ichervonyi@albany.edu,~olunin@albany.edu}

\end{center}

\begin{abstract}

Motivated by the search for new backgrounds with integrable string theories, we classify the D--brane geometries leading to integrable geodesics. Our analysis demonstrates that the Hamilton--Jacobi equation for massless geodesics can only separate in elliptic or spherical coordinates, and all known integrable backgrounds are covered by this separation. In particular, we identify the standard parameterization of AdS$_p\times $S$^q$ with elliptic coordinates on a flat base. We also find new geometries admitting separation of the Hamilton--Jacobi equation in the elliptic coordinates. Since separability of this equation is a necessary condition for integrability of strings, our analysis gives severe restrictions on the potential candidates for integrable string theories.

\b

\end{abstract}

\newpage

{\small
\tableofcontents
}

\newpage

\section{Introduction and summary}
\renewcommand{\theequation}{1.\arabic{equation}}
\setcounter{equation}{0}

\bigskip

Over the last two decades, AdS/CFT correspondence \cite{AdSCFT} has led to great advances in our understanding of gauge theories and string theory on Ramond--Ramond backgrounds. A special role in this progress has been played by integrability, a surprising property of field theories, which allows one to compute spectrum, correlation functions, and scattering amplitudes \cite{review_AdSCFT_int} using an infinite set of conserved charges \cite{DNW}. Originally integrable structures were discovered in the ${\cal N}=4$ super--Yang--Mills theory, but they have been  extended to other systems\footnote{The examples include the marginal deformation of ${\cal N}=4$ super--Yang--Mills \cite{margin,Frolov}, the three dimensional Chern--Simmons theory \cite{Chern} and two--dimensional CFT \cite{AdS3CFT2}.}, and it is important 
to classify field theories admitting integrability. A promising approach to such classification, which is based on analyzing behavior of strings on a dual background, has led to ruling out integrability for the superconformal theory on a quiver \cite{Zayas} and for a certain deformation of ${\cal N}=4$ SYM \cite{StepTs}\footnote{See \cite{Zayas1} for further discussion of non-integrability and chaos in the context of AdS/CFT correspondence.}. In this paper we will analyze integrability of strings on a large class of Ramond--Ramond backgrounds, rule out integrability for a wide range of field theories, and identify the potential candidates for integrable models. 

To put our results in perspective, let us briefly review the status of integrability in ${\cal N}=4$ SYM (or in string theory on $AdS_5\times S^5$). In the planar limit, the field theory can be solved by the Bethe ansatz \cite{bethe}, and the spectrum of strings on the gravity side can be found by solving the Landau--Lifshitz model \cite{TsLL}. The agreement between these two exact solutions provides a highly nontrivial check of the AdS/CFT correspondence. The methods of 
\cite{bethe,TsLL} are applicable only to the light states, whose conformal dimension obeys the relation 
\bea\label{Planar}
\Delta\ll N,
\eea
While the techniques of \cite{bethe} are not applicable when inequality (\ref{Planar}) is violated, the integrability might still persist in this case, at least for some sectors of the theory. Violation of (\ref{Planar}) implies that excitations of $AdS_5\times S^5$ might contain D--branes in addition to the fundamental strings, and generic excitations of this type are very complicated. Fortunately, some states violating the condition (\ref{Planar}) still have very simple behavior: these are the BPS states with\footnote{Here $\Delta$ is a conformal dimension of the state, and $J$ is its R charge. For simplicity we are focusing on 1/2--BPS states, but condition (\ref{NearBPS}) can be easily generalized to BPS states with lower amount of supersymmetry.} $\Delta=J$. On the gravity side of the correspondence, the BPS states are represented by supergravity modes or by D--branes, depending on the value of $J$. To have interesting dynamics, one can introduce some fundamental strings in addition to these BPS branes and to replace (\ref{Planar}) by
\bea\label{NearBPS}
\Delta-J\ll N.
\eea
As we already mentioned, the planar techniques of \cite{bethe} are not applicable to the states (\ref{NearBPS}) which violate (\ref{Planar}), and in this paper we will use alternative methods to study integrability of such states. 

The most useful version of the AdS/CFT duality involves field theory on $R\times S^3$, then the bulk configurations satisfying  (\ref{NearBPS}) are represented by fundamental strings in the presence of giant gravitons \cite{giant}. The interactions between these objects can be very complicated \cite{BalFeng}, but additional simplifications occur for semiclassical configurations of giant gravitons with $J\sim N^2$, which can be viewed as classical geometries \cite{LLM}. In this regime of parameters, integrability of the sector (\ref{NearBPS}) reduces to integrability of strings on the bubbling geometries constructed in \cite{LLM}. If the CFT is formulated on $R^{3,1}$, the counterparts of the bubbling geometries are given by brane configurations describing the Coulomb branch of ${\cal N}=4$ SYM \cite{KrausLarsen}. In the latter case, one can introduce an additional deformation which connects $AdS_5\times S^5$ asymptotics to flat space and see whether integrability persists for such configurations. 

It turns out that the answer to the last question is no, and this result discovered in \cite{StepTs} was the main motivation for our investigation. As demonstrated in \cite{StepTs}, addition of one to the harmonic function describing a single stack of D3 branes destroys integrability of the closed strings on a new asymptotically-flat background. Since continuation to the flat asymptotics destroys the dual field theory, this procedure appears to be more drastic than a transition to the Coulomb branch, which corresponds to a normalizable excitations, so the latter might have a chance to remain integrable. In this paper we focus on geometries dual to the Coulomb branch of ${\cal N}=4$ SYM (either on $R^{3,1}$ or on $R\times S^3$) and on similar geometries involving other D branes. A different class of theories, which involves putting D branes on singular manifolds, was explored in \cite{Zayas}, where it was demonstrated that strings are not integrable on the conifold. From the point of view of field theory, this result pertains to the vacuum of ${\cal N}=1$ SYM with a quiver gauge group, which is complementary to our analysis of excited states in ${\cal N}=4$ SYM.

To identify the backgrounds leading to integrable string theories, one has to analyze the equations of motion for the sigma model and to determine whether they admit an infinite set of conserved quantities. Instead of solving this complicated problem, we will focus on necessary conditions for integrability and demonstrate that strings are not integrable on a large class of backgrounds created by D--branes.
Integrability on a given background should persist for string of arbitrary size, and in the limit of point-like strings it 
leads to integrability of null geodesics\footnote{In this paper we focus on Ramond--Ramond backgrounds produced by D--branes, but in the presence of the NS--NS $B$ field, pointlike strings could carry additional charges, which modify equations for the geodesics.}, which implies that the motion of a particle is characterized by 10 conserved quantities, matching the number of the degrees of freedom $x^i$. Massless geodesics can be found by solving the Hamilton--Jacobi (HJ) equation,
\bea\label{HJone}
g^{MN}\frac{\d S}{\d X^M}\frac{\d S}{\d X^N}=0,
\eea 
where $S$ is the action of a particle, and $g_{MN}$ is the background metric. The system is called integrable if the HJ equation separates 
\cite{Arnold}, i.e., if there exists a new set of coordinates $Y^M$, such that 
\bea\label{SeparIntro}
S(Y_0,\dots Y_9)=\sum_{I=0}^9 S_I(Y_I).
\eea
This also implies that the HJ equation has ten independent integrals of motion. Non--trivial examples of geometries leading to integrable geodesics  include Kerr--Neumann black hole \cite{Carter} and its generalizations to Kerr--NUT--AdS spacetimes in higher dimensions \cite{FrolovKerrNUTAdS}. To rule out integrability of geodesics on a particular background, it is sufficient to demonstrate that separation (\ref{SeparIntro}) cannot be accomplished in any set of coordinates. 

In this paper we will analyze the motion of  massless particles in the geometries produced by stacks of parallel Dp branes and identify the distributions of branes which lead to integrable HJ equation (\ref{HJone}). Specifically, we will focus on supersymmetric configurations of D$p$--branes with flat worldvolume\footnote{In section \ref{Bubbles} we will also discuss a special class of spherical branes.}, and assume that Ramond--Ramond $(p+1)$--form sourced by the branes is the only nontrivial flux in the geometry. This implies that metric $g_{ij}$ has the form
\bea\label{PreGeomT}
g_{ij}dx^idx^j=\frac{1}{\sqrt{H}}\eta_{\mu\nu}dx^\mu dx^\nu+
\sqrt{H}ds^2_{base},
\eea
where the first term represents $(p+1)$--dimensional Minkowski space parallel the branes, and $H$ is a harmonic function on the $(9-p)$--dimensional base space. We will further assume that the base space is flat.

For a single stack of D$p$--branes, the HJ equation separates in spherical coordinates, and this well-known case is reviewed in section 
\ref{Examples}. This section also includes another example, separation in elliptic coordinates, which plays an important role in the subsequent discussion. 

Section \ref{SecGnrGds} describes our main procedure, which is subsequently used to study geodesics on a variety of backgrounds. 
In subsection \ref{RedTo2D} we demonstrate that the HJ equation (\ref{HJone}) does not separate unless the metric on the base space has the form 
\bea\label{BaseMetrIn}
ds_{base}^2=dr_1^2+dr_2^2+r_1^2d\Omega_{d_1}^2+
r_2^2d\Omega_{d_2}^2,
\eea
and $H$ depends only on $r_1$ and $r_2$. The most general harmonic function $H$ leading to integrable HJ equation is derived in section 
\ref{SectPrcdr}, and equations (\ref{TheSolnMthree})-(\ref{TheSolnMtwo}) summarize the main result of this paper for branes with flat worldvolume. The brane configurations giving rise to geometries 
(\ref{TheSolnMthree})--(\ref{TheSolnMtwo}) are analyzed in section \ref{Sources}. The results of section 
\ref{SecGnrGds} imply that $(Y_0,\dots,Y_{10})$ leading to separation (\ref{SeparIntro}) must reduce to the elliptic coordinates discussed in section \ref{Examples}.

Section \ref{SecKillWave} discusses physical properties of the geometries leading to integrable geodesics. We demonstrate that separability persists for the wave equation beyond the eikonal approximation, a property that have been observed earlier for various black holes \cite{Carter,cvetLars}. In section \ref{SecKill} we show that separability of the wave equation is associated with a hidden symmetry of the background, and we construct the conformal Killing tensor associated with this symmetry. In section \ref{NonintStr} we apply the techniques of  \cite{Zayas,StepTs} to demonstrate that most backgrounds with separable wave equation do not lead to integrable string theories. 

In section \ref{DpDp4} our results are generalized to D$p$--branes dissolved in D$(p+4)$--branes, the system which plays an important role in understanding the physics of black holes \cite{StrVafa}. We find that in asymptotically--flat space there are no integrable solutions apart from the spherically--symmetric distribution of branes. However, there are several separable configurations in the near--horizon limit of D$(p+4)$--branes, and the most general D$p$--D$(p+4)$ configurations leading to separable HJ equation are presented in (\ref{H1D1D5}), (\ref{H2D1D5}). 

In section \ref{Rotations} we consider another generalization by allowing the branes to rotate, i.e., by breaking the Poincare symmetry on the brane worldvolume. Although the general analysis of rotating branes is beyond the scope of this paper, we consider the special class of rotating solutions which cover all microscopic states of the D1--D5 black hole \cite{lmPar,lmm}. Such solutions are parameterized by curves in eight--dimensional space, and our analysis demonstrates that HJ equation is not separable unless this curve is a simple circle. For such configuration the separable coordinates have been found before \cite{MultiStr}, and we will demonstrate that these coordinates reduce to a special case of the general elliptic coordinates discussed in section \ref{SectPrcdr}.

In section \ref{Bubbles} we consider a different class of rotating solutions, which describes all half--BPS states of IIB supergravity supported by the five--form field strength \cite{LLM}. We demonstrate that there are only three bubbling solutions leading to separable HJ equation for the geodesics: 
AdS$_5\times$S$^5$, pp--wave and a geometry dual to a single M2 brane. In section \ref{StdPar} we discuss the equation for geodesics on bubbling geometries in M theory, and we demonstrate that the elliptic coordinates emerging from the separation of variables in the geometries of \cite{LLM} coincide with standard parameterization of AdS$_5\times$S$^5$, AdS$_7\times$S$^4$, and AdS$_4\times$S$^7$.

\section{Examples: spherical and elliptic coordinates}\label{Examples}
\renewcommand{\theequation}{2.\arabic{equation}}
\setcounter{equation}{0}

We begin with discussing two known examples of brane configurations which lead to integrable equations for the geodesics. First we recall the situation for a single stack of Dp branes. In this case the metric has the form
\bea\label{DbrnMetr}
ds^2&=&H^{-1/2}\eta_{\mu\nu}dx^\mu dx^\nu+H^{1/2}(dr^2+r^2 d\Omega_{8-p}^2),
\eea
where
\bea\label{HrmFncSph}
 H=a+\frac{Q}{r^{7-p}}
\eea
and $d\Omega_{8-p}^2$ is the metric on a $(8-p)$--dimensional sphere:
\bea\label{8dSphere}
d\Omega_{8-p}^2&=&h_{ij}dy^i dy^j.
\eea
The Hamilton--Jacobi equation for a particle propagating in the geometry (\ref{DbrnMetr}) has the form
\bea
\left(\frac{\d S}{\d r}\right)^2+\frac{h^{ij}}{r^2}\frac{\d S}{\d y^i}\frac{\d S}{\d y^j}+H\eta^{\mu\nu}\frac{\d S}{\d x^\mu}\frac{\d S}{\d x^\nu}=0,
\eea
and variables in this equation separate:
\bea\label{HJact}
S=p_\mu x^\mu+S_{L}(y)+R(r).
\eea 
Here
\bea\label{RHJ}
(R')^2+\frac{L^2}{r^2}+ H p_\mu p^\mu=0,\\
\label{AngHJ}
{h^{ij}}\frac{\d S_L}{\d y^i}\frac{\d S_L}{\d y^j}=L^2.
\eea
Solution (\ref{HJact}) has $(p+1)$ integrals of motion $p_\mu$, $8-p$ independent integrals coming from $S_L$ (this is ensured by the isometries of the sphere, the explicit form of $S_L$ is given in appendix \ref{AppSphere}), and one integration constant coming from the differential equation (\ref{RHJ}). 
This implies that action (\ref{HJact}) can be written in terms of $10$ conserved quantities, so the geometry (\ref{DbrnMetr}) leads to integrable geodesics. As demonstrated in \cite{StepTs}, integrability does not persist for strings, unless $a=0$ in (\ref{HrmFncSph}). Spherical coordinates (\ref{DbrnMetr}) will play an important role in our construction since any localized distribution of D branes leads to a harmonic function which approaches (\ref{HrmFncSph}) at infinity. Thus, any set of separable coordinates must reduce to (\ref{DbrnMetr}) far away from the branes. 

\begin{figure}[h!]
\centering
\includegraphics[width=0.8\textwidth]{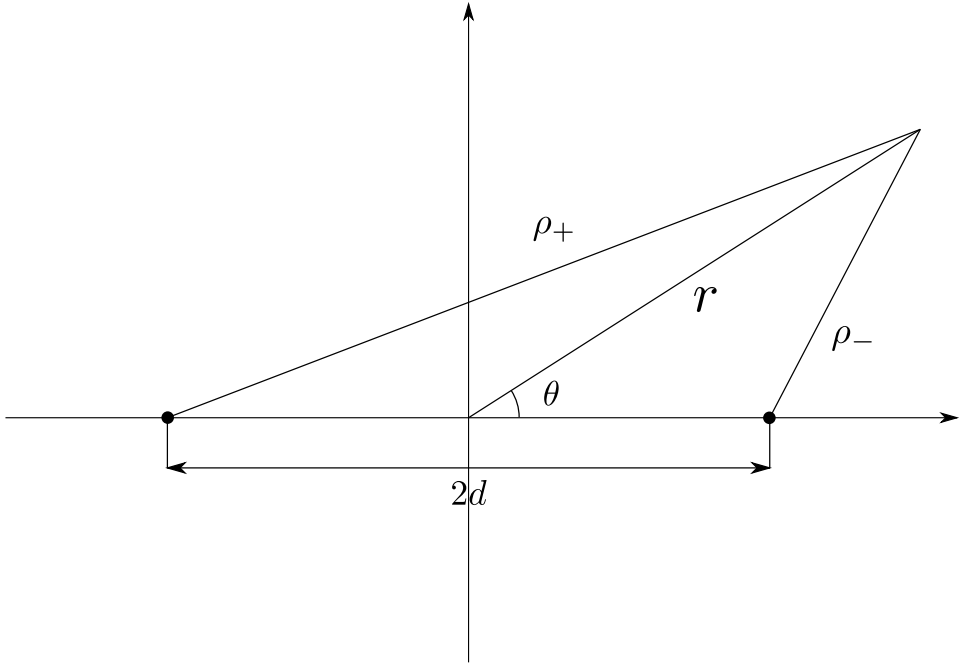} 
\caption{Geometrical meaning of $\rho_+$ and $\rho_-$ appearing in the definition of the elliptic coordinates (\ref{DefEllCoord}).}
\label{Fig:EllCoor}
\end{figure}

Our second example deals with two stacks of Dp branes separated by a distance $2d$ (see figure \ref{Fig:EllCoor}). The metric produced by this configuration is given by
\bea\label{2cntrMetr}
ds^2&=&H^{-1/2}\eta_{\mu\nu}dx^\mu dx^\nu+H^{1/2}(dr^2+r^2d\theta^2+r^2\sin^2\theta d\Omega_{7-p}^2),
\eea
where
\bea\label{EllHarm}
 H=a+\frac{Q}{\rho_+^{6-p}}+\frac{Q}{\rho_-^{6-p}},\qquad
\rho_\pm=\sqrt{r^2+d^2\pm 2rd\cos\theta}
\eea
Introducing coordinates $y_i$ on the $(7-p)$--dimensional sphere and separating variables in the action,
\bea\label{HJactEl}
S=p_\mu x^\mu+S_{L}(y)+R(r,\theta),
\eea 
we can rewrite (\ref{HJone}) as a PDE for $R(r,\theta)$:
\bea\label{EllHJ}
(\d_r R)^2+\frac{1}{r^2}(\d_\theta R)^2+\frac{L^2}{r^2\sin^2\theta}+ H p_\mu p^\mu=0
\eea
This equation describes the motion of a particle in a two--center potential, and it is well--known that for $p=5$ it can be separated further by introduction of the elliptic coordinates \cite{Landau}:
\bea\label{DefEllCoord}
\xi=\frac{\rho_++\rho_-}{2d},\qquad \eta=\frac{\rho_+-\rho_-}{2d},
\eea
In appendix \ref{SprElliptic} we review the separation procedure that leads to (\ref{HJactEl}) and write down the explicit form of $R$ (see (\ref{RActEll})).
For future reference we quote the asymptotic relation between elliptic and spherical coordinates: 
\bea\label{AsympEll}
\xi=\frac{r}{d}+\frac{d}{2r}\sin^2\theta + O\left(\frac{1}{r^3}\right),\qquad \eta=\cos\theta+\frac{d^2}{2r^2}\cos\theta\sin^2\theta
+O\left( \frac{1}{r^4} \right).
\eea

This completes our review of spherical and elliptic coordinates, and in the remaining part of this paper we will investigate whether the separation of variables persists for more general geometries produced by Dp--branes.

\section{Geodesics in D--brane backgrounds}
\label{SecGnrGds}
\renewcommand{\theequation}{3.\arabic{equation}}
\setcounter{equation}{0}

We now turn to the main topic of this paper: the analysis of geodesics in the geometry produced by D--branes:
\bea\label{PreGeom}
g_{MN}dX^MdX^N&=&\frac{1}{\sqrt{H}}\eta_{\mu\nu}dx^\mu dx^\nu+
\sqrt{H}ds^2_{base},\\
\nabla^2_{base}H&=&0\nonumber
\eea
We will further assume that the base space is flat:
\bea\label{FlatBase}
ds_{base}^2=dx^jdx^j.
\eea
The massless geodesics in the background (\ref{PreGeom}) are integrable if and only if the HJ equation (\ref{HJone}) separates in some coordinates $(Y_0,\dots,Y_{9})$, as in (\ref{SeparIntro}).
In section \ref{RedTo2D} we will use this separation and the Laplace equation 
\bea\label{LaplAA}
\frac{\d^2 H}{\d x_j\d x_j}=0
\eea
to demonstrate that function $H$ can only depend on two of the ten coordinates $(Y_0,\dots,Y_{9})$. The further analysis presented in section \ref{SectPrcdr} demonstrates that $(Y_0,\dots,Y_{9})$ must reduce to a slight generalization of the elliptic coordinates presented in section \ref{Examples}. Although the Laplace equation (\ref{LaplAA}) is satisfied away from the branes, any nontrivial function $H$ must have sources, and the brane configurations leading to a separable HJ equation are analyzed in section \ref{Sources}.

\subsection{Reduction to two dimensions}
\label{RedTo2D}

Let us assume the separation (\ref{SeparIntro}) in the HJ equation (\ref{HJone}) and explore the consequences for the harmonic function $H$ appearing in (\ref{PreGeom}). Metric (\ref{FlatBase}) is invariant under $SO(9-p)$ transformations, but part of this rotational symmetry might be broken by the harmonic function $H$. Let $SO(d_1+1)\times SO(d_2+1)\times SO(d_k+1)$ be the maximal subgroup of $SO(9-p)$ preserved by $H$, then the metric on the base space can be written as
\bea\label{BaseMetr}
ds_{base}^2=dr_1^2+r_1^2d\Omega_{d_1}^2+\dots+
dr_k^2+r_k^2d\Omega_{d_k}^2,
\eea
and $H$ becomes a function of $(r_1,\dots,r_k)$.  Moreover, since all rotational symmetries have been isolated, we conclude that\footnote{If this relation is not satisfied for $i=1$, $j=2$, then $SO(d_1+1)\times SO(d_2+1)$ is enhanced to $SO(d_1+d_2+2)$.}
\bea\label{NoRotation}
(r_i\d_j-r_j\d_i)H\ne 0.
\eea
If the branes are localized in a compact region, then at sufficiently large values of 
$$R=\sqrt{r_1^2+\dots+r_k^2}$$ 
function $H$ satisfies the Laplace equation
\bea\label{LaplRk}
\frac{1}{r_1^{d_1}}\frac{\d}{\d r_1}\left[r_1^{d_1}\frac{\d H}{\d r_1}\right]+\dots+
\frac{1}{r_k^{d_k}}\frac{\d}{\d r_k}\left[r_k^{d_k}\frac{\d H}{\d r_k}\right]=0,
\eea
and asymptotic behavior of function $H$ is given by 
\bea\label{Hasymt}
H\sim a+\frac{Q}{(r_1^2+\dots+r_k^2)^{q}}.
\eea
Here $a$ is a parameter, which is equal to zero for the near--horizon geometries and which can be set to one for asymptotically--flat solutions. 

Our goal is to classify the backgrounds (\ref{PreGeom}) that lead to separable HJ equations for geodesics, and we begin with separating variables associated with symmetries. Poincare invariance of (\ref{PreGeom}) and rotational invariance of the base metric (\ref{BaseMetr}) allow us to write the action appearing in the HJ equation (\ref{BaseMetr}) as
\bea\label{GenAction}
S=p_\mu x^\mu+\sum_{i=1}^k S^{(d_i)}_{L_i}(\Omega_{d_i})+
{\tilde S}(r_1,\dots r_k).
\eea
Here $p_\mu$ is the momentum of the particle in $p+1$ directions longitudinal to the branes, and 
$S^{(d_i)}_{L_i}(\Omega_{d_i})$ is the part of the action that depends on coordinates $y_1,\dots,y_{d_i}$ of the sphere $\Omega_{d_i}$. The label $L_i$ represents the angular momentum of a particle along this sphere, and it is defined by relation
\bea\label{AngHJaa}
{h^{ij}}\frac{\d S_L}{\d y^i}\frac{\d S_L}{\d y^j}=L_i^2.
\eea
An explicit construction of $S^{(d)}_{L}(\Omega)$ is presented in appendix \ref{AppSphere}. 

Substitution of (\ref{GenAction}) in the HJ equation (\ref{HJone}) leads to equation for ${\tilde S}$:
\bea\label{HJaa}
\sum_{i=1}^k\left[\left(\frac{\d {\tilde S}}{\d r_i}\right)^2+\frac{L_i^2}{r_i^2}\right]+ H p_\mu p^\mu=0.
\eea
A special case of this equation with $k=1$ separates in spherical coordinates (see section \ref{Examples}), and we will now prove the equation (\ref{HJaa}) does not separate if $k>2$. The separation of (\ref{HJaa}) for $k=2$ will be discussed in section \ref{SectPrcdr}. 

Separability of equation (\ref{HJaa}) should persist for all values of angular momenta, so we begin with setting all $L_j$ to zero and $p_1=\dots=p_p=0$. The resulting equation (\ref{HJaa}) can be viewed as a HJ equation on an effective $(k+1)$--dimensional space
\bea\label{trrMetr}
ds^2=-H^{-1}dt^2+dr_1^2+\dots+dr_k^2
\eea
A general theory of separable HJ equations on curved backgrounds has a long history (see \cite{MorseFesh}), and a complete classification is presented in \cite{KalMil1, KalMil2}. In particular, this theory distinguishes between ignorable directions (which correspond to Killing vectors) and non-ignorable ones. Clearly, the time direction in (\ref{trrMetr}) is ignorable, but (\ref{trrMetr}) does not have additional Killing vectors which commute with $\d_t$. Indeed, any such vector would be a Killing vector of the $k$--dimensional flat space, so it must be a combination of translations and rotations in $r_i$. However, the asymptotic behavior of function $H$ (\ref{Hasymt}) breaks translational symmetry, and our assumption (\ref{NoRotation}) destroys the rotational Killing vectors, so if the HJ equation for the metric (\ref{trrMetr}) separates in some coordinates 
$(t,x_1,\dots x_k)$, only one of them (specifically, $t$) can correspond to an ignorable direction. Moreover, the discrete symmetry $t\rightarrow -t$ of  (\ref{trrMetr}) guarantees that $t$ does not mix with 
$(x_1,\dots x_k)$ in the metric, and such orthogonality leads to simplifications in the general analysis of \cite{KalMil1, KalMil2}.

Specifically, according to theorem 6 of \cite{KalMil2} separation of variables in (\ref{trrMetr}) implies that\footnote{The discussion of \cite{KalMil1,KalMil2} is more general: it allows mixing between ignorable and essential coordinates.} 
\begin{enumerate}[(1)]
\item{There exist $k$ independent conformal Killing tensors $A^{(a)}$ with components $(A^{(a)}_{ij},A^{(a)}_{tt})$.}
\item{Each of the one--forms $dx^l=\omega^l_i dr^i$ is a simultaneous eigenform of all $A_{(a)}^{ij}$ with eigenvalues 
$\rho_{(a,l)}$. This implies that  a projector ${P^{(l)i}}_j$ onto $dx^l$ satisfies equation
\bea\label{KMeigen}
(A_{(a)}^{ij}-\rho_{(a,l)}g^{ij}){P^{(l)j}}_k=0.
\eea
}
\item{The metric in coordinates $dx^l$ is diagonal, so projectors ${P^{(l)i}}_j$ commute with $A_{(a)}^{ij}$ and 
$g^{ij}$, and projectors with different values of $l$ project onto orthogonal subspaces. Since the number of projectors is equal to the number of coordinates, we arrive at the decomposition
\bea\label{KMdecomp}
A_{(a)}^{ij}=\sum_l^k \rho_{(a,l)}h_l P_{(l)}^{ij},\quad g^{ij}=\sum_l^k h_l P_{(l)}^{ij}
\eea }
\item{The components of $A_{(a)}$ along the Killing direction 
satisfy an overdefined system of differential equations:
\bea\label{KMProj}
\d_i\left[A_{(a)}^{tt}\right]-\sum_l^k 
\rho_{(a,l)}{P_i}^{(l)j}\d_j g^{tt}=0
\eea
}
\end{enumerate}
Notice that in \cite{KalMil2} the theorem is formulated in terms of coordinates $x_i$, so it does not use the projectors. For our purposes the covariant formulation given above is more convenient, in particular, to rule our the separation of variables, we will have to work in the original coordinates $r_i$ and demonstrate that the required Killing tensors $A_{(a)}$ do not exist. Notice that equation (\ref{KMProj}) can be rewritten in the form which does not refer to projectors (and thus to coordinates $x_i$): multiplying this equation by $g^{ij}$ and using (\ref{KMdecomp}), we find
\bea\label{kkk}
g^{ij}\d_j(A_{(a)}^{tt})-A_{(a)}^{ij}\d_j g^{tt}=0.
\eea
This relation is equivalent to $(tti)$ component of the equation for the Killing tensor $A_{(a)}$. 

The theorem quoted above implies that $A^{(a)}_{ij}$ are conformal Killing tensors on the $k$--dimensional base of (\ref{trrMetr}), and, for the flat base, all such tensors can be written as quadratic combinations of $k(k+3)(k+4)(k+5)/12$ Killing vectors \cite{Weir}\footnote{The explicit form of the conformal Killing vectors is given in appendix \ref{AppFltKil}.}:
\bea
A^{(a)}_{ij}=\sum_{m,n}b^{(a)}_{m,n} V^{(m)}_i
V^{(n)}_j
\eea
Equations (\ref{KMeigen}) and (\ref{kkk}) give severe restrictions on coefficients $b^{(a)}_{m,n}$, but fortunately the consequences of (\ref{KMeigen}) have been analyzed elsewhere. Indeed, equation (\ref{KMeigen}) does not involve $g_{tt}$, so it remains the same for $H=1$, when (\ref{trrMetr}) gives the flat space, and the corresponding HJ equation gives an eikonal approximation for the standard wave equation. It is well-known that in $3+1$ dimensions ($k=3$) the latter can only be separated in ellipsoidal coordinates and their special cases \cite{MorseFesh}, and generalization of this result to $k>3$ is presented in \cite{KalMil1, KalMil2}. This leads to the conclusion that the HJ equation in the metric (\ref{trrMetr}) with $k>2$ can only separate in ellipsoidal coordinates or in the degenerate form thereof. Before ruling out this possibility, we briefly comment on the peculiarities of the two--dimensional base. In this case the conformal group becomes infinite-dimensional, 
so the base space admits an infinite number of the conformal Killing tensors. This situation will be analyzed in section \ref{SectPrcdr}.

To summarize, we concluded that for $k>2$, the HJ equation can only be separable in some special case of ellipsoidal coordinates $(x_1,\dots,x_k)$, which are defined by \cite{Jacobi,Miller4}
\bea\label{ref317}
&&r_i^2=-\left[\prod_j(a_i^2+x_j)\right]\left[\prod_{j\ne i}\frac{1}{(a_j^2-a_i^2)}\right].
\eea
Here $(a_1,\dots,a_k)$ is the set of positive constants, which specify the ranges of variables $x_i$:
\bea
x_1\ge -a_1^2\ge x_2\ge -a_2^2\ge \dots x_k\ge -a_k^2,
\eea 
Rewriting the metric (\ref{trrMetr}) in terms of $x_i$ and substituting the result into (\ref{HJaa}), we find the HJ equation in ellipsoidal coordinates (see appendix \ref{AppEllips} for detail):
\bea\label{HJelps}
\sum_{i=1}^k\left[\frac{1}{h_i^2}\left(\frac{\d {\tilde S}}{\d x_i}\right)^2+\frac{L_i^2}{r_i^2}\right]+ H p_\mu p^\mu=0.
\eea
Here $h_i$ is defined by 
\bea
h_i^2=\frac{1}{4}\left[\prod_{j\ne i}(x_i-x_j)\right]\left[\prod_j\frac{1}{a_j^2+x_i}\right].
\eea
Function $H$ appearing in (\ref{HJelps}) must satisfy equation (\ref{LaplRk}) away from the sources, and appendix \ref{AppEllips} we demonstrate that for such functions equation 
(\ref{HJelps}) never separates in ellipsoidal coordinates. This shows that the HJ equation can only be integrable for $k=1$ (the situation considered in section \ref{Examples}) and for $k=2$, which will be analyzed in the next subsection.

\subsection{Separation of variables and elliptic coordinates}
\label{SectPrcdr}

In the last subsection we have demonstrated that the HJ equation (\ref{HJaa}) is not integrable unless the flat base has the form (\ref{BaseMetr}) with $k\le 2$ and $H$ is a function of $r_1$ and $r_2$ only. In section \ref{Examples} we have already discussed $k=1$ and this subsection is dedicated to the analysis of $k=2$. To simplify some formulas, we slightly deviate from the earlier notation and write the metric (\ref{PreGeom}) with the base (\ref{BaseMetr}) for $k=2$ as
\bea\label{SSmetr}
ds^2=\frac{1}{\sqrt{H}}\eta_{\mu\nu}dx^\mu dx^\nu+
\sqrt{H}(dr^2+r^2d\theta^2+r^2\cos^2\theta d\Omega_{m}^2+r^2\sin^2\theta d\Omega_{n}^2),
\eea
The connection to coordinates of (\ref{BaseMetr}) is obvious:
\bea
r_1=r\cos\theta,\quad r_2=r\sin\theta,
\eea
with $d_1=m, d_2=n$. The arguments presented in the last subsection ensure that $H$ appearing in (\ref{SSmetr}) can only depend on $r$ and $\theta$, i.e., the distribution of Dp branes that sources this harmonic function is invariant under $SO(m+1)\times SO(n+1)$ rotations. Notice that
\bea\label{MNasPQ}
p=7-m-n.
\eea

The Poincare and $SO(m+1)\times SO(n+1)$ symmetries of (\ref{SSmetr}) lead to a partial separation of the Hamilton--Jacobi equation (\ref{HJone}) for geodesics (this equation is a counterpart of (\ref{GenAction})): 
\bea\label{GenSeparAct}
S=p_\mu x^\mu+S^{(m)}_{L_1}(y)+S^{(n)}_{L_2}({\tilde y})+R(r,\theta),
\eea
where $S^{(m)}_{L_1}$ and $S^{(n)}_{L_2}$ satisfy differential equations
\bea\label{HJSphere}
{h^{ij}}\frac{\d S^{(m)}_{L_1}}{\d y^i}\frac{\d S^{(m)}_{L_1}}{\d y^j}=L_1^2,\qquad
{{\tilde h}^{ij}}\frac{\d S^{(n)}_{L_2}}{\d {\tilde  y}^i}\frac{\d S^{(n)}_{L_2}}{\d{\tilde y}^j}=L_2^2,
\eea
and $L_1$, $L_2$ are angular momenta on the spheres. An explicit solution of equations (\ref{HJSphere}) is presented in appendix 
\ref{AppSphere}.

Substitution of (\ref{GenSeparAct}) into (\ref{HJone}) leads to the equation for $R(r,\theta)$:
\bea\label{TheHJ}
(\d_r R)^2+\frac{1}{r^2}(\d_\theta R)^2+
\frac{L_1^2}{r^2\cos^2\theta}+
\frac{L_2^2}{r^2\sin^2\theta}+
 p_\mu p^\mu H(r,\theta)=0.
\eea
We recall that $H(r,\theta)$ is a harmonic function describing the distribution of D branes, so away from the sources it satisfies 
the Laplace equation on the base of the ten--dimensional metric (\ref{SSmetr}):
\bea\label{LaplEqn}
\frac{1}{r^{m+n+1}}\d_r(r^{m+n+1}\d_r H)+\frac{1}{r^2\sin^{n}\theta\cos^{m}\theta}
\d_\theta(\sin^{n}\theta\cos^{m}\theta\d_\theta H)=0.
\eea

Let us assume that the massless HJ equation (\ref{TheHJ}) separates in coordinates $(x_1,x_2)$. In particular, this implies that the metric in $(x_1,x_2)$ must have a form \cite{Stackel,Eisenhart}
\bea\label{StackMetr}
dr^2+r^2d\theta^2=A(x_1,x_2)\left[e^{g_1}(dx_1)^2+
e^{g_2}(dx_2)^2\right],
\eea
where $g_1(x_1,x_2)$ and $g_2(x_1,x_2)$ satisfy the St\"ackel conditions:
\bea\label{Stackel}
\d_j\d_i g_l-\d_j g_l\,\d_i g_l+\d_j g_l\,\d_i g_j+
\d_i g_l\,\d_j g_i=0,\quad i\ne j
\eea
We wrote the St\"ackel conditions (\ref{Stackel}) for an arbitrary number of coordinates to relate to our discussion in section 
\ref{RedTo2D}, but in the present case ($k=2$) equation 
(\ref{Stackel}) gives only two relations ($i=1$, $j=2$, $l=1,2$):
\bea\label{Stack2D}
\d_2\d_1 g_1+\d_2 g_1\,\d_1 g_2=0,\qquad
\d_2\d_1 g_2+\d_1 g_2\,\d_2 g_1=0
\eea
In particular, we find that
\bea
g_1-g_2=f_1(x_1)-f_2(x_2),
\eea
so by adjusting function $A$ in (\ref{StackMetr}) we can set
\bea
g_1=f_1(x_1),\qquad g_2=f_2(x_2).
\eea
We can further redefine variables, 
$x_1\rightarrow {\tilde x}_1(x_1)$, $x_2\rightarrow {\tilde x}_2(x_2)$  to set $g_1=g_2=0$, at least locally.\footnote{Notice that separation in variables $(x_1,x_2)$ implies separation in 
$({\tilde x}_1,{\tilde x}_2)$.} Introducing $x={\tilde x}_1$, $y={\tilde x}_2$, we rewrite (\ref{StackMetr}) as 
\bea\label{RThAxY}
dr^2+r^2d\theta^2=A(x,y)\left[dx^2+dy^2\right].
\eea
To summarize, we have demonstrated that in two dimensions the St\"ackel conditions (\ref{Stackel}) imply that coordinates $x_i$ separating the HJ equation are essentially the same as the conformally--Cartesian coordinates $(x,y)$ in (\ref{RThAxY})\footnote{Although any two--dimensional metric can be written as $A[(d{\tilde x}_1)^2+(d{\tilde x}_2)^2]$ in {\it some} coordinates, a priori the HJ equation does not have to separate in $({\tilde x}_1,{\tilde x}_2)$. It is the St\"ackel conditions (\ref{Stack2D}) that guarantee that any set of separable coordinates can be rewritten in a conformally--flat form without destroying the separation.}. In higher dimensions, conditions (\ref{Stackel}) are less stringent than the requirement for coordinates $x_i$ to be conformally--Cartesian: for example, conditions (\ref{Stackel}) are satisfied by spherical coordinates that have
\bea
g_1=0,\quad g_2=2\ln  x_1,\quad g_2=2\ln (x_1\sin x_2),
\eea
but there is no change of coordinates of the form 
${\tilde x}_i(x_i)$ that allows one to write 
\bea
(dx_1)^2+x_1^2dx_2^2+[x_1\sin x_2]^2 (dx_3)^2=
A\sum (d{\tilde x}_i)^2.
\eea
Moreover, for $k>2$, a relation 
\bea\label{Jul14}
\sum_{i=1}^k (dr_i)^2=A\sum_{i=1}^k (d{\tilde x}_i)^2
\eea
implies that $A$ must be equal to constant, so if conformally--Cartesian coordinates ${\tilde x}_i$ separate the HJ equation, then 
${\tilde x}_i$ must be Cartesian\footnote{To see this, one has to evaluate the Riemann tensor for both sides of (\ref{Jul14}).}.

Returning to $k=2$, we will now find restrictions on $A(x,y)$ and $H(r,\theta)$. First we define ${\tilde R}(x,y)$ by
\bea
{\tilde R}(x,y)\equiv R(r,\theta).
\eea
Then equation (\ref{RThAxY}) can be used to rewrite (\ref{TheHJ}) in terms of ${\tilde R}$, and we find the necessary and sufficient conditions for the Hamilton--Jacobi equation (\ref{TheHJ}) to be separable:
\bea\label{XYdiff}
&&(\d_r R)^2+\frac{1}{r^2}(\d_\theta R)^2=\frac{1}{A(x,y)}\left[(\d_x {\tilde R})^2+(\d_y {\tilde R})^2\right],
\\
\label{XYalg}
&&\frac{L_1^2}{r^2\cos^2\theta}+
\frac{L_2^2}{r^2\sin^2\theta}
-M^2 H(r,\theta)=\frac{1}{A(x,y)}\left[U_1(x)+U_2(y)\right].
\eea
Here $M$ is an effective mass in $(p+1)$ dimensions defined by
\bea
M^2=-p_\mu p^\mu.
\eea

The construction of the most general harmonic function 
$H(r,\theta)$ that admits the separation of variables (\ref{XYdiff})--(\ref{XYalg}) will be performed in three steps:
\begin{enumerate}
\item{Determine the restrictions on function $A(x,y)$ imposed by equation (\ref{XYdiff}).}
\item{Use equation (\ref{XYalg}) to find $H(r,\theta)$ corresponding to a given $A(x,y)$.}
\vskip -1.5cm
\bea\label{ThreeSteps}
\eea
\item{\vskip -0.2cm Use the Laplace equation (\ref{LaplEqn}) to find further restrictions on $A(x,y)$.}
\end{enumerate}
To implement the first step, it is convenient to introduce complex variables
\bea\label{defCmplVar}
z=x+iy,\qquad w=\ln \frac{r}{l}+i\theta.
\eea  
Here $l$ is a free parameter which has dimension of length. 
Rewriting equation (\ref{XYdiff}) in terms of complex coordinates, 
\bea\label{FromZtoW}
\d_w R\d_{\bar w}R=\frac{l^2e^{w+{\bar w}}}{A}\d_z {\tilde R}\d_{\bar z}{\tilde R}.
\eea
we conclude  that\footnote{The alternative solution, $\d_z{\tilde R}=\d_{\bar z}{\tilde R}=0$, leads to ${\tilde R}=R=\mbox{const}$, which does not solve (\ref{TheHJ}).}
\bea
\frac{\d z}{\d w}\frac{\d { z}}{\d {\bar w}}=0,
\eea
so $z(w,{\bar w})$ is either holomorphic or anti--holomorphic. Without loss of generality we assume 
that
\bea\label{holomZ}
z=h(w),
\eea 
then equation (\ref{FromZtoW}) gives an expression for $A(x,y)$:
\bea
A=\frac{l^2e^{w+{\bar w}}}{|h'|^2}=\frac{r^2}{|h'|^2}
\eea

The second step amounts to rewriting equation (\ref{XYalg}) as
\bea\label{HarmRTheta}
H(r,\theta)=
\frac{1}{M^2}\left[
\frac{L_1^2}{r^2\cos^2\theta}+
\frac{L_2^2}{r^2\sin^2\theta}-
\frac{|h'|^2}{r^2}\left[
U_1\left(\frac{h+{\bar h}}{2}\right)+
U_2\left(\frac{h-{\bar h}}{2i}\right)\right]
\right].
\eea

Implementation of the third step amounts to finding expressions for $h(w)$ and $U_1(x)$, $U_2(y)$ which are consistent with Laplace equation (\ref{LaplEqn}) for function $H$. Physically interesting configurations correspond to branes distributed in a compact spacial region, so at large values of $r$ function $H$ behaves as 
\bea\label{LeadHarm}
H=a+\frac{Q}{r^{7-p}}+O(r^{p-8}).
\eea
Here $Q$ is the total brane charge, and $a$ is a parameter, which can be set to zero for asymptotically flat space, and which is equal to zero for the near--horizon geometry of branes (cf. equation (\ref{HrmFncSph})). Keeping only the two leading terms in (\ref{LeadHarm}), we conclude that the Hamilton--Jacobi equation (\ref{TheHJ}) separates in coordinates $(r,\theta)$, and the solution is  similar to the first example discussed in section \ref{Examples}. The subleading corrections in (\ref{LeadHarm}) obstruct the separation in spherical coordinates, but for some harmonic functions $H$ the Hamilton--Jacobi equation (\ref{TheHJ}) separates in a different coordinate system, as in the second example discussed in section \ref{Examples}. Notice that at large values of $r$ the new coordinate system must approach spherical coordinates since (\ref{LeadHarm}) depends only on $r$, and for elliptic coordinates such asymptotic reduction was given by equation (\ref{AsympEll}). 

In appendix \ref{AppPrcdr} we construct the most general expressions for $h(w)$, $U_1(x)$, $U_2(y)$ by starting with asymptotic relations (\ref{LeadHarm}) and
\bea\label{BoundCond}
z=w+O\left(\frac{l}{r}\right),
\eea
and writing expansions in powers of $l/r$. Notice that these asymptotics lead to the unique value of $l$ for a given configuration of branes (for example, $l=d/2$ in (\ref{ElliptCase})). Requiring that function (\ref{HarmRTheta}) satisfies the Laplace equation 
(\ref{LaplEqn}), the boundary condition (\ref{LeadHarm}), and remains regular at sufficiently large $r$, we find three possible expressions for $h$ and $U_1$, $U_2$:\footnote{To avoid unnecessary complications in (\ref{TheSolnMthree})--(\ref{TheSoln67}) we set $a=0$ in these expressions, but constant $a$ can be added to the harmonic function without destroying the separation (\ref{XYalg}). This leads to minor changes in $U_1$ and $U_2$.}
\bea
\label{TheSolnMthree}
&&\hskip -1.4cm \mbox{I}:h(w)=w,\quad
H=\frac{Q}{r^{m+n}},\quad
{U}_1(r)=-\frac{QM^2}{r^{m+n-2}},\quad 
{U}_2(\theta)=\frac{L_1^2}{\cos^2 \theta}+\frac{L_2^2}{\sin^2 \theta}.\\
&&\phantom{\frac{e^x}{r^t}}\nonumber\\
\label{TheSolnMone}
&&\hskip -1.4cm \mbox{II}:h(w)=\ln\left[\frac{1}{2}\left\{e^w+\sqrt{e^{2w}-4}\right\}\right],\quad
H=\frac{1}{(\cosh^2 x-\cos^2 y)}
\frac{\tilde Q}{\sinh^{n-1}x
\cosh^{m-1}x}.
\nonumber\\
&&\hskip -1.2cm{U}_1(x)=-\frac{{\tilde Q}(Md)^2}{\sinh^{n-1}x
\cosh^{m-1}x}-\frac{L_1^2}{\cosh^2 x}+
\frac{L_2^2}{\sinh^2 x},\quad 
{U}_2(y)=\frac{L_1^2}{\cos^2 y}+\frac{L_2^2}{\sin^2 y}.\\
&&\phantom{\frac{e^x}{r^t}}\nonumber\\
\label{TheSolnMtwo}
&&\hskip -1.4cm \mbox{III}:h(w)=\ln\left[\frac{1}{2}\left\{e^w+\sqrt{e^{2w}+4}\right\}\right],\quad
H=\frac{1}{(\cosh^2 x-\sin^2 y)}
\frac{\tilde Q}{\sinh^{m-1}x
\cosh^{n-1}x}.
\nonumber\\
&&\hskip -1.2cm{U}_1(x)=-\frac{{\tilde Q}(Md)^2}{\sinh^{m-1}x
\cosh^{n-1}x}-\frac{L_2^2}{\cosh^2 x}+
\frac{L_1^2}{\sinh^2 x},\quad 
{U}_2(y)=\frac{L_1^2}{\cos^2 y}+\frac{L_2^2}{\sin^2 y}.
\eea
~\\
The derivation of these constraints is presented in appendix \ref{AppPrcdr}. 

For six-- and seven--branes\footnote{Condition (\ref{MNasPQ}) as well as restrictions $m,n\ge 0$ imply that ansatz (\ref{SSmetr}) covers only $p\le 7$} (i.e., for $m+n<2$), harmonic function is slightly more general:
\bea\label{TheSoln67}
\mbox{I}:&&H=-\frac{1}{M^2r^2}\left[\frac{C_1}{r^{m+n-2}}+\frac{C_3}{\sin^{n-1}\theta\cos^{m-1}\theta}\right.\nonumber\\
&&\qquad\qquad\qquad\left.
+\frac{C_4}{\sin^{n-1}\theta} F\left(\frac{m-1}{2},
\frac{3-n}{2},\frac{m+1}{2};\cos^2\theta\right)\right]\nonumber
 \nonumber\\
\mbox{II}:&&H=\frac{1}{(\cosh^2 x-\cos^2 y)}\left[
\frac{\tilde Q}{\sinh^{n-1}x
\cosh^{m-1}x}+\frac{P}{\sin^{m-1}y\cos^{n-1}y}\right],\\
\mbox{III}:&&H=\frac{1}{(\cosh^2 x-\sin^2 y)}\left[
\frac{\tilde Q}{\sinh^{m-1}x
\cosh^{n-1}x}+\frac{P}{\cos^{m-1}y\sin^{n-1}y}\right].\nonumber
\eea
The terms proportional to $P$ are ruled out by the boundary condition (\ref{LeadHarm}) for $p<6$, but they are allowed for $p=6,7$. 

\bigskip

Notice that case III can be obtained from case II by interchanging the spheres $S^m$  and $S^n$, so, without loss of generality, we can focus on solutions I and II. Case I corresponds to spherical coordinates, and case II corresponds to elliptic coordinates discussed in the second example of section \ref{Examples}: as demonstrated in appendix \ref{AppPrcdr}, expression (\ref{TheSolnMone}) for $h(w)$,
\bea\label{HolomElpt}
h(w)=\ln\left[\frac{1}{2}\left\{e^w+\sqrt{e^{2w}-4}\right\}\right],
\eea
is equivalent to 
\bea\label{ElliptCase}
\cosh x=\frac{\rho_++\rho_-}{2d},\quad \cos y=\frac{\rho_+-\rho_-}{2d},\quad
\rho_\pm=\sqrt{r^2+d^2\pm 2rd\cos\theta},\quad
d=2l.
\eea
Thus $(x,y)$ are equivalent to the elliptic coordinates $(\xi,\eta)$ defined by (\ref{DefEllCoord}). For future reference, we rewrite the metric (\ref{SSmetr}) in terms of $x$ and $y$ 
(see (\ref{metrAA})):
\bea\label{SSmetrAA}
ds^2&=&\frac{1}{\sqrt{H}}\eta_{\mu\nu}dx^\mu dx^\nu+
\sqrt{H}ds_{base}^2,
\\
ds_{base}^2&=&
4l^2\left[(\cosh^2x-\cos^2y)(dx^2+dy^2)+
\sinh^2 x\sin^2 y d\Omega_{n}^2+
\cosh^2 x\cos^2 y d\Omega_{m}^2\right].\nonumber
\eea

\subsection{Properties of the brane sources}\label{Sources}

In section \ref{SectPrcdr} we have classified the geometries which lead to separable Hamilton--Jacobi equations for geodesics. The Laplace equation (\ref{LaplEqn}) played an important role in our construction, but this equation is only satisfied away from the sources. In this subsection we will analyze the solutions (\ref{TheSolnMthree})--(\ref{TheSolnMtwo}) to find the sources of the Poisson equation and to identify the corresponding distribution of branes. As already discussed in section \ref{Examples}, the spherically symmetric distribution (\ref{TheSolnMthree}) corresponds to a single stack of Dp branes. 

Since solution (\ref{TheSolnMtwo}) can be obtained from (\ref{TheSolnMone}) by interchanging $S^m$ and $S^n$, it is sufficient to discuss only (\ref{TheSolnMone}). Function $H$ defined by (\ref{TheSolnMone}) becomes singular at
\bea\label{SingOne}
x=0,\quad y=0
\eea
for all values of $(m,n)$, at\footnote{Recall that coordinate $\theta$ in the metric (\ref{SSmetr}) is bounded by $\pi$ if $m=0$ or by $\pi/2$ if $m>0$, and ranges of $y$ and $\theta$ are the same.}
\bea\label{SingTwo}
x=0,\quad y=\pi\mbox{\ \ if\ \ }m=0,
\eea
and at 
\bea\label{SingThree}
x=0\mbox{\ \ if\ \ }n>1.
\eea
The first condition (\ref{SingOne}) implies that $e^z=e^{x+iy}=1$, and since $z=h(w)$,  it can be rewritten as
\bea
e^w+\sqrt{e^{2w}-4}=2
\eea
using $h(w)$ from (\ref{TheSolnMone}). Solving this equation and recalling the definition of $w$ (\ref{defCmplVar}), we find the first singular locus, which is present for all values of $(m,n)$:
\bea\label{EndPoints}
r=2l,\quad \theta=0.
\eea

For $m=0$, we have an additional locus (\ref{SingTwo}), and repeating the steps above, we find a counterpart of (\ref{EndPoints}):
\bea\label{EndPointOne}
m=0:\quad r=2l,\quad \theta=\pi.
\eea
Equations (\ref{EndPoints}) and (\ref{EndPointOne}) describe two point--like sources, and we
have already encountered these points in the original elliptic coordinates discussed in section \ref{Examples}. 
In the remaining part of this section we will focus on $m>0$. 

For $n=0,1$, the $m$--dimensional sphere described by (\ref{EndPoints}) is the only singularity of the harmonic function, and for $n>1$ there is an additional locus given by (\ref{SingThree}). To formulate (\ref{SingThree}) in terms of $r$ and $\theta$, we first use the expression 
(\ref{TheSolnMone}) for $h(w)$, to rewrite  (\ref{SingThree}) as
\bea\label{Jun11}
x=\mbox{Re}\left\{\ln\left[\frac{1}{2}\left\{e^w+\sqrt{e^{2w}-4}\right\}\right]\right\}=0
\eea
The last relation can be rewritten as
\bea
\left|e^w+\sqrt{e^{2w}-4}\right|=2\quad
\Rightarrow\quad
\left|re^{i\theta}+\sqrt{r^2e^{2i\theta}-4l^2}\right|=2l,
\eea
so there must exist an angle $\psi$, such that
\bea
r+\sqrt{r^2-4l^2e^{-2i\theta}}=2le^{i\psi}.
\eea
Solving the last equation for $r$, we find
\bea
r=l(e^{i\psi}+e^{-2i\theta-i\psi})
\eea
The right--hand side of the last equation must be real, 
this implies that (\ref{SingThree}) is equivalent to
\bea\label{Rpsi}
n>1:\qquad\theta=0,\qquad r=2l\cos\psi
\eea
Substituting this relation into (\ref{ElliptCase}) and recalling that $d=2l$, we conclude that $\psi=y$. 

\begin{figure}
\begin{minipage}[h!]{0.4\linewidth}
\center{\includegraphics[width=1\linewidth]{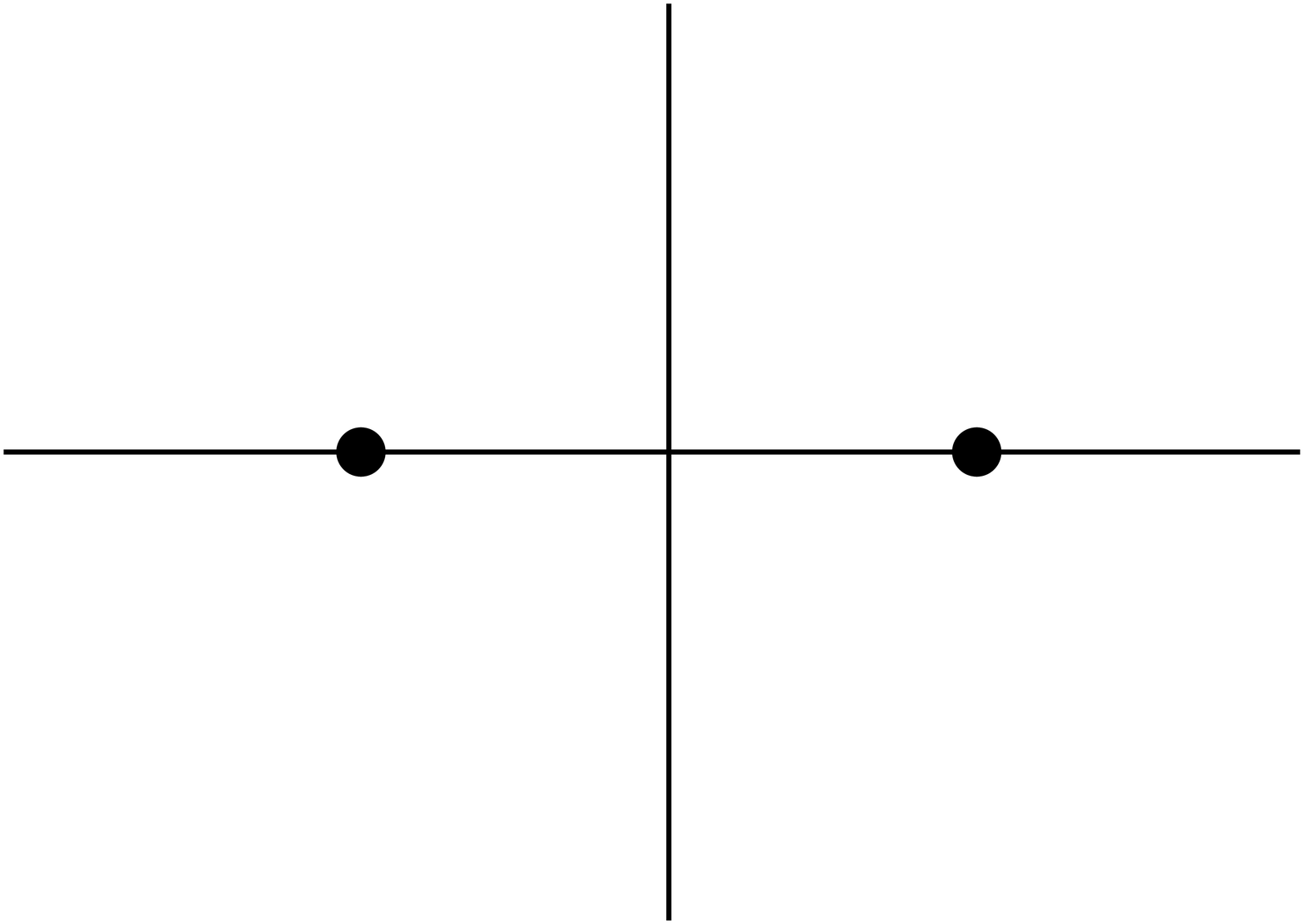}} (a) $m=0,n\le1$ \\
\end{minipage}
\hfill
\begin{minipage}[h!]{0.4\linewidth}
\center{\includegraphics[width=1\linewidth]{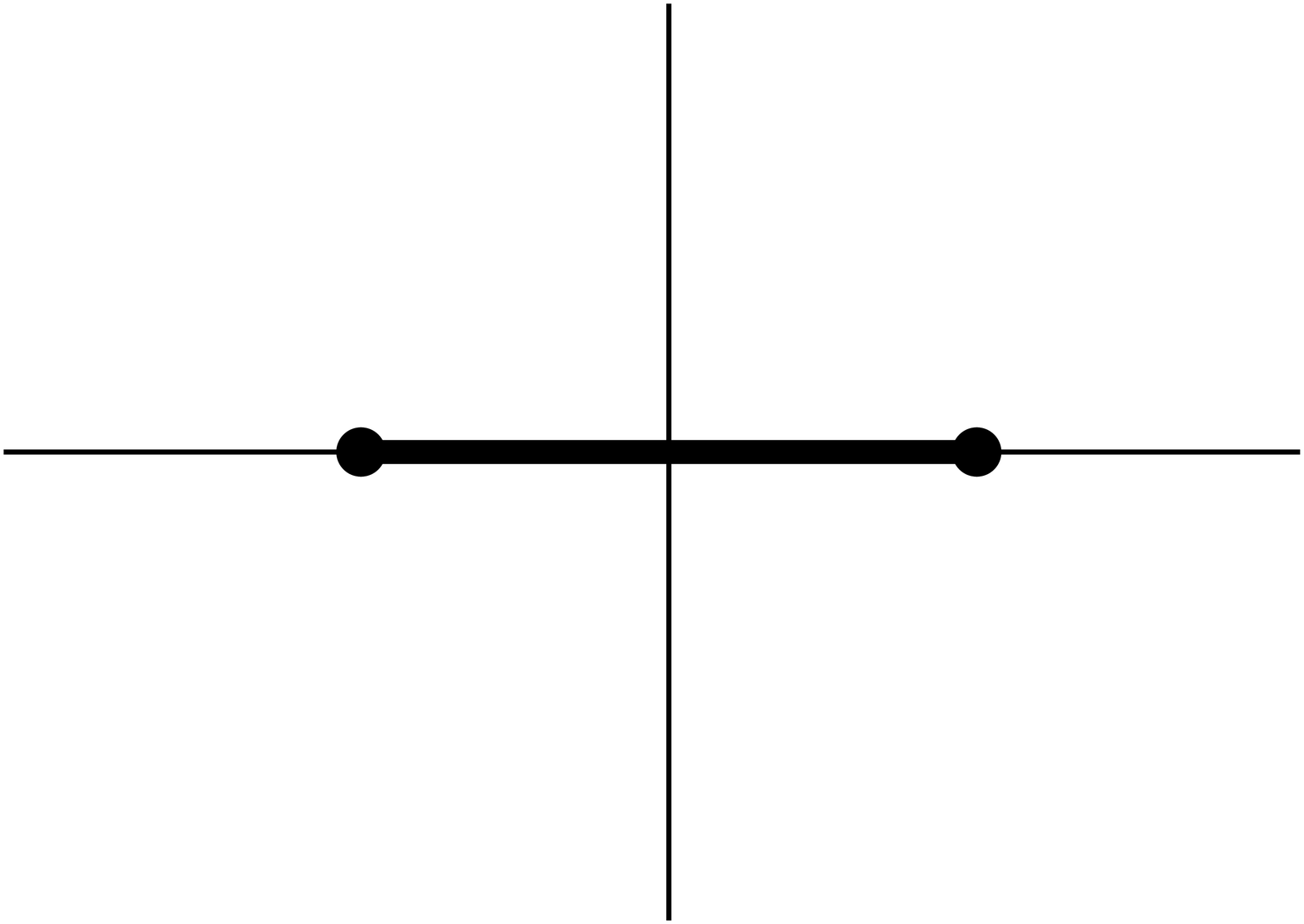}} (b) $m=0,n>1$ \\
\end{minipage}
\begin{center}
\hskip 1.2cm
\end{center}
\begin{minipage}[h!]{0.4\linewidth}
\center{\includegraphics[width=1\linewidth]{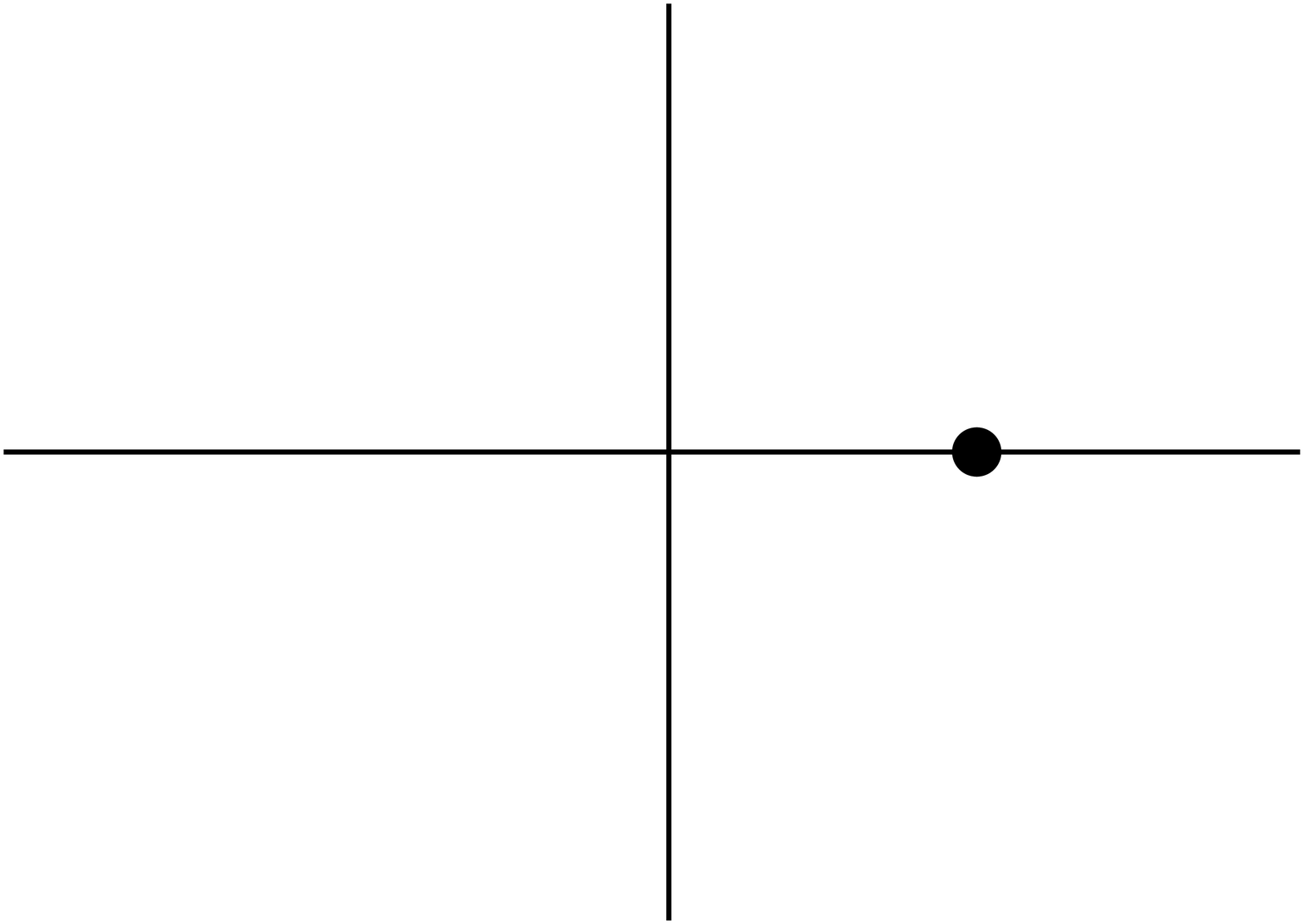}} (c) $m>0,n\le1$ \\
\end{minipage}
\hfill
\begin{minipage}[h!]{0.4\linewidth}
\center{\includegraphics[width=1\linewidth]{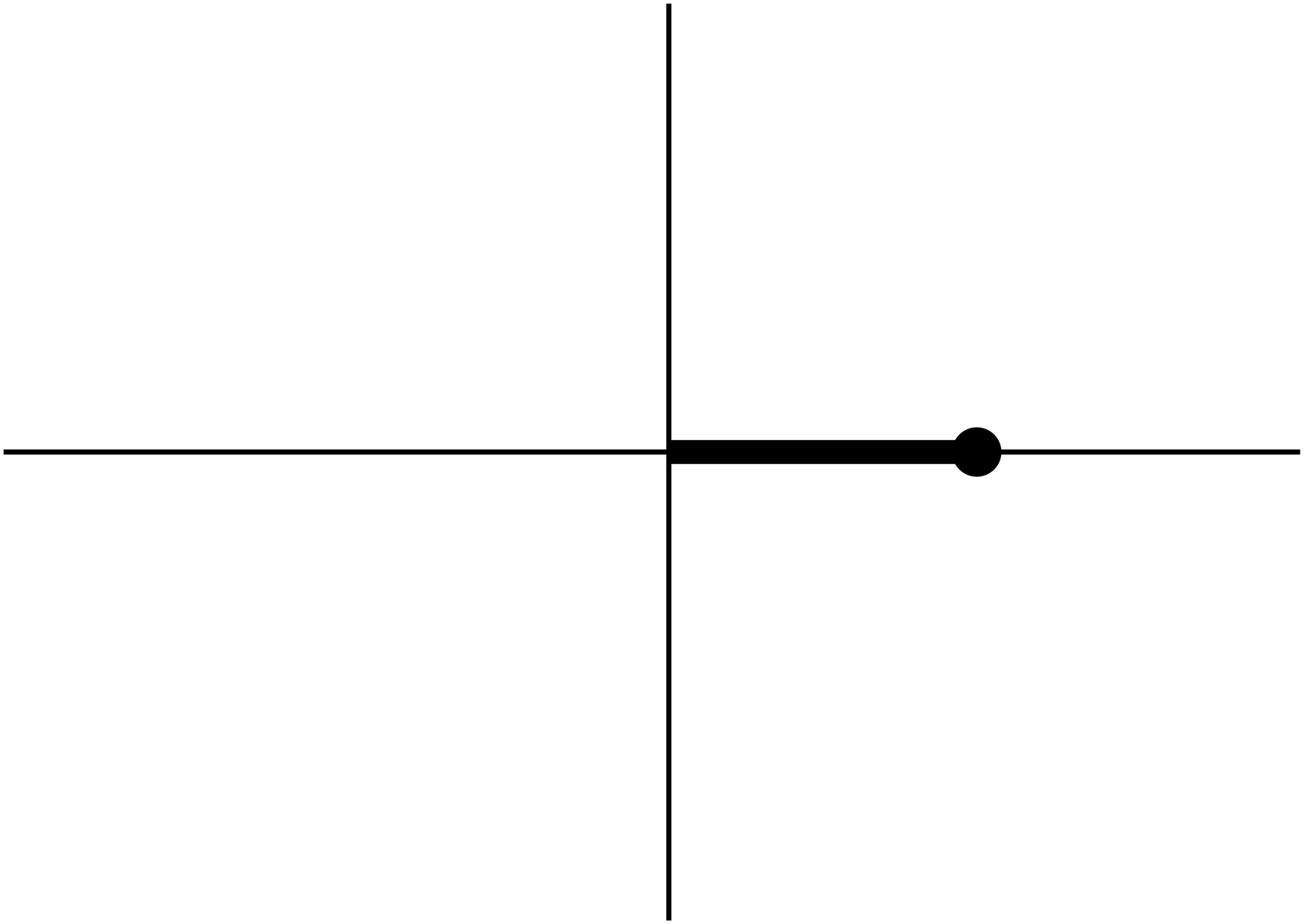}} (d) $m>0,n>1$ \\
\end{minipage}
\caption{Distribution of branes for various values of  $(m,n)$.}
\label{Fig:Sources}
\end{figure}

To give a geometric interpretation of (\ref{Rpsi}), we recall that, for $m>0$, $y$ ranges from zero to $\pi/2$, so (\ref{Rpsi}) describes a line connecting $r=0$ with $r=2l$. The geometry (\ref{SSmetr}) has an $m$-dimensional sphere attached to every of this line, so the singular locus (\ref{Rpsi}) has a topology of $(m+1)$--dimensional disk. For $m=0$, 
$y$, and $\theta$ range from zero to $\pi$, so (\ref{Rpsi}) represents a line connecting two singular points (\ref{EndPoints}), 
(\ref{EndPointOne}). The pictorial representation of singular loci in $(r,\theta)$ plane is given in figure \ref{Fig:Sources}.

\bigskip

\noindent
We will now combine (\ref{EndPoints}) and (\ref{Rpsi}) to analyze the brane distribution for $m>0$.

\begin{enumerate}[(a)]
\item{$n=0$.\\
In this case, the $m$--dimensional sphere described by (\ref{EndPoints}) is the only singularity of the harmonic function, and in the vicinity of this singularity we find
\bea\label{Jun16}
H=\frac{{\tilde Q}x}{x^2+y^2}+regular
\eea
To give a geometric interpretation of this expression, we consider the base metric (\ref{SSmetr}) in the vicinity of singularity (\ref{EndPoints}):
\bea\label{Jul15}
ds^2_{9-p}\approx (2l)^2d\Omega_m^2+dr^2+(2l)^2d\theta^2.
\eea
It is convenient to introduce polar coordinates $(R,\Phi)$ by
\bea\label{RPhiCrd}
r=2l+R\cos\Phi,\quad 2l\theta=R\sin\Phi.
\eea
Recalling that $\theta$ varies from $-\frac{\pi}{2}$ to $\frac{\pi}{2}$ when $n=0$, we conclude that $\Phi\in[0,2\pi)$ (or $\Phi\in[-\pi,\pi)$, see below), as long as $R$ remains small. Then metric (\ref{Jul15}) takes the standard form
\bea\label{Jun16ac}
ds^2_{9-p}\approx (2l)^2d\Omega_m^2+dR^2+R^2d\Phi^2.
\eea
To write $(x,y)$ in terms of $(R,\Phi)$, we use the real form (\ref{ElliptCase}) of the holomorphic map (\ref{TheSolnMone}). In particular, for small 
$x$ and $y$ we find
\bea\label{Jun16b}
&&\hskip -1.7cm x^2+y^2\approx 2(\cosh x-\cos y)=\frac{2\rho_-}{2l}=
\frac{2}{2l}\sqrt{(r-2l)^2+4rl(1-\cos\theta)}
\approx\frac{R}{l}\\
&&\hskip -1.7cm x^2\approx 2(\cosh x-1)\approx\frac{1}{2l}\left[(r-2l)+\sqrt{(r-2l)^2+4rl(1-\cos\theta)}\right]
\approx\frac{R}{l}\cos^2\frac{\Phi}{2}.
\nonumber
\eea
Extraction of the square root from the last expression should be done carefully: positivity of the harmonic function (\ref{Jun16}) (or (\ref{TheSolnMone})) requires that $x>0$. Thus we can write
\bea\label{Jun16a}
x=\sqrt{\frac{R}{l}}\cos\frac{\Phi}{2},
\eea 
as long as $\Phi\in[-\pi,\pi)$. The range $\Phi\in[0,2\pi)$ is equivalent from the point of view of (\ref{RPhiCrd}), but it leads to a more complicated counterpart of (\ref{Jun16a}), and it will not be explored further. Substitution of (\ref{Jun16b}) and (\ref{Jun16a}) into equation (\ref{Jun16}) leads to a simple expression for the harmonic function in the vicinity of the sources:
\bea\label{Jun16ab}
H={\tilde Q}\sqrt{\frac{l}{R}}\cos\frac{\Phi}{2}+regular
\eea
We recall that $R=0$ corresponds to the $m$--dimensional sphere (\ref{EndPoints}) with radius $2l$. Notice that this expression never becomes negative since 
$\Phi\in[-\pi,\pi)$. 
}
\item{$n=1$.\\
Here again, the $m$--dimensional sphere described by (\ref{EndPoints}) is the only singularity of the harmonic function, and in the vicinity of this singularity we find
\bea\label{Jun16e}
H=\frac{A}{x^2+y^2}+regular=\frac{Al}{R}+regular
\eea
As before, we used (\ref{Jun16b}) to rewrite the harmonic function in terms of coordinates (\ref{RPhiCrd}), and in this case
\bea
ds^2_{9-p}\approx (2l)^2d\Omega_m^2+dR^2+R^2d\Phi^2+
R^2\sin^2\Phi d\phi^2,\qquad 0\le\Phi<\pi\nonumber
\eea
Formula (\ref{Jun16e}) gives a standard harmonic function in three dimensions transverse to $S^m$, so this configuration corresponds to D--branes uniformly distributed over $S^m$.
}
\item{$n>1$.\\
In this case the singularity consists of a line (\ref{Rpsi}) connecting $r=0$ and (\ref{EndPoints}) with sphere $S^m$ fibered over it. In the vicinity of $\theta=0$, the base of the metric (\ref{SSmetr}) becomes
\bea\label{Jun18}
ds^2_{9-p}\approx dr^2+r^2d\Omega_m^2 +r^2d\theta^2+
r^2\theta^2d\Omega_n^2,
\eea
and the locus (\ref{Rpsi}) is an $m+1$--dimensional ball with metric
\bea
ds^2_{sing}\approx  dr^2+r^2d\Omega_m^2,\qquad 0\le r<2l
\eea
The singularity corresponds to $x=0$ (recall (\ref{SingOne}) and (\ref{SingThree})), and the leading contribution to the harmonic function (\ref{TheSolnMone}) for small $x$ is
\bea\label{Jun18b}
H=\frac{1}{x^2+\sin^2 y}
\frac{\tilde Q}{x^{n-1}}+\dots
\eea
For small $x$, metric (\ref{SSmetrAA}) becomes
\bea\label{Jun18a}
ds^2_{base}\approx 4l^2\cos^2 y d\Omega_m^2+
4l^2(x^2+\sin^2 y)\left[dx^2+dy^2\right]+
4l^2x^2\sin^2 y d\Omega_n^2.
\eea
Away from $y=0$, we can neglect $x^2$ in comparison with $\sin^2y$, then function (\ref{Jun18b}) describes the Coulomb potential produced by D--branes uniformly distributed over $S^m$. Rewriting (\ref{Jun18a}) as
\bea\label{Jun18ab}
ds^2_{base}\approx dR^2+R^2 d\Omega_m^2+
(4l^2-R^2)\left[dx^2+x^2d\Omega_n^2\right],
\eea 
we find the charge density $\rho$:
\bea\label{DensNgr1}
H&=&\frac{\rho}{[(4l^2-R^2)x^2]^{(n-1)/2}}+\dots,\nonumber\\
\rho&=&4l^2{\tilde Q}(4l^2-R^2)^\frac{n-3}{2}=
(2l)^{n-1}{\tilde Q}\sin^{n-3}y
\eea
As expected, the charge density vanishes on the boundary (\ref{EndPoints}) of the ball, where $y=0$. 
}
\end{enumerate}
To summarize, we found that for $n\le 1$ the brane sources are localizes on the $m$--dimensional sphere, they produce a Coulomb potential (\ref{Jun16e}) for $n=1$ and a potential (\ref{Jun16ab}) with a fractional power of the radial coordinate $R$ for $n=0$. For $n>1$, the branes are located on a line connecting $r=0$ and (\ref{EndPoints}) with sphere $S^m$ fibered over it ($(R,\Omega_m)$ subspace of (\ref{Jun18ab})). These sources produce a Coulomb potential in $(n+1)$ transverse directions with charge density (\ref{DensNgr1}).

\section{Beyond geodesics: wave equation, Killing tensors, and strings}
\label{SecKillWave}
\renewcommand{\theequation}{4.\arabic{equation}}
\setcounter{equation}{0}

The main goal of this paper is identification of backgrounds which can potentially lead to integrable string theories. As discussed in the introduction, integrability can be ruled out by looking at relatively simple equations for the light modes of strings (massless particles), and in the last section we demonstrated that classical equations of motion for such particles are integrable only for the harmonic functions given by (\ref{TheSolnMthree})--(\ref{TheSolnMtwo}). It is natural to ask whether separability of the HJ equation (\ref{HJone}) in the elliptic coordinates (\ref{ElliptCase}) persists at the quantum level and whether it is related to some hidden symmetry of the system. In section \ref{SecWave} we analyze integrability of the wave equation, a quantum counterpart of (\ref{HJone}), and in section \ref{SecKill} we identify the Killing tensor responsible for the separation. Finally in section \ref{NonintStr} we investigate the question whether integrability of geodesics persists for finite size strings.

\subsection{Separability of the wave equation}
\label{SecWave}

In this subsection we will analyze the wave equation which governs dynamics of a minimally--coupled massless scalar:
\bea\label{WaveEqn}
\frac{1}{\sqrt{-g}}\partial_{M}\left( g^{MN}
\sqrt{-g}\partial_{N} \Psi \right)=0.
\eea
Most D$p$--branes generate a nontrivial dilaton, so the last equation would look differently in the string and in the Einstein frames\footnote{Unlike (\ref{WaveEqn}), the HJ equation (\ref{HJone}) is invariant under conformal rescaling of the metric, so the results of section \ref{SecGnrGds} are valid in both the string and the Einstein frames.}, and here we will focus on the most interesting case of the Einstein frame:
\bea\label{EinstFrame}
g^{(E)}_{MN}dx^Mdx^N=
e^{-\Phi/2}\left[
\frac{1}{\sqrt{H}}\eta_{\mu\nu}dx^{\mu}dx^{\nu}+
\sqrt{H}ds^2_{base}\right],\quad e^{2\Phi}=H^{(3-p)/2}.
\eea
Since the HJ equation (\ref{HJone}) arises in the eikonal approximation of (\ref{WaveEqn}), the arguments presented in sections 
\ref{SecGnrGds} imply that (\ref{WaveEqn}) is not integrable unless the metric $ds^2_{base}$ has the form (\ref{SSmetrAA}) and $H$ is given by (\ref{TheSolnMone})\footnote{Solution (\ref{TheSolnMthree}) leads to a trivial separation in spherical coordinates, and solution 
(\ref{TheSolnMtwo}) reduces to (\ref{TheSolnMone}).}, although these conditions are not sufficient for integrability of (\ref{WaveEqn})\footnote{For example, as we will see below, equation (\ref{WaveEqn}) does not separate if $g_{MN}$ is a metric in the string frame, although it still reduces to (\ref{HJone}) in the eikonal approximation.}. 

To write the wave equation in the geometry (\ref{EinstFrame}), we recall the metric on the base (\ref{SSmetrAA}),
\bea\label{WWbase}
ds_{base}^2=(\cosh^2x-\cos^2y)(dx^2+dy^2)+\sinh^2x\sin^2y d\Omega_n^2+\cosh^2x\cos^2y d\Omega_m^2,
\eea
and introduce a convenient notation:
\bea
A=(\cosh^2x-\cos^2y),\quad X=\sinh^nx\cosh^mx,\quad
Y=\sin^ny\cos^my. 
\eea
Evaluating the determinant of the metric,
\bea\label{detEinst}
\sqrt{-g^{(E)}}= e^{-5\Phi/2}H^{2-p/2}X(x)Y(y)A,
\eea
and substituting (\ref{EinstFrame}), (\ref{WWbase}), (\ref{detEinst}) into (\ref{WaveEqn}), we find the wave equation:
\bea\label{WaveEinst}
&&H\partial_{\mu}\partial^{\mu}\Psi +
\frac{1}{\sinh^2x\sin^2y}\Delta_{\Omega_n}\Psi+\frac{1}{\cosh^2x\cos^2y}\Delta_{\Omega_m}\Psi\nonumber\\
&&+\frac{H^{(p-3)/2}e^{2\Phi}}{AXY}\left\{\partial_x \left[ H^{(3-p)/2}e^{-2\Phi}XY\partial_x \Psi \right] + \partial_y \left[ H^{(3-p)/2}e^{-2\Phi}XY \partial_y \Psi \right] \right\}=0
\eea
This equation separates since $H^{(3-p)/2}e^{-2\Phi}=1$ (see (\ref{EinstFrame})). Specifically, if we write\footnote{Here $Y_{k}(\Omega_m)$ and 
$Y_{l}(\Omega_n)$ are standard spherical harmonics with angular momenta $k$ and $l$. For example, $\Delta_{\Omega_n}Y_l(\Omega_n)=-l(l+n-1)Y_l(\Omega_n)$.}
\bea\label{MultSepar}
\Psi=e^{ip_\mu x^\mu}Y_{k}(\Omega_m)
Y_{l}(\Omega_n)F(x)G(y),
\eea
then (\ref{WaveEinst}) splits into two ordinary differential equations with separation constant $\Lambda$: 
\bea\label{SeprWave1}
\left[-\left(a\cosh^2 x+\frac{{\tilde Q}}{\sinh^{n-1}x\cosh^{m-1}x}\right)p_\mu p^\mu -
\frac{l(l+n-1)}{\sinh^2x}+\frac{k(m+k-1)}{\cosh^2x}\right]F\nonumber\\
+\frac{1}{X}\left[XF' \right]'=\Lambda F
\eea
\bea\label{SeprWave2}
ap_\mu p^\mu\cos^2 y -\left[
\frac{l(l+n-1)}{\sin^2y}+\frac{k(m+k-1)}{\cos^2y}\right]G+\frac{1}{Y}
\left[YG' \right]'=-\Lambda G
\eea
We used the expression for the harmonic function,
\bea\label{KGharm}
H=a+\frac{1}{(\cosh^2 x-\cos^2 y)}
\frac{{\tilde Q}}{\sinh^{n-1}x\cosh^{m-1}x}
\eea
found in section \ref{SecGnrGds} (see (\ref{TheSolnMone}), (\ref{TheSolnMtwo})). 

Let us now demonstrate that the wave equation (\ref{WaveEqn}) does not separate in the string metric unless 
$p=3$. The string--frame counterpart of (\ref{WaveEinst}) can be obtained by formally setting $e^{2\Phi}=1$ in that equation:
\bea\label{WaveStr}
&&H\partial_{\mu}\partial^{\mu}\Psi +
\frac{1}{\sinh^2x\sin^2y}\Delta_{\Omega_n}\Psi+\frac{1}{\cosh^2x\cos^2y}\Delta_{\Omega_m}\Psi\nonumber\\
&&+\frac{H^{(p-3)/2}}{AXY}\left\{\partial_x \left[ H^{(3-p)/2}XY\partial_x \Psi \right] + \partial_y \left[ H^{(3-p)/2}XY \partial_y \Psi \right] \right\}=0
\eea
Clearly, the multiplicative separation (\ref{MultSepar}) does not work for this equation unless $p=3$. Since coordinates $(x,y)$ are uniquely fixed by the discussion of the HJ equation (which is an eikonal limit of (\ref{WaveStr})) presented in section \ref{SecGnrGds}, to rule out the separation, it is sufficient to show that a substitution of
\bea\label{PsdMultSepar}
\Psi=e^{ip_\mu x^\mu}Y_{k}(\Omega_m)
Y_{l}(\Omega_n)F(x)G(y)P(x,y)
\eea
into (\ref{WaveStr}) does not lead to separate equations for $F$ and $G$ for any {\it fixed} function $P(x,y)$. To demonstrate this, we perform such substitution and rewrite the result as
\bea\label{ToSeparate}
&&-A\left[Hp_{\mu}p^{\mu}+
\frac{l(l+n-1)}{\sinh^2x\sin^2y}+\frac{k(k+m-1)}{\cosh^2x\cos^2y}\right]\nonumber\\
&&\qquad+\frac{H^{(p-3)/2}}{P}\left\{
\frac{1}{X}\partial_x \left[ H^{(3-p)/2}X\partial_x P \right]+
\frac{1}{Y}\partial_y \left[ H^{(3-p)/2}Y\partial_y P \right]\right\}\nonumber\\
&&\qquad+\frac{1}{F}\left\{F'' + 
F'\partial_x \ln\left[ H^{(3-p)/2}X P^2 \right] \right\}\\
&&\qquad
+\frac{1}{G}\left\{G'' + 
G'\partial_y \ln\left[ H^{(3-p)/2}Y P^2 \right] \right\}=0\nonumber
\eea
Since $H$ and $P$ are fixed functions, the first two lines of (\ref{ToSeparate}) remain the same for all $F$ and $G$, so equation (\ref{ToSeparate}) does not separate unless its third line is only a function of $x$ and the forth line is only a function of $y$. This implies that
\bea\label{tmpWaveSep}
\d_x\d_y\ln\left[ H^{(3-p)/2}P^2 \right]=0\quad\Rightarrow
\quad P=H^{(p-3)/4}P_1(x)P_2(y).
\eea
Functions $P_1$ and $P_2$ can be absorbed into $F$ and $G$ (recall (\ref{PsdMultSepar})), so we set $P_1=P_2=1$. Direct substitution into (\ref{ToSeparate}) shows that the third line of that equation obstructs separation unless $p=3$.

To summarize, we have demonstrated that while the HJ equation is integrable in the elliptic coordinates (\ref{ElliptCase}), its quantum version may or may not be separable depending on the frame. In particular, the equation for the minimally--coupled massless scalar separates in the Einstein, but not in the string frame.

\subsection{Killing tensor}
\label{SecKill}

Separation the Hamilton--Jacobi and Klein--Gordon equations implies an existence of nontrivial conserved charges which are associated with symmetries of the background. In the simplest case, such symmetries are encoded in the Killing vectors, which correspond to invariance of the metric under reparametrization
\bea
x^M\rightarrow x^M+V^M(x),
\eea
where $V^M(x)$ satisfies the equation 
\bea
V_{M;N}+V_{N;M}=0.
\eea
This symmetry guarantees a conservation of the charge
\bea
Q_V=V^M\frac{\d S}{\d x^M},
\eea
and momenta $p_\mu$ appearing in (\ref{GenAction}) were examples of such charges. Although not every separation of variables can be associated with Killing vector (separation between $x$ and $y$ coordinates found in sections \ref{SecGnrGds} and \ref{SecWave} is our prime example), the general theory developed in 
\cite{Carter,KalMil1,KalMil2,Miller3} guarantees that any such separation is related to a symmetry of the background, which is encoded by a (conformal) Killing tensor. In this subsection we will discuss the conformal Killing tensor associated with separation (\ref{TheSolnMone}) and (\ref{SeprWave1})--(\ref{SeprWave2}).

The conformal Killing tensor of rank two satisfies equation
\bea\label{CKTdef}
\nabla_{(M}K_{NL)}=\frac{1}{2}W_{(M}g_{NL)},
\eea
which is solved in appendix \ref{AppKill}. Here we will deduce the same solution by using the separation of variables found in section \ref{SecGnrGds}. A conformal Killing tensor $K_{MN}$ always implies that 
\bea\label{IntMtnKil}
I=K^{MN}\d_M S\d_N S
\eea
is an integral of motion of the massless HJ equation\footnote{A Killing tensor $K_{MN}$, which has $V_M=0$ in (\ref{CKTdef}),  implies that (\ref{IntMtnKil}) is an integral of motion of the massive HJ equation.}, so $K^{MN}$ can be extracted from the know separation. Going to the eikonal approximation in (\ref{WaveEinst}) with harmonic function (\ref{KGharm}), we find
\bea\label{WaveKill}
&&\left[a\cosh^2x+\frac{{\tilde Q}}{\sinh^{n-1}x\cosh^{m-1}x}\right]
\d_{\mu}S\d^{\mu}S+
\frac{1}{\sinh^2x}h^{ij}\d_i S\d_j S\nonumber\\
&&\qquad\qquad-
\frac{1}{\cosh^2x}{\tilde h}^{ij}{\tilde \d}_i S{\tilde\d}_j S
+\d_x S\d_x S\\
&&=-a\d_{\mu}S\d^{\mu}S-\frac{1}{\sin^2y}h^{ij}\d_i S\d_j S-\frac{1}{\cos^2y}{\tilde h}^{ij}{\tilde \d}_i S{\tilde\d}_j S
-\d_y S\d_y S\nonumber
\eea
For separable solutions, both sides of this equation must be constant, and identifying this constant with $-I$ in (\ref{IntMtnKil}), we find
\bea\label{Jul25}
K^{MN}p_Mp_N=ap_\mu p^\mu+\frac{1}{\sin^2y}h^{ij}p_ip_j+\frac{1}{\cos^2y}{\tilde h}^{ij}
{\tilde  p}_i {\tilde p}_j +p_yp_y
\eea
In appendix \ref{AppKill} this expression is derived in a geometrical way by solving the equation (\ref{CKTdef}) for the Killing tensor. 

\subsection{Non--integrability of strings}\label{NonintStr}

In section \ref{SecGnrGds} we have classified all supersymmetric configurations of Dp--branes that lead to integrable equations for null geodesics. Specifically, we demonstrated that a metric (\ref{PreGeom})--(\ref{FlatBase}) leads to a separable HJ equation (\ref{HJone}) if and only if it has the form 
\bea\label{IntGeodMetr}
ds^2=\frac{1}{\sqrt{H}}\eta_{\mu\nu}dx^{\mu}dx^{\nu}+\sqrt{H}(dr^2+r^2d\theta^2+r^2\cos^2\theta d\Omega_m^2 +r^2\sin^2\theta d\Omega_n^2)
\eea
with the harmonic function $H$ from (\ref{TheSolnMthree})--(\ref{TheSolnMtwo}). In this subsection we will investigate whether integrability of geodesics extends to strings with finite size. Our discussion will follow the logic presented in \cite{StepTs}, and to compare our results with ones from that paper we rewrite the metric in terms of a new function $f=H^{1/4}$ so the metric (\ref{IntGeodMetr}) becomes:
\bea\label{MetricStrInt}
ds^2=\frac{1}{f^2}\eta_{\mu\nu}dx^{\mu}dx^{\nu}+f^2(dr^2+r^2d\theta^2+r^2\cos^2\theta d\Omega_m^2+r^2\sin^2\theta d\Omega_n^2).
\eea
To demonstrate integrability of strings on a particular background, one has to find an infinite set of integrals of motions, and this ambitious problem has only been solved for very few geometries \cite{RoibPolch,Frolov}. However, to {\it rule out} integrability of sigma model on a given background, it is sufficient to start with a particular solution and look at linear perturbations around it. If such linearized problem  has no integrals of motion, one concludes that the original system is not integrable. This approach has been used in \cite{Zayas,StepTs} to rule out integrability of strings on the conifold and on the asymptotically--flat geometry produced by a  single stack of Dp--branes. The analysis presented in this subsection is complimentary to  \cite{StepTs}: we still focus on the near--horizon limit (where strings are known to be integrable for a single stack), but allow a nontrivial distribution of sources. 

The equation for linear perturbations around a given solution of a dynamical system is known as Normal Variational Equation (NVE) \cite{ziglinMor}, and to determine whether NVE is integrable, one can use the Kovacic algorithm\footnote{The Kovacic algorithm is implemented in Maple and one can use the function kovacicsols to check integrability of particular physical systems.}\cite{Kovacic_alg}.
Thus to demonstrate that the string theory on a particular background is not integrable one needs to perform the following steps:
\begin{enumerate}
\item write down the equations of motion
\item compute the variational equations
\item choose a particular solution and consider the normal equations (NVE)
\item algebrize NVE (rewrite equations as differential equations with rational coefficients) and transform them to normal form to make NVE be suitable for using the Kovacic algorithm
\item apply the Kovacic algorithm to the obtained NVE, if it fails the system is non--integrable.
\end{enumerate}
Now we apply this method to check integrability of strings in the background (\ref{MetricStrInt}). We begin with looking at the Polyakov action
\begin{equation}
S=-\frac{1}{4 \pi \alpha^{'}} \int d\sigma d\tau G_{MN}(X) \partial_a X^{M} \partial^{a} X^{N}.
\end{equation}
supplemented by the Virasoro constraints
\bea\label{Virasoroo}\label{Virasoro11}
G_{MN}\dot{X^M} X^{'N}=0,\\
\label{Virasoro21}
G_{MN}(\dot{X^M} \dot{X^N}+X^{'M} X^{'N})=0.
\eea

For a specific string ansatz on 2--sphere,
\begin{equation}\label{StringAnsatz}
x^{0}=t(\tau),\quad r=r(\tau), \quad \phi=\phi(\sigma), \quad \theta=\theta(\tau),
\end{equation}
the system has an effective Lagrangian density
\begin{equation}
{\cal L}=-f^{-2}\dot{t}^2+f^2\dot{r}^2+f^2r^2(-\sin^2\theta\phi^{'2}+\dot{\theta}^2).
\end{equation}
Equations of motion for cyclic variables $t,\phi$ lead to two integrals of motion ($E$ and $\nu$), and combining this with Virasoro constraint (\ref{Virasoro21}) we find
\begin{eqnarray}
\dot{t}&=&Ef^2,\nonumber\\
\phi'&=&\nu=const,\\
E^2&=&\dot{r}^2+r^2\dot{\theta}^2+\nu^2r^2\sin\theta^2.
\nonumber
\end{eqnarray}
These equations can be derived from the effective Lagrangian
\begin{equation}
L=f^2(\dot{r}^2+r^2\dot{\theta}^2-r^2\sin^2\theta\nu^2-E^2).
\end{equation}
Expanding around a particular solution,
\bea
 r=\frac{E}{\nu}\sin{\nu\tau},\quad \theta=\frac{\pi}{2},
\eea
we find the following NVE for $\eta\equiv \delta\theta$ (see appendix \ref{NonIntStrsDBr} for detail):
\begin{equation}\label{NVERes}
\eta^{''}+\left[ \frac{\ddot{r}}{\dot{r}^2}+2\left(\frac{f'_r}{f}+\frac{1}{r}\right) \right] \eta' - \left[ \frac{f^{''}_{\theta\theta}}{f} \frac{1}{r^2} -3\left(\frac{f'_{\theta}}{f}\right)^2\frac{1}{r^2}-\left(\frac{f'_{\theta}}{f}\right)^2\frac{\nu^2}{\dot{r}^2}-\frac{f^{''}_{\theta\theta}}{f}\frac{\nu^2}{\dot{r}^2}+\frac{\nu^2}{\dot{r}^2} \right]\eta=0.
\end{equation}

To proceed we need to choose a particular configuration corresponding to the specific function $f$. In section \ref{SecGnrGds} we have demonstrated that equations for geodesics are integrable only if function $H$ is given by 
(\ref{TheSolnMthree})--(\ref{TheSoln67}), and here we consider (\ref{TheSolnMone})  ignoring the $P$--term in (\ref{TheSoln67})\footnote{We will discuss NVE associated with this term in appendix \ref{AppNintGen}.}:
\bea\label{HGen}
H&=&f^4=\frac{\tilde{Q}}{\cosh^2x-\cos^2y}\sinh^{1-n}x\cosh^{1-m}x\nonumber\\&&=\frac{d^2\tilde{Q}}{\rho_+\rho_-}\left[ \frac{\rho_++\rho_-}{2d} \right]^{1-m}\left[ \left(\frac{\rho_++\rho_-}{2d}\right)^2-1\right]^{\frac{1-n}{2}},\\
\rho_\pm&=&\sqrt{r^2+d^2\pm 2rd\cos\theta}\nonumber.
\eea
Here we used the map (\ref{ElliptCase}) between the coordinates $(x,y)$ and $(\rho_{+},\rho_{-})$.
After carrying out all calculations one obtains the following NVE
\bea\label{NVEGen}
\eta^{''}&+&\frac{U}{D}\eta=0,\nonumber\\
U&=&E^4 \left[-d^4(n-1)^2-2 d^2 r^2 \left((m-3) n-5 m+n^2+2\right)-r^4
   (m+n-4) (m+n)\right] \nonumber\\
   &&+2 E^2 r^2 \nu^2 \Big[d^4 ((n-2) n+5)+2 d^2 r^2 (m (n-5)+(n-4) n+9)\nonumber\\
   &&+r^4 \left(m^2+2 m (n-3)+(n-6) n+4\right)\Big]-r^4 \nu^4 \Big[d^4 (n-3) (n+1)\\&&+2 d^2
   r^2 \left((m-5) n-5 m+n^2+4\right)+r^4 \left(m^2+2 m (n-4)+(n-8) n-4\right)\Big],\nonumber\\
D&=&16 r^2 \left(d^2+r^2\right)^2 [E^2-(r \nu)^2]^2.\nonumber
\eea
Application of the Kovacic algorithm to (\ref{NVEGen}) shows that the system is not integrable unless $d=0$, $m+n=4$ (this corresponds to AdS$_5\times$S$^5$). Detailed description of the method used in this section and complete calculations are presented in appendices \ref{NVEReview}--\ref{AppNintGen}.

To summarize, we have demonstrated that integrability of geodesics discussed in section \ref{SecGnrGds} does not persist for classical strings, and AdS$_5\times$S$^5$ is the only static background produced by a single type of D--branes placed on a flat base that leads to an integrable string theory\footnote{Analysis presented in this subsection does not rule out integrability on backgrounds containing NS--NS fluxes in addition to D--branes or on geometries produced by several types of branes, such as D$p$--D$(p+4)$ system discussed in the next section.}.

\section{Geodesics in static D$p$--D$(p+4)$ backgrounds}\label{DpDp4}

\renewcommand{\theequation}{5.\arabic{equation}}
\setcounter{equation}{0}

In the sections \ref{Examples}-\ref{SecGnrGds} we analyzed geodesics in the backgrounds produced by a single type of D branes. However, some of the most successful applications of string theory to black hole physics \cite{StrVafa} and to study of strongly coupled gauge theories \cite{HanWitt} involve intersecting branes, and in this section our analysis will be extended to a particular class of brane intersections. Specifically, we will extend the results of sections \ref{SecGnrGds} to 1/4--BPS configurations involving D$p$ and D$(p+4)$ branes. The geometries produced by such ``branes inside branes" continue to play an important role in understanding the physics of black holes, and a progress in understanding of the infall problem and Hawking radiation requires a detailed analysis of geodesics and waves on the backgrounds produced by D$p$--D$(p+4)$ systems. In this section we will continue to explore static configurations, and a large class of stationary solutions produced by D1--D5 branes will be analyzed in the next section. 

Let us consider massless geodesics in the geometry produced by D$p$ and D$(p+4)$ branes:
\bea\label{D1D5mGn}
ds^2=\frac{1}{\sqrt{H_1H_2}}\eta_{\mu\nu}dx^\mu dx^\nu+
\sqrt{H_1H_2}ds^2_{base}+\sqrt{\frac{H_1}{H_2}}dz_4^2.
\eea
The first and the second terms describe the spaces 
parallel/transverse to the entire D$p$--D$(p+4)$ system, and the four--dimensional torus represented by $dz_4^2$ is wrapped by D$(p+4)$ branes. We assume that D$p$ branes are smeared over the torus\footnote{Some localized solutions are also known \cite{CherkHash}, but we will not discuss them here.}. Metric (\ref{D1D5mGn}) contains two harmonic functions, $H_1$ and $H_2$, which are sourced by D$p$ and D$(p+4)$ branes. Away from the sources, these functions satisfy the Laplace equation on the $(5-p)$--dimensional flat base with metric $ds^2_{base}$.

Let us assume that geometry (\ref{D1D5mGn}) leads to a separable Hamilton--Jacobi equation. Then arguments presented in section \ref{SectPrcdr} imply that $H_2$ can only depend on two coordinates, $(r_1,r_2)$, where metric $ds^2_{base}$ has the form (\ref{BaseMetr}). To see this, we separate the Killing directions in the action
\bea
S=p_\mu x^\mu+q_i z^i+S_{base},
\eea
and rewrite the HJ equation (\ref{HJone}) as
\bea\label{HJD1D5w}
(\nabla S_{base})^2+p_\mu p^\mu H_1H_2+q_iq_iH_2=0.
\eea
Here we define
\bea
M^2=-p^\mu p_\mu,\qquad N^2=q_iq_i
\eea
If a given distribution of branes corresponds to integrable geodesics, then equation (\ref{HJD1D5}) should be separable for all allowed values of $M$ and 
$N$. Since equation (\ref{HJD1D5}) is analytic in these parameters, separability must persist even in the unphysical region where $M=0$ and $N$ is 
arbitrary\footnote{For asymptotically--flat solutions, $H_1$ and $H_5$ go to one at infinity, so $M>N$.}. In this region, equation (\ref{HJD1D5w}) reduces to  
(\ref{TheHJ}) with $H\rightarrow H_2$, $p_\mu p^\mu\rightarrow N^2$,  then arguments presented in section \ref{RedTo2D} reduce the problem to 
$H_2(r,\theta)$ with base space
\bea\label{Jul16}
ds_{base}^2=dr^2+r^2d\theta^2+r^2\cos^2\theta d\Omega_m^2+
r^2\sin^2\theta d\Omega_n^2,
\eea
and the analysis of section \ref{SectPrcdr} leads to three possible solutions ((\ref{TheSolnMthree}), (\ref{TheSolnMone}), (\ref{TheSolnMtwo})) for $(x,y)$ and $H_2$. 

Let us now demonstrate that $H_1$ can only be a function of $r$ and $\theta$. Indeed, separation for $M=0$ implies that  
\bea\label{GenSeparActw}
S_{base}={\tilde S}(y,{\tilde y})+R(r,\theta),
\eea
where $y_i$ are coordinates on $S^m$, and ${\tilde y}_j$ are coordinates on $S^n$. Substituting (\ref{Jul16}), (\ref{GenSeparActw}) into (\ref{HJD1D5w}) and differentiating the result with respect to $y_k$, we find
\bea
\frac{\d}{\d y_k}\left[
\frac{1}{r^2\cos^2\theta}h^{ij}
\left(\frac{\d{\tilde S}}{\d y_i}\right)
\left(\frac{\d{\tilde S}}{\d y_j}\right)+
\frac{1}{r^2\sin^2\theta}{\tilde h}^{ij}
\left(\frac{\d{\tilde S}}{\d {\tilde y}_i}\right)
\left(\frac{\d{\tilde S}}{\d {\tilde y}_j}\right)-H_1H_2M^2\right]=0.
\eea
Rewriting this relation as 
\bea
\frac{\d H_1}{\d y_k}=\frac{1}{M^2H_2}
\frac{\d}{\d y_k}
\left[\frac{1}{r^2\cos^2\theta}h^{ij}
\left(\frac{\d{\tilde S}}{\d y_i}\right)
\left(\frac{\d{\tilde S}}{\d y_j}\right)+
\frac{1}{r^2\sin^2\theta}{\tilde h}^{ij}
\left(\frac{\d{\tilde S}}{\d {\tilde y}_i}\right)
\left(\frac{\d{\tilde S}}{\d {\tilde y}_j}\right)\right],\nonumber
\eea
we conclude that $H_1$ develops unphysical singularities at 
$\theta=0,\frac{\pi}{2}$ for arbitrarily large $r$ unless 
$(\d H_1/\d y_k)=0$. Similar argument demonstrates that 
$(\d H_1/\d {\tilde y}_k)=0$, so $H_1$ can only depend on 
$(r,\theta)$.

To summarize, separability of the HJ equation (\ref{HJD1D5w}) requires the functions $H_1$ and $H_2$ to depend only on 
$(r,\theta)$, then the action has the form (\ref{GenSeparAct}),
\bea\label{GenSeparAbb}
S=p_\mu x^\mu+q_i z^i+S^{(m)}_{L_1}(y)+S^{(n)}_{L_2}({\tilde y})+R(r,\theta),
\eea
where $S^{(m)}_{L_1}$ and $S^{(n)}_{L_2}$ satisfy equations (\ref{HJSphere}). This results in the HJ equation
\bea\label{HJD1D5}
(\d_r R)^2+\frac{1}{r^2}(\d_\theta R)^2+
\frac{L_1^2}{r^2\cos^2\theta}+
\frac{L_2^2}{r^2\sin^2\theta}-H_1H_2M^2+N^2H_2=0,
\eea
and our analysis of separation at $M=0$ leads to three possible solutions ((\ref{TheSolnMthree}), (\ref{TheSolnMone}), (\ref{TheSolnMtwo})) for $(x,y)$ and $H_2$. As already discussed in section \ref{SectPrcdr}, solutions (\ref{TheSolnMone}) and  
(\ref{TheSolnMtwo}) are related by interchange of two spheres, so without loss of generality, we will focus on (\ref{TheSolnMthree}) and (\ref{TheSolnMone}).

Spherical coordinates (\ref{TheSolnMthree}) lead to separable equation (\ref{HJD1D5}) if an only if $H_1$ and $H_2$ do not depend on $\theta$, and since these functions have to be harmonic we find
\bea
H_1=a+\frac{Q_p}{r^{m+n}},\qquad H_2=a+\frac{Q_{p+4}}{r^{m+n}}.
\eea 
Here $a=1$ for asymptotically--flat space, and $a=0$ for the near--horizon solution. The corresponding metric 
(\ref{D1D5mGn}) gives the geometry produced by a single stack of D$p$--D$(p+4)$ branes \cite{HorStr}. 

Separation in the elliptic coordinates (\ref{TheSolnMone}) leads to 
\bea\label{H2D1D5}
&&\qquad\qquad H_2=a+\frac{1}{(\cosh^2 x-\cos^2 y)}
\frac{A}{\sinh^{n-1}x
\cosh^{m-1}x},\\
\label{ElCrdD1D5}
&&\cosh x=\frac{\rho_++\rho_-}{2d},\quad \cos y=\frac{\rho_+-\rho_-}{2d},\quad
\rho_\pm=\sqrt{r^2+d^2\pm 2rd\cos\theta},
\eea
and now we will determine the corresponding function $H_1$. Equation (\ref{HJD1D5}) separates in coordinates (\ref{ElCrdD1D5}) 
if and only if
\bea
N^2H_2-M^2H_1H_2=\frac{1}{(\cosh^2 x-\cos^2 y)}\left[V_1(x)+V_2(y)\right].
\eea
Since $H_2$ is already given by (\ref{H2D1D5}), the last relation implies that\footnote{Notice that $M\ne 0$ due to equation (\ref{HJD1D5}).}
\bea\label{H1H2z}
H_1H_2=\frac{1}{(\cosh^2 x-\cos^2 y)}\left[{\tilde V}_1(x)+{\tilde V}_2(y)\right].
\eea
First we set $a=0$ in (\ref{H2D1D5}), then equation (\ref{H1H2z}) becomes
\bea\label{H1H2y}
H_1=\frac{1}{\tilde Q}{\sinh^{n-1}x
\cosh^{m-1}x}\left[{\tilde V}_1(x)+{\tilde V}_2(y)\right].
\eea
To determine ${\tilde V}_1(x)$ and ${\tilde V}_2(y)$, we recall that, away from the sources, function $H_1$ must satisfy the Laplace equation (\ref{LaplInXY}):
\bea\label{LaplInD1D5}
\frac{1}{\sinh^{n}x\cosh^{m}x}\frac{\d}{\d x}\left[
\sinh^{n}x\cosh^{m}x\frac{\d H_1}{\d x}
\right]+
\frac{1}{\sin^{n}y\cos^{m}y}\frac{\d}{\d y}\left[
\sin^{n}y\cos^{m}y\frac{\d H_1}{\d y}
\right]=0
\eea
Substituting (\ref{H1H2y}) into (\ref{LaplInD1D5}) and performing straightforward algebraic manipulations, we find the most general solutions for $H_1$: 
\bea\label{H1D1D5}
\mbox{D0-D4:}&&(m,n)=(3,0)\quad H_1=C_1+C_2\left\{\arctan\left[\tanh\frac{x}{2}\right]+\frac{\sinh x}{2\cosh^2x}\right\}\nonumber\\
&&(m,n)=(2,1)\quad H_1=C_1+\frac{C_2}{\cosh x}\left\{1+\cosh x\ln\left[\tanh\frac{x}{2}\right]\right\}\nonumber \\
&&(m,n)=(1,2)\quad H_1=C_1+\nonumber\\&&\frac{C_2}{2\sqrt{2}\sinh x}\left\{ -1 +2\sinh x\left( \mathrm{arctanh}\left[\tanh\frac{x}{2}\right] + 2\Pi\left[ -1; -\arcsin\left[\tanh\frac{x}{2}\right] |1 \right] \right) \right\} \nonumber\\
&&(m,n)=(0,3)\quad H_1=C_1+C_2\left\{\coth x+\ln\left[\tanh\frac{x}{2}\right]\right\}\nonumber\\
\mbox{D1-D5:}&&(m,n)=(2,0)\quad H_1=C_1+C_2\tanh x\nonumber\\
&&(m,n)=(1,1)\quad C_1+C_2\ln[\tanh x]+C_3\ln[\tan y]+C_4\ln[\sin(2y)\sinh(2x)]\nonumber\\
&&(m,n)=(0,2)\quad H_1=C_1+C_2\coth x\nonumber\\
\mbox{D2-D6:}&&(m,n)=(1,0)\quad H_1=C_1+C_2\arctan\left[\tanh\frac{x}{2}\right]\nonumber\\
&&(m,n)=(0,1)\quad H_1=C_1+C_2\ln\left[\tanh\frac{x}{2}\right]\nonumber\\
\mbox{D3-D7:}&&(m,n)=(0,0)\quad H_1=C_1+C_2 x
\eea
Here $\Pi[n;\phi|m]$ is the incomplete elliptic integral.

So far we have assumed that $a=0$ in $H_2$. The case $a=1$, ${\tilde Q}=0$ corresponds to D$p$ branes only, so it is covered by discussion in section \ref{SecGnrGds}. Solutions with nonzero $a$ and ${\tilde Q}$ can be analyzed by looking at formal perturbation theory in $a$, and it turns out that $H_1$ must be constant for such solutions.

\section{Geodesics in D1--D5 microstates}\label{Rotations}

\renewcommand{\theequation}{6.\arabic{equation}}
\setcounter{equation}{0}

In the last three sections we have analyzed geodesics in a variety of static backgrounds produced by D--branes. In general, supersymmetric geometries are guaranteed to have a time--like (or light--like) Killing vector, so they must be stationary, but not necessarily static.  In particular, an interest in stationary geometries produced by the D1--D5 branes has been generated by the fuzzball proposal for resolving the black hole information paradox 
\cite{lmPar,fuzz}. According to this picture, microscopic states accounting for the entropy of a black hole have nontrivial structure that extents to the location of the na\"ive horizon, and the black hole geometry emerges as a course graining over such structures. Although the vast majority of fuzzballs is expected to be quantum, some fraction of microscopic states should be describable by classical geometries, and study of this subset has led to important insights into qualitative properties of generic microstates \cite{3charge}.

The fuzzball program has been particularly successful in identifying the microscopic states corresponding to D1--D5 black hole, where all microstate geometries have been constructed in \cite{lmPar,lmm}. Moreover, a strong support for the fuzzball picture came from analyzing the properties of these metrics \cite{tube}, and success of this study was based on separability of the wave equation on a special classes of metrics. In this paper we have been  focusing on separability of the HJ equation as necessary condition for integrability of strings, but such separability also implies separability of the wave equation. This provides an additional motivation for studying the HJ equation for microscopic states in the D1--D5 system.

In this section we will mostly focus on the HJ equation for particles propagating on metrics constructed in \cite{lmPar}, and extension to the geometries for the remaining microstates of D1--D5 black holes \cite{lmm} will be discussed in the end. The solutions of \cite{lmPar},
\bea\label{GenD1D5J}
ds^2&=&\frac{1}{\sqrt{H_1H_2}}\left[-(dt-A_idx^i)^2+(du+B_idx^i)^2\right]+
\sqrt{H_1H_2}dx_idx_i+\sqrt{\frac{H_1}{H_2}}dz_4^2,
\nonumber\\
&&d(\star_x dH_1)=d(\star_x dH_2)=0,\qquad dB=-\star_x dA,
\eea
generalize the static metric (\ref{D1D5mGn}) with $p=1$ by allowing the branes to vibrate on the four dimensional base, which is transverse to D1 and D5, and geometries of \cite{lmm} account for fluctuations on the torus. While the metric (\ref{GenD1D5J}), supplemented by the appropriate matter fields given in \cite{lmPar}, always gives a supersymmetric solution of supergravity away from the sources, the bound states of D1 and D5 branes, which are responsible for the entropy of a black hole, have additional relations between $H_1$, $H_2$ and $A$. Such bound states are uniquely specified by a closed contour ${F}_i(v)$ in four non--compact directions, and the harmonic functions are given by \cite{lmPar}\footnote{Relations (\ref{MicroHarm}) contain a constant parameter $\alpha$. Solutions with $\alpha=1$ correspond to asymptotically--flat geometries, and metrics with $\alpha=0$ asymptote to AdS$_3\times$S$^3$.}
\bea\label{MicroHarm}
H_1=\alpha+\frac{Q_5}{L}\int_0^L\frac{|\dot{\bf F}|^2dv}{|{\bf x}-{\bf F}|^2},\quad
H_2=\alpha+\frac{Q_5}{L}\int_0^L\frac{dv}{|{\bf x}-{\bf F}|^2},\quad
A_i=-\frac{Q_5}{L}\int_0^L\frac{\dot{ F}_idv}{|{\bf x}-{\bf F}|^2}.
\eea 
Remarkably, the resulting metric (\ref{GenD1D5J}) is completely smooth and horizon--free in spite of an apparent coordinate singularity at 
the location of the contour \cite{lmm}. To avoid unnecessary complications, we will focus on a special case 
$|{\dot {\bf F}}|=1$ (which leads to $H_1=H_2\equiv H$), although our results hold for arbitrary ${\dot {\bf F}}$. 

Applying the arguments presented in section \ref{RedTo2D} to metric (\ref{GenD1D5J}), we conclude that this geometry must preserve $U(1)\times U(1)$ symmetry of the base space, i.e., the profile ${F}_i(v)$ must be invariant under shifts of $\phi$ and $\psi$ in\footnote{This is a counterpart of (\ref{BaseMetr}) with $d_1=d_2=1$ and the base in (\ref{SSmetr}) with $m=n=1$.}
\bea
dx_idx_i=dr^2+r^2d\theta^2+r^2\cos^2\theta d\psi^2+
r^2\sin^2\theta d\phi^2.
\eea
This implies that the singular curve, $x_i=F_i(v)$ can only contain concentric circles with radii $r=(R_1,R_2,\dots,R_n)$ in the $\theta=\frac{\pi}{2}$ plane and concentric circles with radii $r=({\tilde R}_1,{\tilde R}_2,\dots,{\tilde R}_l)$ in 
the $\theta=0$ plane. First we focus on circles in the $\theta=\frac{\pi}{2}$ plane and demonstrate that separability of the HJ equation implies that $n=1$. Then we will show that $n=1$ also implies that $l=0$. 

\begin{enumerate}[1.]
\item{{\bf Circles in the $(x_1,x_2)$ plane.}

For a single circular contour $(r=R_1,\theta=\frac{\pi}{2})$, the integrals (\ref{MicroHarm}) have been evaluated in \cite{MultiStr}, and superposition of these results gives the harmonic functions for several circles:
\bea\label{ExplsD1D5J}
H&=&\alpha+\sum_{i=1}^n\frac{Q_i}{f_i},\quad A=\sum_{i=1}^n \frac{2Q_iR_i s_i}{f_i}\frac{r^2\sin^2\theta}{r^2+R_i^2+f_i}d\phi,\nonumber\\
f_i&=&\sqrt{(r^2+R_i^2)^2-4r^2R_i^2\sin^2\theta}.
\eea
Here $Q_i$ is a five--brane charge of a circle with radius $R_i$, and $s_i$ is a sign that specifies the direction for going around this circle.  

Let us assume the the HJ equation (\ref{HJone}) separates in some coordinates. Metric (\ref{GenD1D5J}) has eight Killing directions ($t,u,\phi,\psi$ and the torus), they can be separated in the action:
\bea
S=p_tt+p_u u+J_\phi\phi+J_\psi\psi+q_iz_i+{\tilde S}(r,\theta),
\eea
Our assumption of integrability amounts to further separation of ${\tilde S}(r,\theta)$ in some coordinates $(x,y)$. 
In particular, for $J_\psi=J_\phi=p_u=0$, the HJ equation (\ref{HJone}) in the metric (\ref{GenD1D5J}) can be written as
\bea\label{HJD1D5J}
(\d_r {\tilde S})^2+\frac{1}{r^2}(\d_\theta {\tilde S})^2-p_t^2\left[H^2-\frac{(A_\phi)^2}{r^2\sin^2\theta}\right]=0.
\eea
This equation looks very similar to (\ref{TheHJ}), but in practice it is easier to analyze: we don't need to impose the Laplace equation (as we did for (\ref{TheHJ})), since the explicit forms of $H$ and $A$ are known (see (\ref{ExplsD1D5J})). The discussion of section 
\ref{SectPrcdr} implies that (\ref{HJD1D5J}) should be viewed as a relation between $h(w)$ (which is defined by (\ref{defCmplVar}) and (\ref{holomZ})) and $(R_1,\dots,R_n)$. In particular, substitution of the perturbative expansion (\ref{AppPert}) for $h(w)$ leads to an infinite set of constraints on $(R_1,\dots,R_n)$. To write these constraints, we introduce a convenient notation: 
\bea
D_k=\sum_{j=1}^n Q_j (s_jR_j)^k.
\eea
Then separation in $(k+2)$-rd order of perturbation theory gives a constraint
\bea\label{ConstrD1D5J}
(D_0)^{k-1} D_{k}=(D_1)^{k} \quad k\ge1.
\eea
Already the first nontrivial relation ($k=2$) implies that $R_1=R_2=\dots=R_n$, so it is impossible to have more than one circle (see below). We conclude that the HJ equation does not separate on the background (\ref{GenD1D5J}), (\ref{ExplsD1D5J}) unless $n=1$. The remaining constraints (\ref{ConstrD1D5J}) for $k>2$ are automatically satisfied for this case. 

We will now prove that equation (\ref{ConstrD1D5J}) with $k=2$ implies that $R_1=\dots=R_n$, and the readers who are not interested argument can go directly to part 2. Let us order the radii by $R_1\ge R_2\ge\dots\ge R_n$ and define a function 
\bea
G(R_1,\dots R_n)\equiv D_0D_2-(D_1)^2.
\eea
Then the derivative
\bea
\frac{\d G}{\d R_1}=2Q_1(D_0 R_1-s_1D_1)=
2Q_1\sum_{j=1}^n Q_j(R_1-s_1s_jR_j)
\eea
is positive unless $R_1=\dots=R_n$ and $s_1=\dots=s_n$, so function $G$ reaches its minimal value when all radii are equal, and it is this value, 
\bea
G_{min}=G(R_1,\dots R_1)=\left[\sum Q_j\right]
\left[\sum Q_j R_1^2\right]-\left[\sum Q_j s_jR_1\right]^2=0,
\eea
that gives (\ref{ConstrD1D5J}) for $k=2$. We conclude the equation $G=0$ implies that $R_1=\dots=R_n$.}
\item{{\bf Circles in orthogonal planes.}

Having established that separation requires to have at most one circle in $\displaystyle\theta=\frac{\pi}{2}$ plane, we conclude that the same must be true about 
$\theta=0$ plane, but in principle it is possible to have one circle in each of the two planes. In this case we find
\bea\label{MuntiString}
&&H=\alpha+\sum_{i=1}^2\frac{Q_i}{f_i},\quad A=\frac{2Q_1R_1r^2\sin^2\theta}{f_1(r^2+R_1^2+f_1)}d\phi+
\frac{2Q_2R_2 r^2\cos^2\theta}{f_2(r^2+R_2^2+f_2)}d\psi,\nonumber\\
&&f_1=\sqrt{(r^2+R_1^2)^2-4r^2R_1^2\sin^2\theta},\quad f_2=\sqrt{(r^2+R_2^2)^2-4r^2R_2^2\cos^2\theta}
\eea
and for $J_\psi=J_\phi=p_y=0$ the HJ equation becomes
\bea
(\d_r {\tilde S})^2+\frac{1}{r^2}(\d_\theta {\tilde S})^2-p_t^2\left[H^2-\frac{(A_\phi)^2}{r^2\sin^2\theta}-\frac{(A_\psi)^2}{r^2\cos^2\theta}\right]=0
\eea
Fifth order of perturbation theory gives a relation $Q_1Q_2=0$, which implies that there is no separation in the geometry produced by two orthogonal circles. 
}
\item{{\bf Separable coordinates.}

The perturbative procedure implemented in part 1 also gives the expression for $h(w)$  in terms of $D_k$ for the configurations satisfying (\ref{ConstrD1D5J}):
\bea\label{Jul17}
h(w)&=&\ln\left[\frac{1}{2}\left( e^{w}+\sqrt{e^{2w}+\frac{D_1}{lD_0}} \right) \right].
\eea
We have already encountered this holomorphic function in (\ref{TheSolnMone})--(\ref{TheSolnMtwo}) (depending on the sign of $D_1$), and demonstrated that it corresponds to elliptic coordinates (\ref{ElliptCase}) or (\ref{RepCase3}). For completeness we present the expression for $H^2$ that clearly demonstrates the separation of variables in (\ref{HJD1D5J}):
\bea
H^2&=&\frac{1}{A}\left[  2\alpha D_0+\frac{16e^{2x}D_0^6l^4}{(4e^{2x}D_0^2l^2+D_1^2)^2} +\alpha^2\left( e^{2x}+\frac{e^{-2x}D_1^4}{16l^4D_0^4} \right) + \frac{\alpha^2}{2} \left( \frac{D_1}{lD_0} \right)^2 \cos2y  \right],
\nonumber\\
&&A=l^2(\sinh^2 x+\cos^2y).
\eea
}
\item{{\bf Separation with non--vanishing angular momenta.}

So far we have demonstrated that the HJ equation with $J_\psi=J_\phi=p_u=0$ separates only for microstate whose harmonic 
functions are given by (\ref{ExplsD1D5J}) with $k=1$, and separation takes place in the elliptic coordinates (\ref{Jul17}), 
(\ref{ElliptCase}). A direct check demonstrates that this separation persists for all values of momenta, when the relevant HJ equation is 
\bea\label{D1D5JhjGen}
(\d_r S)^2+\frac{1}{r^2}(\d_\theta S)^2+\frac{(B_\psi p_u-J_\psi)^2}{r^2\cos^2\theta}+
\frac{(J_\phi + A_\phi p_t)^2}{r^2\sin^2\theta}+ H^2(p_u^2-p_t^2)=0
\eea
and\footnote{To compare with \cite{MultiStr}, we replaced $R_1$ in (\ref{ExplsD1D5J}) by $a$ and $f_1$ by $f$.}
\bea\label{D1D5Jaa}
H&=&\alpha+\frac{Q}{f},\quad A_\phi=\frac{2Qa}{f}\frac{r^2\sin^2\theta}{r^2+a^2+f},\quad
B_\psi=\frac{2Qa}{f}\frac{r^2\cos^2\theta}{r^2+a^2+f}\\
f&=&\sqrt{(r^2+a^2)^2-4r^2a^2\sin^2\theta}.\nonumber
\eea
}
\end{enumerate}
To summarize, we have demonstrated that the HJ equation (\ref{HJone}) separates for the stationary D1--D5 geometry 
(\ref{GenD1D5J})--(\ref{MicroHarm}) with $|{\dot{\bf F}}|=1$ if and only if the string profile is circular, i.e., the harmonic functions 
are given by (\ref{D1D5Jaa}). 
\bigskip

Notice that integrability of (\ref{D1D5JhjGen}) follows from separability of the wave equation in the background (\ref{GenD1D5J}), 
(\ref{D1D5Jaa}), which has been discovered long time ago \cite{cvetLars,tube}. Let us clarify the relation between variables used in these papers and the elliptic coordinates  (\ref{ElliptCase}), (\ref{Jul17}). 

To separate the wave equation in the metric (\ref{GenD1D5J}) with harmonic functions (\ref{D1D5Jaa}), one can use the coordinates $(r',\theta')$ which appear naturally if the D1--D5 solution is viewed as an extremal limit of a black hole \cite{cvet,cvetLars,tube}:
\bea\label{D1D5Jprime}
ds^2&=&-\frac{1}{H}\left(dt-\frac{aQ}{f}\sin^2\theta' d\phi\right)^2+\frac{1}{H}\left(du+\frac{aQ}{f}\cos^2\theta' d\psi\right)^2+dz_idz_i
\nonumber\\
&&+Hf\left[\frac{(dr')^2}{(r')^2+a^2}+(d\theta')^2\right]+H\left[(r'\cos\theta')^2d\psi^2+
(r^2+a^2)\sin^2\theta'd\phi^2\right],\nonumber\\
f&=&(r')^2+a^2\cos^2\theta',\qquad H=\alpha+\frac{Q}{f}.
\eea
The relation between $(r,\theta)$ and $(r',\theta')$ was found in \cite{MultiStr}\footnote{This comes from interchanging $(r,\theta)$ with 
$(r',\theta')$ in formula (4.7) of \cite{MultiStr} and (\ref{D1D5Jprime}) is obtained from setting $Q_1=Q_5$ in equation (5.12) in that paper.}
\bea\label{RThTrnls}
r=\sqrt{(r')^2+a^2\sin^2\theta'},\qquad \cos\theta=\frac{r'\cos\theta'}{\sqrt{(r')^2+a^2\sin^2\theta'}}.
\eea
Looking at the $(r',\theta')$ sector of the metric:
\bea
ds^2=(Q+f)\left[\frac{(dr')^2}{(r')^2+a^2}+(d\theta')^2\right]
\eea
we arrive at natural ``conformally--Cartesian" coordinates $(x,y)$:
\bea\label{RYprime}
r'=a\sinh x,\qquad \theta'=y.
\eea
Substituting this into (\ref{RThTrnls}) we find the expressions for $(r,\theta)$ in terms of $(x,y)$:
\bea
r=a\sqrt{\sinh^2 x+\sin^2y},\quad \cos\theta=\frac{\sinh x\cos y}{\sqrt{\sinh^2 x+\sin^2y}},
\eea
Using the definitions (\ref{defCmplVar}), we conclude that 
\bea
w&\equiv&\ln r+i\theta=\frac{1}{2}\ln\left[\sinh^2 x+\sin^2y\right]+i\arccos\left[\frac{\sinh x\cos y}{\sqrt{\sinh^2 x+\sin^2y}}\right]\nonumber\\
&=&\ln(\sinh z).
\eea
is a holomorphic function of $z=x+iy$, as expected from the general analysis presented in section \ref{SectPrcdr}. Inverting the last expression, we recover the relation (\ref{HolomElpt}). 
\bea
z=\ln\left[\frac{1}{2}\left(e^{w}+ \sqrt{e^{2{w}}+1}\right)\right].
\eea
Thus we conclude that coordinates (\ref{RYprime}) used in \cite{cvet,tube} are essentially the elliptic coordinates up to a minor redefinition of $r'$. We will come back to this feature in section \ref{StdPar}.

\section{Geodesics in bubbling geometries.}\label{Bubbles}

\renewcommand{\theequation}{7.\arabic{equation}}
\setcounter{equation}{0}
 
Given integrability of sigma model on $AdS_5\times S^5$, it is natural to look at deformations of this 
background which might preserve integrable structures. In particular, reference \cite{StepTs} demonstrated that deformation of $AdS_5\times S^5$ to asymptotically-flat geometry by adding one to the harmonic function destroys integrability of sigma model, although the Hamilton--Jacobi equation for geodesics remains separable. The extension from $AdS_5\times S^5$ to flat geometry is only possible if one choses flat metric on the worldvolume of D3 branes\footnote{This corresponds to field theory living on $R^{3,1}$, which is dual to the Poincare patch of $AdS_5$}, and in section \ref{SecGnrGds} we analyzed several classes of geometries produced by flat Dp branes. From the point of view of AdS/CFT correspondence, it is equally interesting to look at field theories on $R\times S^3$, which are dual to geometries produced by spherical D3 branes. The most symmetric geometry of this type is a direct product of global $AdS_5$ and a five dimensional sphere, but less symmetric examples are also known \cite{LLM,quarter}. In this section we will apply the techniques developed in section \ref{SecGnrGds} to identify the most general geometries of \cite{LLM} with separable geodesics. 
 
We begin with recalling the metrics of $1/2$--BPS geometries constructed in \cite{LLM}\footnote{These metrics are supported by the five--form field strength, and expression for $F_5$ can be found in \cite{LLM}. Notice that our notation in (\ref{GenBubble}) slightly differs from one in \cite{LLM}: we  replaced $y$ of \cite{LLM} by $Y$ to avoid the confusion with coordinate $y$ introduced in section \ref{SectPrcdr}.}:
\bea\label{GenBubbleM}
ds^2&=&-h^{-2}(dt+V_i dx^i)^2+h^2(dY^2+dx_1^2+dx_2^2)+Ye^G d\Omega_3^2+Ye^{-G} d{\tilde\Omega}_3^2,\\
h^{-2}&=&2Y\cosh G,\qquad 
YdV=\star_3 dz,\qquad
z=\frac{1}{2}\tanh G\nonumber
\eea
The solutions are parameterized by one function
$z(x_1,x_2,Y)$ that satisfies the Laplace equation,
\bea\label{bbLplM}
\d_i\d_i z+Y\d_Y\left(Y^{-1}\d_Y z\right)=0,
\eea
and obeys the boundary conditions 
\bea\label{DropletM}
z(Y=0)=\pm \frac{1}{2}.
\eea
The regions with $z=\frac{1}{2}$ form droplets in $(x_1,x_2)$ plane, and any configuration of droplets leads to the unique regular geometry. Solutions (\ref{GenBubbleM}) are dual to half--BPS states in $N=4$ super--Yang--Mills theory, and droplets in  $(x_1,x_2)$ plane correspond to eigenvalues of the matrix model describing such states 
\cite{beren} (see \cite{LLM} for detail).

\begin{figure}[]
\begin{minipage}[h!]{0.4\linewidth}
\center{\includegraphics[width=1\linewidth]{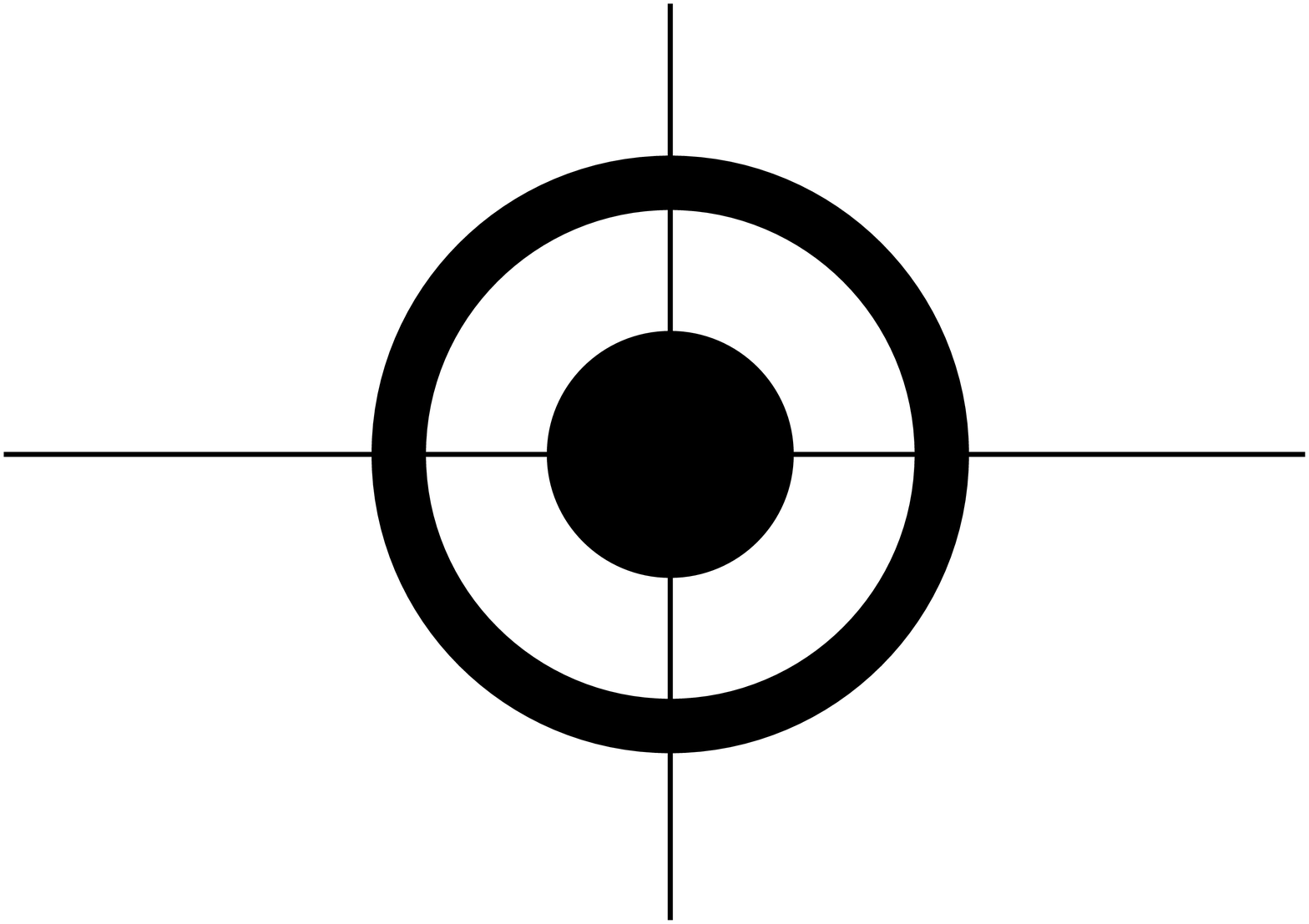}} (a) \\
\end{minipage}
\hfill
\begin{minipage}[h!]{0.25\linewidth}
\center{\includegraphics[width=1\linewidth]{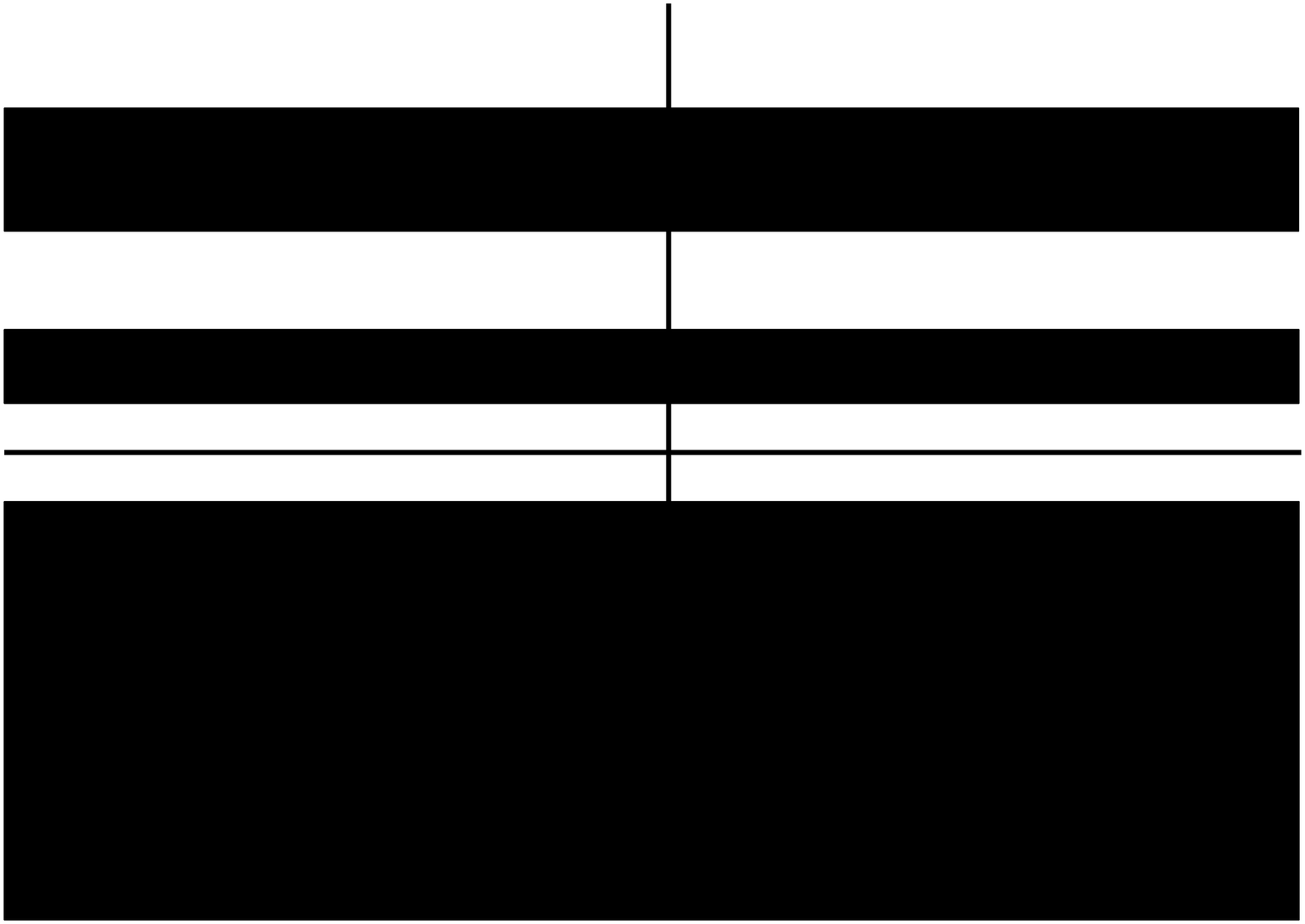}} (b) \\
\end{minipage}
\hfill
\begin{minipage}[h!]{0.25\linewidth}
\center{\includegraphics[width=1\linewidth]{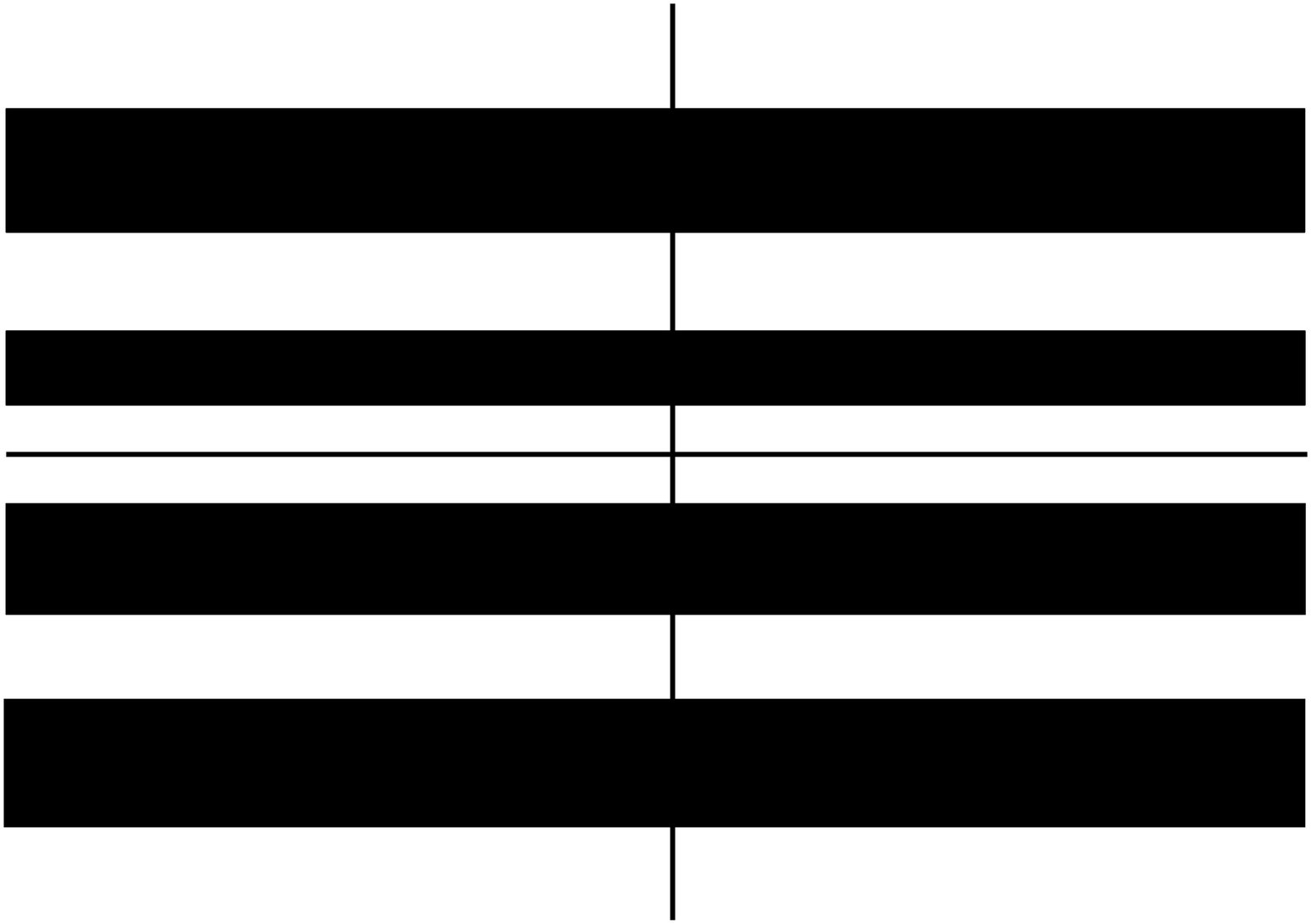}} (c) \\
\end{minipage}
\caption{Boundary conditions in the $Y=0$ plane corresponding to geometries with rotational and translational isometries, (\ref{RingRed}) and (\ref{StripRed}).}
\label{Fig:Bubbles}
\end{figure}

\begin{figure}
\centering
\begin{minipage}[h!]{0.4\linewidth}
\center{\includegraphics[width=1\linewidth]{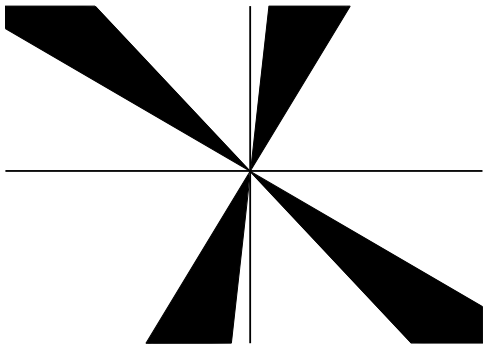}}
\end{minipage}
\caption{Boundary conditions in the $Y=0$ plane corresponding to geometries with $\d_r z=0$.}
\label{Fig:Wedges}
\end{figure}

A generic distribution of droplets in $(x_1,x_2)$ plane leads to solution (\ref{GenBubbleM}), which has a nontrivial dependence upon three coordinates 
$(x_1,x_2,Y)$. Repeating the arguments presented in section \ref{RedTo2D}, one can show that geodesics can only be integrable if at least one of these coordinates corresponds to a Killing direction. Such configurations can be obtained by performing a dimensional reduction of (\ref{GenBubbleM}) along one of the directions in $(x_1,x_2)$ plane. Only two such reductions are possible \footnote{There are also counterparts of (\ref{RingRed}) with $\d_r z=0$, which correspond to wedges in the $(x_1,x_2)$ plane ( figure \ref{Fig:Wedges}). However, such configurations lead to singular geometries, see \cite{Jabb} for further discussion. The counterpart of (\ref{StripRed}) with $\d_2 z=0$ is related to (\ref{StripRed}) by rotation.}:
\bea\label{RingRed}
\d_\phi z=0:&& x_1+ix_2=r\cos\theta e^{i\phi},\quad
Y=r\sin\theta,\quad 0\le\theta<\pi,\\
\label{StripRed}
\d_{1} z=0:&& x_2=r\cos\theta,\quad
Y=r\sin\theta,\quad 0\le\theta<\pi.
\eea
Reduction (\ref{RingRed}) corresponds to concentric rings in the $(x_1,x_2)$--plane (see figure \ref{Fig:Bubbles}(a)), and it describes excitations of $AdS_5\times S^5$. Reduction (\ref{StripRed}) corresponds to parallel strips in $(x_1,x_2)$--plane (see figure \ref{Fig:Bubbles}(b,c)), which can describe either excitations of the pp--wave or states of Yang--Mills theory on $S^1\times R$. The three cases depicted in figure \ref{Fig:Bubbles} are analyzed in the appendix \ref{AppBbl}, and here we just summarize the results. 

\begin{enumerate}[(a)]
\item{{\bf Geometries with 
AdS$_5\times$S$^5$ asymptotics.}\\
The boundary conditions depicted in figure \ref{Fig:Bubbles}(a) lead to geometries (\ref{GenBubbleM}), which are invariant under rotations in $(x_1,x_2)$ plane, and such solutions are conveniently formulated in terms of coordinates introduced in 
(\ref{RingRed}):
\bea
ds^2=-h^{-2}(dt+V_\phi d\phi)^2+h^2(dr^2+r^2d\theta^2+r^2\cos^2\theta d\phi^2)+
Ye^G d\Omega_3^2+Ye^{-G} d{\tilde\Omega}_3^2.
\nonumber
\eea
The complete solution of the Laplace equation (\ref{bbLplM}) and expression for $V_\phi$ for this case were found in \cite{LLM}:
\bea\label{zVa}
{z}=\frac{1}{2}+\frac{1}{2}\sum_{i=1}^n(-1)^{i+1}\left[\frac{r^2-R_i^2}{\sqrt{(r^2+R_i^2)^2-4R_i^2r^2\cos^2\theta}}-1\right],
\nonumber \\
V_\phi=-\frac{1}{2}\sum_{i=1}^n(-1)^{i+1}\left[ \frac{r^2+r_i^2}{\sqrt{(r^2+R_i^2)^2-4R_i^2r^2\cos^2\theta}} - 1 \right],
\eea 
Summation in (\ref{zVa}) is performed over $n$ circles with radii $R_i$, and following conventions of \cite{LLM} we will take $R_1$ to be the radius of the largest circle. For example, a disk corresponds to one circle, a ring to two circles, and so on. 

The HJ equation for the solutions specified by (\ref{zVa}) is analyzed in the appendix \ref{BubblesCircles}, where it is demonstrated that integrability leads to an infinite set of relations between radii $R_i$. Specifically, the expressions defined by 
\bea\label{RiRingA}
D_{p}\equiv\sum_{j=1}^n (-1)^{j+1} (R_{j})^p
\eea
must satisfy the relations (\ref{ReqnRing}):
\bea\label{ReqnRingA}
(D_2)^{k-1}D_{2(k+1)}=(D_4)^k.
\eea
As demonstrated in appendix \ref{BubblesCircles}, this requirement implies that $n<2$ in (\ref{zVa}), so variables separate only for flat space ($n=0$) and for AdS$_5\times$S$^5$ ($n=1$) (figure \ref{Fig:BubblesInt}(a)). Moreover, construction presented in the appendix \ref{BubblesCircles} gives the unique set of separable coordinates (\ref{BblElptCrdA}) for AdS$_5\times$S$^5$
\bea\label{BblElptCrdM}
\cosh x=\frac{\rho_++\rho_-}{2d},\quad \cos y=\frac{\rho_+-\rho_-}{2d},\quad
\rho_\pm=\sqrt{r^2+R_1^2\pm 2rR_1\cos\theta},
\eea
and their relation with standard parameterization of this manifold will be discussed in section \ref{StdPar}. 

}
\item{
{\bf Geometries with pp--wave asymptotics}\\
We will now discuss the geometries with translational $U(1)$ symmetry (\ref{StripRed}), which correspond to parallel strips in the $(x_1,x_2)$ plane (see figure \ref{Fig:Bubbles}(b,c)). It is convenient to distinguish two possibilities: $z$ can either approach different values $x_2\rightarrow \pm\infty$ (as in figure \ref{Fig:Bubbles}(b) or approach the same value on both sides (as in figure \ref{Fig:Bubbles}(c)). Here we will focus on the first option, which corresponds to geometries with plane wave asymptotics, and the second case will be discussed in part (c).

Pp--wave can be obtained as a limit of $AdS_5\times S^5$ geometry by taking the five--form flux to infinity \cite{BMN}. This limit has a clear representation in terms of boundary conditions in $(x_1,x_2)$ plane: taking the radius of a disk (figure \ref{Fig:BubblesInt}(a)) to infinity, we recover a half-filled plane corresponding to the pp--wave (see figure \ref{Fig:BubblesInt}(c)). Taking a similar limit for a system of concentric circles (figure \ref{Fig:Bubbles}(a)), we find excitations of pp--wave geometry by 
a system of parallel strips (see figure \ref{Fig:Bubbles}(b)). Since strings are integrable on the pp--wave geometry 
\cite{BMN}, it is natural to ask whether such integrability persists for the deformations represented in figure \ref{Fig:Bubbles}(b). We will now rule out integrability on the deformed backgrounds by demonstrating that even equations for massless geodesics are not integrable. 

Solutions of the Laplace equation (\ref{bbLplM}) corresponding to the boundary conditions depicted in figure \ref{Fig:Bubbles}(b) were found in \cite{LLM}, and their explicit form is given by (\ref{zVpwave}). Such solutions are parameterized by the strip boundaries, and the black strip number $i$ is located at $d_{2i-1}<x_2<d_{2i}$. In appendix \ref{StripsPlane} we use the techniques developed in section \ref{SecGnrGds} to demonstrate that the HJ equation can only be separable when $n=0$ in (\ref{zVpwave}), i.e., when the solution represents an unperturbed pp--wave:
\bea
ds^2&=&-2dt dx_1-(x^2+y^2)dt^2+dx^2+x^2d\Omega^2+dy^2+y^2d{\tilde\Omega}^2,\\
&&r_1=x,\quad r_2=y.\nonumber
\eea
Interestingly, some special solutions also separate for the pp--wave with an additional strip (see figure \ref{Fig:BubblesInt}(d)). Specifically, the geodesics which do not move on the spheres and along $x_2$ direction (i.e., geogesics with $p=0$, $L_1=L_2=0$ in (\ref{F26})) separate in coordinates $(x,y)$ defined by (\ref{hOneStripHalfPlane}):
\bea
x+iy&=&w+\ln\left[\frac{1}{2}\left(\sqrt{1-\frac{d_0}{le^w}}+1\right)\right] +\ln\left[\frac{1}{2}\left(\sqrt{1-\frac{d_2}{le^w}}+1\right)\right],\\
w&=&\ln\frac{r}{l}+i\theta.\nonumber
\eea

\begin{figure}[]
\begin{minipage}[h!]{0.2\linewidth}
\center{\includegraphics[width=1\linewidth]{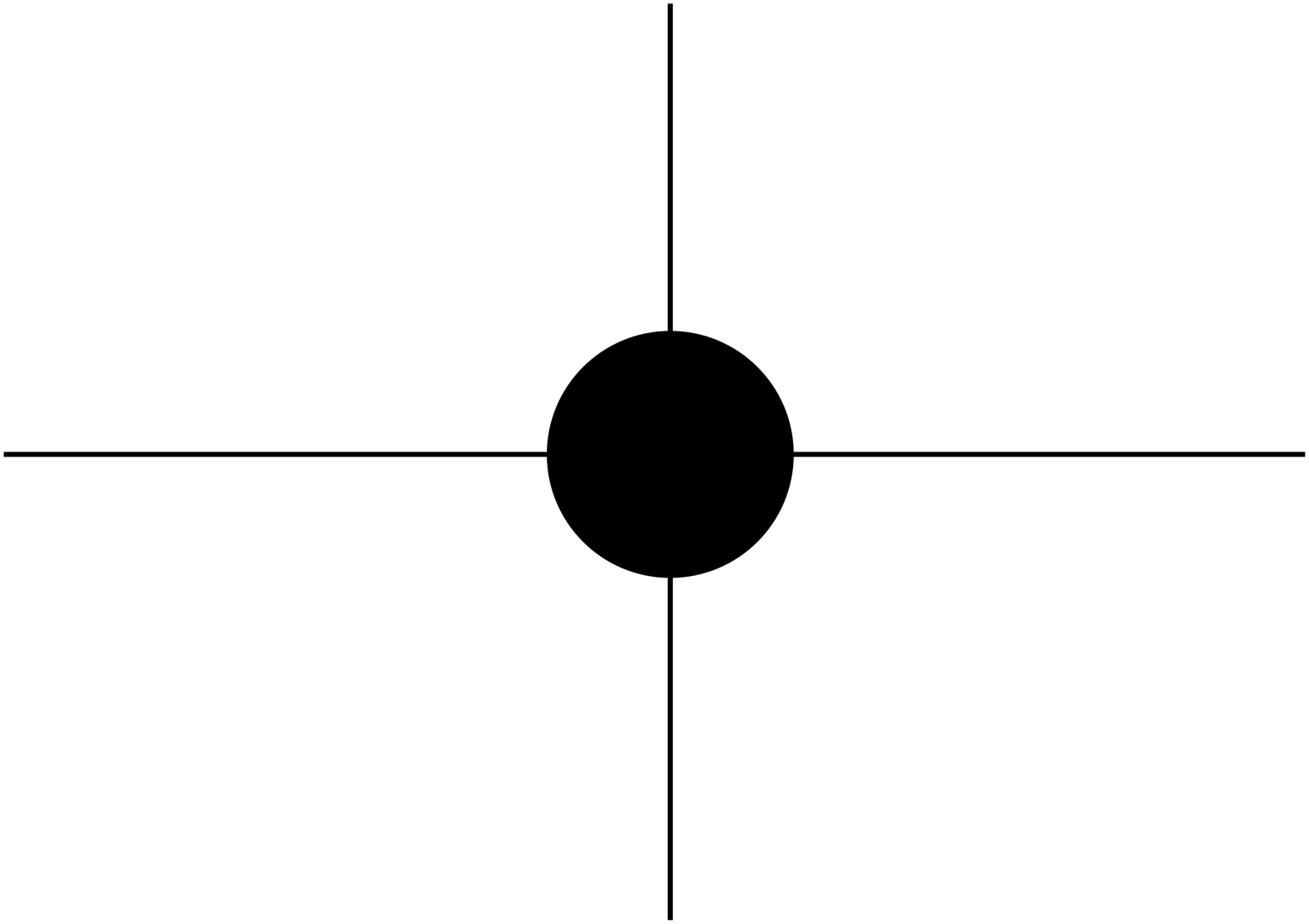}} (a) \\
\end{minipage}
\hfill
\begin{minipage}[h!]{0.2\linewidth}
\center{\includegraphics[width=1\linewidth]{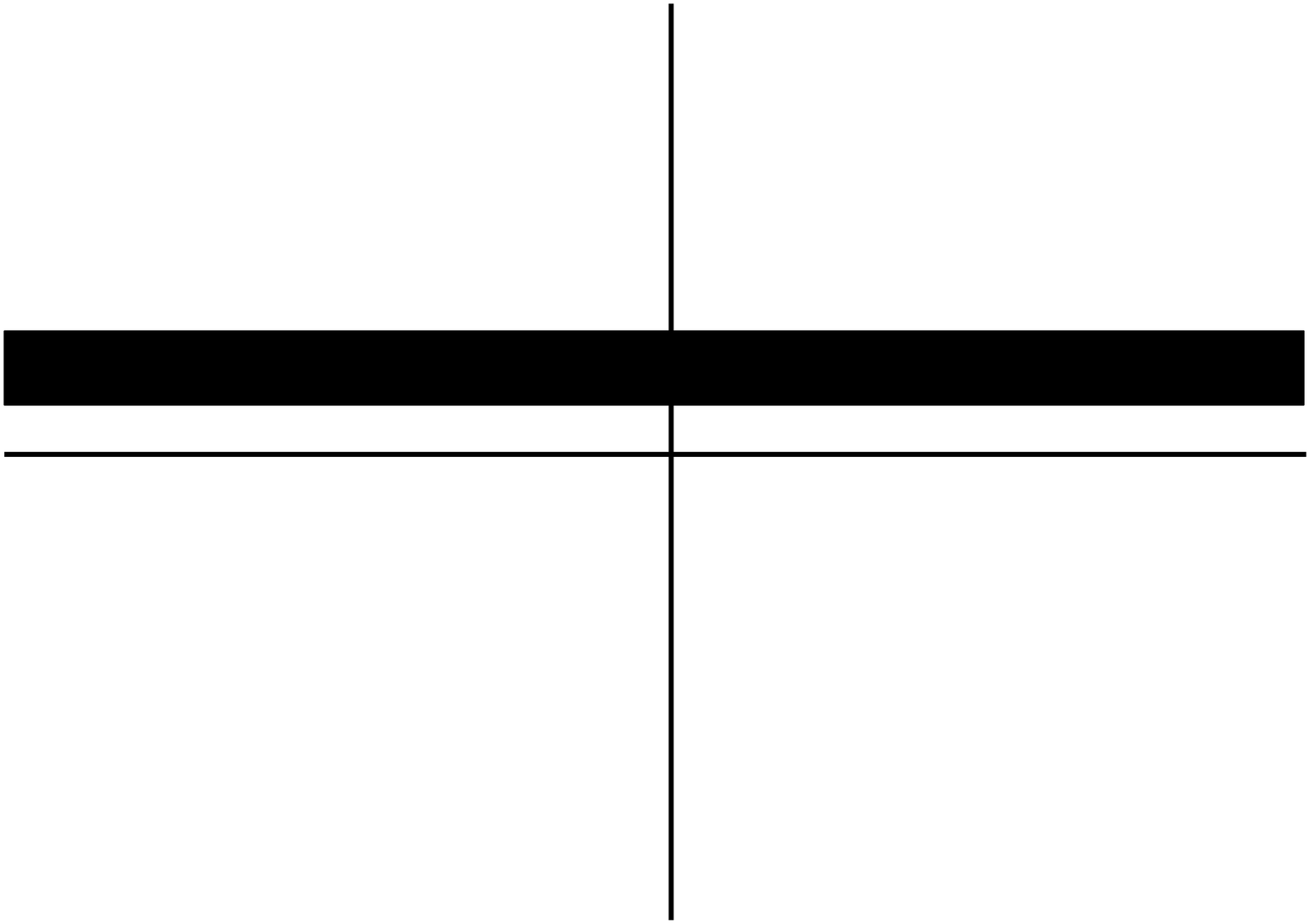}} (b) \\
\end{minipage}
\hfill
\begin{minipage}[h!]{0.2\linewidth}
\center{\includegraphics[width=1\linewidth]{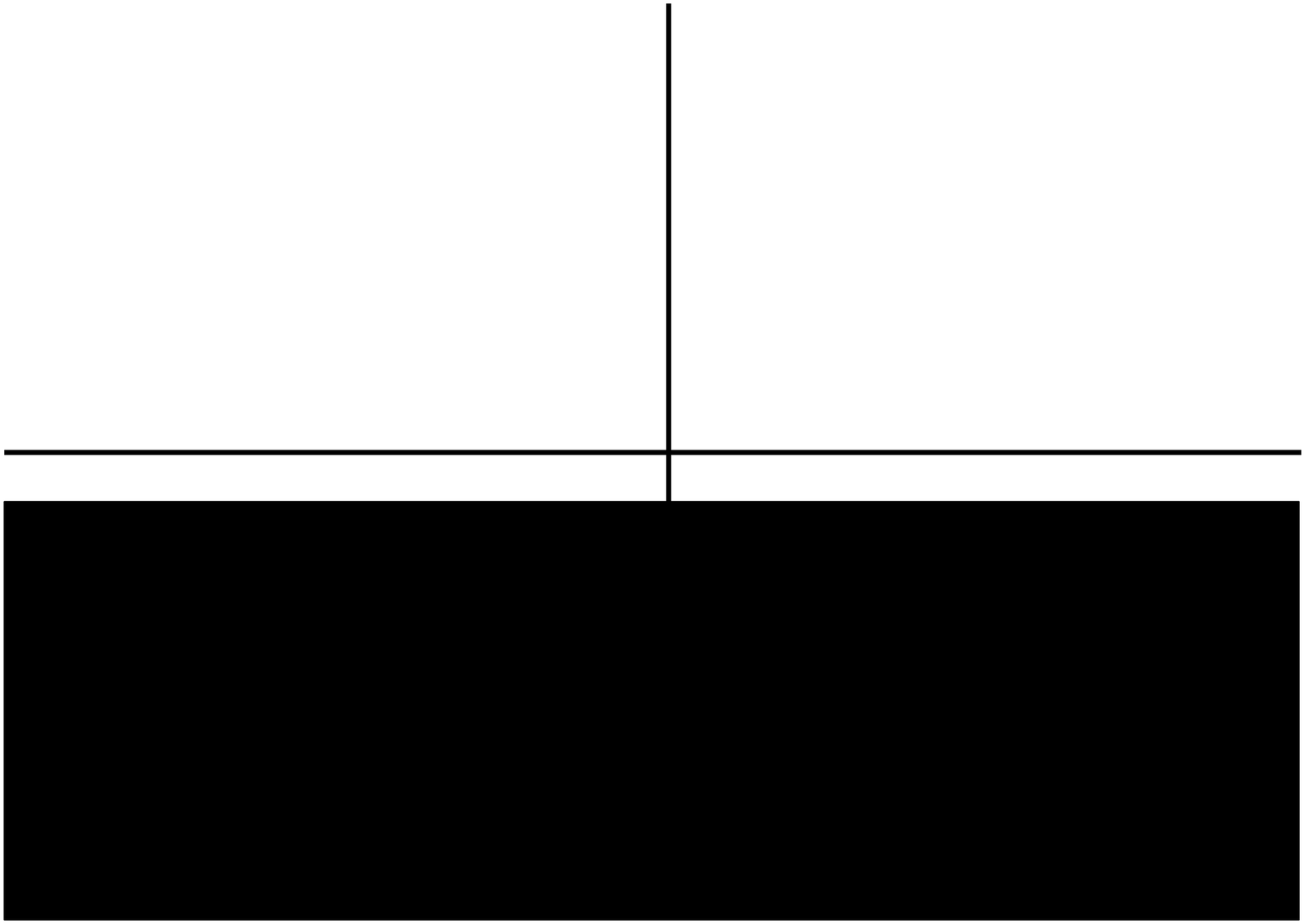}} (c) \\
\end{minipage}
\hfill
\begin{minipage}[h!]{0.2\linewidth}
\center{\includegraphics[width=1\linewidth]{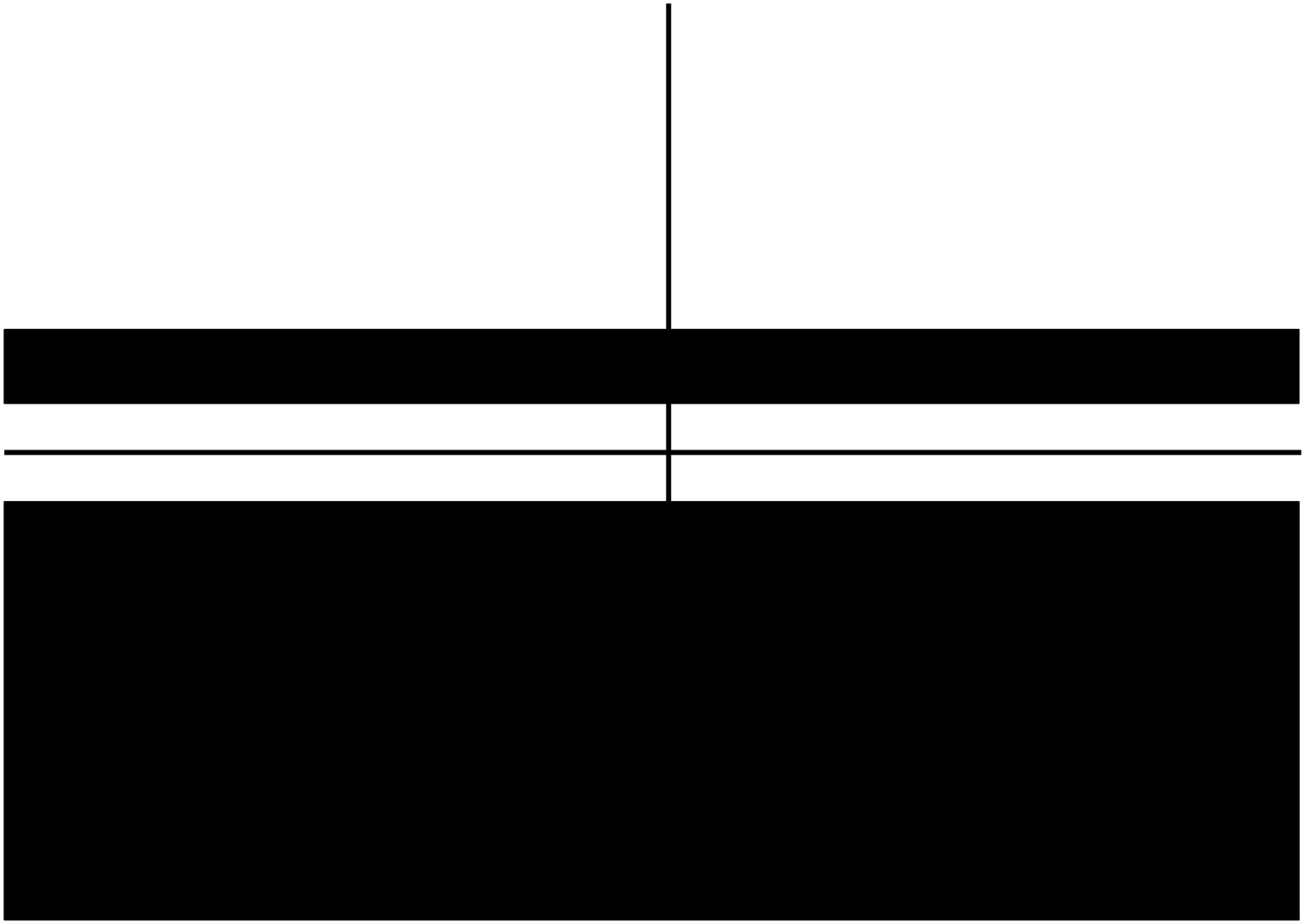}} (d) \\
\end{minipage}
\caption{Boundary conditions in the $Y=0$ plane corresponding to geometries with integrable geodesics.}
\label{Fig:BubblesInt}
\end{figure}

} 

\item{{\bf Geometries dual to SYM on a circle}\\
Finally we consider configuration depicted in figure \ref{Fig:Bubbles}(c). As discussed in \cite{LLM}, these configurations are dual to Yang--Mills theory on $S^3\times S^1\times R$, and since we are only keeping zero modes on the sphere, the solutions (\ref{GenBubbleM}) correspond to BPS states in two--dimensional gauge theory on a circle.   

The solution of the Laplace equations corresponding to figure \ref{Fig:Bubbles}(c) is given by (\ref{zVstrip}), and in appendix \ref{SecStrips}, it is shown that only a single strip (figure \ref{Fig:BubblesInt}(b)) leads to a separable HJ equation. For completeness we present the expression for the natural coordinates:
\bea
x+iy&=&2\ln\left[ \frac{1}{2}
\left(\sqrt{e^w-(d_2/l)}+e^{w/2}\right)\right]\\
w&=&\ln\frac{r}{l}+i\theta.\nonumber
\eea
}
\end{enumerate}
To summarize, we have demonstrated that from the infinite family of the 1/2--BPS geometries constructed in \cite{LLM}, only AdS$_5\times$S$^5$, pp--wave, and a single M2 brane give rise to integrable geodesics (figure \ref{Fig:BubblesInt}). This implies that the short strings can only be integrable on these backgrounds. Notice, however, that our results do not extend to equations of motion for D3 branes (which are expected to be integrable, as least for 1/2--BPS objects) since the HJ equation (\ref{HJone}) did not take into account coupling to the RR field. We also found that the separable coordinates $(x,y)$ given by (\ref{BblElptCrdM}) are the same elliptic coordinates (or their limits) as the one encountered in section \ref{SectPrcdr}, and in the next section we will discuss the relation between $(x,y)$ and the standard parameterization of AdS$_5\times$S$^5$.

\section{Elliptic coordinates and standard parameterization of AdS$_p\times$S$^q$}\label{StdPar}
\renewcommand{\theequation}{8.\arabic{equation}}
\setcounter{equation}{0}

In the last section we have demonstrated that the HJ equation for geodesics on 1/2--BPS geometries of \cite{LLM} separates only for 
AdS$_5\times$S$^5$ and for its pp--wave limit. Moreover, this separation happens in the elliptic coordinates (\ref{BblElptCrdM}). On the other hand, equations for supergravity fields on AdS$_5\times$S$^5$ are usually analyzed in the standard parameterization\footnote{We denoted an azimuthal direction on $S^5$ by $\chi$ to avoid confusion with coordinate $\theta$ introduced in (\ref{RingRed}).}:
\bea\label{StndAdS5}
ds^2=L^2\left[-\cosh^2\rho dt^2+
d\rho^2+\sinh^2\rho d\Omega_3^2+
d\chi^2+\cos^2\chi d{\tilde\phi}^2+\sin^2\chi d{\tilde\Omega}^2\right]
\eea
and the detailed study of \cite{Nvnh} uses the explicit  $SO(4,2)\times SO(6)$ symmetry of this metric to separate the resulting equations and to find the mass spectrum of supergravity modes on AdS$_5\times$S$^5$. This suggests a close relation between the elliptic coordinates and (\ref{StndAdS5}), which will be clarified in this subsection. We will also discuss the elliptic coordinates for AdS$_7\times$S$^4$, AdS$_4\times$S$^7$, and AdS$_3\times$S$^3$.

\bigskip
\noindent
{\bf Standard parameterization of AdS$_5\times$S$^5$.}

To relate the standard parameterization (\ref{StndAdS5}) with elliptic coordinates (\ref{BblElptCrdM}), we first recall the map between (\ref{StndAdS5}) and variables used in (\ref{GenBubbleM}) (see \cite{LLM}):
\bea
x_1+ix_2=L^2\cosh\rho\cos\chi e^{i\phi},\quad Y=L^2\sinh\rho\sin\chi, \quad 
\phi={\tilde\phi}-t.
\eea
Comparing this with (\ref{RingRed}), we relate the standard coordinates (\ref{StndAdS5}) with $(r,\theta)$,
\bea\label{Jul20}
L^2\sinh\rho\sin\chi=r\sin\theta,\qquad
L^2\cosh\rho\cos\chi=r\cos\theta,
\eea
and substitution into (\ref{BblElptCrdM}) gives
\bea\label{XYellAdS5}
\rho_\pm=R_1\left[\cosh\rho\pm\cos\chi\right]
\quad\Rightarrow\quad
x=\rho,\quad y=\chi.
\eea
We conclude that the standard coordinates (\ref{StndAdS5}) on 
AdS$_5\times$S$^5$ can be viewed as elliptic coordinates on the base of the LLM geometries (\ref{GenBubbleM}) in IIB supergravity. 

\bigskip
\noindent
{\bf 1/2--BPS geometries in M-theory.}

Let us now turn to the LLM geometries in M-theory \cite{LLM}:
\bea\label{bblMth}
ds_{11}^2&=&-4e^{2\lambda}(1+Y^2e^{-6\lambda})(dt+V_i dx^i)^2+\frac{e^{-4\lambda}}{1+Y^2e^{-6\lambda}}\left[ dY^2+e^D (dx_1^2+dx_2^2) \right]\nonumber\\
&+&4e^{2\lambda}d\Omega_5^2+Y^2e^{-4\lambda}d\tilde{\Omega}_2^2\nonumber\\
e^{-6\lambda}&=&\frac{\d_Y D}{Y(1-Y\d_Y D)},\qquad
V_i=\frac{1}{2}\epsilon_{ij}\d_jD.\nonumber
\eea
Metric (\ref{bblMth}) is parameterized by one function $D$ satisfying the Toda equation,
\bea\label{Toda}
(\d_1^2+\d_2^2)D+\d_Y^2 e^D=0,
\eea 
on a three--dimensional base,
\bea\label{TodaBase}
ds_{base}^2=dY^2+e^D\left[dx_1^2+dx_2^2\right],
\eea
and some known boundary conditions in the $Y=0$ plane. Although (\ref{Toda}) is much more complicated than the Laplace equation (\ref{bbLplM}), we expect that repetition of the arguments presented in section \ref{Bubbles} ensures that function $D$ can only depend on two rather than three variables, and in this case (\ref{Toda}) can be rewritten as a Laplace equation via a nonlocal change of variables \cite{Ward}. Specifically, for the rotationally--invariant case, it is convenient to rewrite the metric on the base as\footnote{We introduced an obvious notation: $x_1+ix_2=Re^{i\phi}$, $X=\ln R$.}
\bea\label{RotToTrnsl}
ds_{base}^2=dY^2+e^D\left[dR^2+R^2d\phi^2\right]=
dY^2+e^{\tilde D}\left[dX^2+d\phi^2\right]
\eea
Function ${\tilde D}=D+2\ln R$ defined above satisfied the same Toda equation (\ref{Toda}) as $D$, and in terms of $(X,Y)$ this equation can be rewritten as
\bea\label{TrnslToda}
\d_X^2{\tilde D}+\d_Y^2 e^{\tilde D}=0
\eea
The Toda equation with translational invariance along $x_1$ can also be written as (\ref{RotToTrnsl})--(\ref{TrnslToda}) after a replacement $(x_1,x_2,D)\rightarrow (\phi,X,{\tilde D})$. 
A nonlocal change of coordinates \cite{Ward,LLM},
\bea\label{Ward}
e^{\tilde D}=\zeta^2,\quad Y=\zeta\d_\zeta V,\quad
X=\d_\eta V,
\eea
maps the nonlinear Toda equation (\ref{TrnslToda}) into the Laplace equation for function $V$:
\bea
\zeta^{-1}\d_\zeta(\zeta\d_\zeta V)+\d_\eta^2 V=0.
\eea
Unfortunately, the boundary conditions for $V$ are rather complicated \cite{LLM} (see \cite{LinMald} for a detailed discussion), and simple expressions for $D$ and $V_\phi$ similar to (\ref{zVa}) are not known. Nevertheless, the results presented in sections \ref{Rotations} and \ref{Bubbles} strongly suggest that the HJ equation on the geometries (\ref{bblMth}) would only separate on the most symmetric backgrounds: AdS$_7\times$S$^4$, AdS$_4\times$S$^7$, and the pp wave. Let us discuss the relation between the standard parameterizations of these backgrounds and the elliptic coordinates.

Since elliptic coordinates are only defined in flat space, we begin with rewriting the $(X,Y)$ sector of (\ref{RotToTrnsl}) in a conformally--flat form. It turns out that this is accomplished by going to coordinates $(\xi,\eta)$ defined by (\ref{Ward}), and metric (\ref{RotToTrnsl}) becomes
\bea\label{Oct8}
ds_{base}^2=\zeta^2\left\{\left[(\d_\zeta\d_\eta V)^2+(\d^2_\eta V)^2\right]\left(d\zeta^2+d\eta^2\right)+
d\phi^2\right\}
\eea
After introducing the standard polar parameterization $(r, \theta)$,
\bea\label{DefZetaEta}
\zeta=r\sin\theta,\quad \eta=r\cos\theta,
\eea
one can define the elliptic coordinates by (\ref{ElliptCase}).  We will now compare these coordinates with the standard parameterization of AdS$_7\times$S$^4$ and AdS$_4\times$S$^7$.

\bigskip
\noindent
{\bf Standard parameterization of AdS$_7\times$S$^4$.}

Solution corresponding to AdS$_7\times$S$^4$,
\bea\label{AdS7S4}
ds^2=4L^2\left[d\rho^2+\sinh^2\rho d\Omega_5^2-\cosh^2\rho dt^2\right]+L^2\left[d\chi^2+\sin^2\chi d\Omega_2^2+\cos^2\chi d{\tilde\phi}^2\right],
\eea 
is given by equation (3.15) in \cite{LLM} with a replacement 
$r\rightarrow 2\sinh\rho$:
\bea
x_1+ix_2=\cosh^2\rho\cos\chi e^{i\phi},\qquad
Y=\frac{1}{L^3}\sinh^2\rho\sin\chi,\quad 
e^D=\frac{1}{L^{6}}\tanh^2\rho.
\eea
Substitution into (\ref{Ward}) gives the expression for $\zeta$ and equations for $V$: 
\bea\label{WardA}
\zeta=\frac{1}{2L^3}\sinh(2\rho)\cos\chi,\quad 
\d_\zeta V=\tanh\rho\tan\chi,\quad
\d_\eta V=\ln\left[\cosh^2\rho\cos\chi\right].
\eea
Coordinate $\eta$ can be determined using the relation $d\eta=\star_2 d\zeta$ (recall (\ref{Oct8}) and $(\rho,\chi)$ sector of (\ref{AdS7S4})):
\bea\label{etaA}
\eta=\frac{1}{2L^3}\cosh(2\rho)\sin\chi.
\eea
For completeness we also write the expression for $V$, which comes from integrating the differential equations in (\ref{WardA}), although it will not play any role in our discussion:
\bea
V=\frac{1}{2L^3}\sin\chi\left[\cosh(2\rho)\ln(\cos\chi\cosh^2\rho)-1\right]+
\frac{1}{2L^3}\ln\left[\tan\frac{\chi+\frac{\pi}{2}}{2}\right].\nonumber
\eea

To deduce the elliptic coordinates, we begin with finding the counterpart of equations (\ref{Jul20}) by rewriting the left--hand sides of (\ref{DefZetaEta}) in terms of $(\rho,\chi)$:
\bea\label{Jul20a}
\frac{1}{2L^3}\sinh(2\rho)\cos\chi=r\sin\theta,\qquad
\frac{1}{2L^3}\cosh(2\rho)\sin\chi=r\cos\theta.
\eea
Substitution of these relations into the definition (\ref{ElliptCase}) with $R_1=1/(2L^3)$, leads to identification of the elliptic coordinates 
$(x,y)$ with $(\rho,\chi)$ (compare with equation (\ref{XYellAdS5})):
\bea\label{XYellAdS7}
x=2\rho,\quad y=\frac{\pi}{2}-\chi.
\eea
This implies that the standard parameterization of AdS$_7\times$S$^4$ has a simple geometrical meaning: coordinates $(\rho,\chi)$ coincide with elliptic coordinates on the two--dimensional space spanned by $(\eta,\zeta)$.

\bigskip
\noindent
{\bf Standard parameterization of AdS$_4\times$S$^7$.}

Solution corresponding to AdS$_4\times$S$^7$,
\bea
ds^2&=&L^2\left[d\rho^2+\sinh^2\rho d\Omega_2^2-\cosh^2\rho dt^2\right]+4L^2\left[d\chi^2+\sin^2\chi d\Omega_5^2+\cos^2\chi d{\tilde\phi}^2\right],\nonumber
\eea 
is given by equation (3.16) in \cite{LLM} with a replacement 
$r\rightarrow 2\sinh\rho$:
\bea
x_1+ix_2=\sqrt{\cosh\rho}
\cos\chi e^{i\phi},\qquad
Y=\frac{1}{L^3}\sinh\rho\sin^2\chi,\quad 
e^D=\frac{4}{L^{6}}\cosh\rho\sin^2\chi.
\eea
Substitution into (\ref{Ward}) gives the expression for $\zeta$ and $\eta$:
\bea\label{WardB}
\zeta=\frac{1}{L^3}\cosh\rho\sin(2\chi),\quad
\eta=-\frac{1}{L^3}\sinh\rho\cos(2\chi).
\eea
For completeness we also give the equations for $V$ and their solution:
\bea
&&\qquad\qquad\d_\zeta V=\frac{1}{2}\tanh\rho\tan\chi,\quad
\d_\eta V=\ln\left[
\sqrt{\cosh\rho}\cos\chi\right]\nonumber\\
&&V=\frac{1}{2L^3}\left[\sinh\rho-2\arctan\left[\tanh\frac{\rho}{2}\right]-2\cos(2\chi)\sinh\rho\ln\left[\sqrt{\cosh\rho}\cos\chi\right]\right].
\nonumber
\eea

Combining (\ref{WardB}), and analog of (\ref{DefZetaEta})\footnote{We redefined angle $\theta$ in (\ref{DefZetaEta}). Alternatively, one can keep (\ref{DefZetaEta}) and shift $\theta$ in (\ref{ElliptCase}).},
\bea
\zeta=r\cos\theta,\quad \eta=-r\sin\theta.
\eea
and (\ref{ElliptCase}) with $R_1=1/(L^3)$, we identify the elliptic coordinates $(x,y)$ with $(\rho,\chi)$ (compare with equations (\ref{XYellAdS5}) and (\ref{XYellAdS7})):
\bea\label{XYellAdS4}
x=\rho,\quad y=\frac{\pi}{2}-2\chi.
\eea

\bigskip
\noindent
{\bf Standard parameterization of AdS$_3\times$S$^3$.}

As our final example of elliptic coordinates, we consider AdS$_3\times$S$^3$ in global parameterization, which can be obtained by taking the near horizon limit ($H\rightarrow Q/f$) in (\ref{D1D5Jprime}) \cite{BalMM}:
\bea\label{MetrAdS3}
ds^2&=&Q\left[-((r')^2+a^2)\frac{dt^2}{Q^2}+\frac{(dr')^2}{(r')^2+a^2}+\frac{(r'du)^2}{Q^2}+
(d\theta')^2+\sin^2\theta' d{\tilde\phi}^2+\cos^2\theta' d{\tilde\psi}^2\right],\nonumber\\
&&{\tilde\phi}=\phi+\frac{a}{Q}t,\qquad {\tilde\psi}=\psi+\frac{a}{Q}u.
\eea
Rewriting the metric in terms of the elliptic coordinates defined by 
(\ref{RYprime}):
\bea\label{AdS3Elps}
ds^2=Q\left[-\cosh^2 x\frac{(adt)^2}{Q^2}+dx^2+\sinh^2 x\frac{(adu)^2}{Q^2}+
dy^2+\sin^2y d{\tilde\phi}^2+\cos^2y d{\tilde\psi}^2\right],\nonumber\\
\eea
we conclude that these coordinates give the standard parameterization of AdS$_3\times$S$^3$.

\bigskip
\noindent
{\bf Pp--wave limits of AdS$_p\times$S$^q$.}

We conclude this section by commenting on the pp--wave limits of AdS$_p\times$S$^q$. 

The pp--wave limit of AdS$_5\times$S$^5$,
\bea\label{AdS5pp}
ds^2=-2dt dx_1-(r_1^2+r_2^2)dt^2+dr_1^2+r_1^2d\Omega^2+dr_2^2+r_2^2d{\tilde\Omega}^2,
\eea
is obtained by taking
\bea
z=\frac{1}{2}\frac{x_2}{\sqrt{x_2^2+Y^2}},\quad V=\frac{1}{2}\frac{1}{\sqrt{x_2^2+Y^2}}dx_1
\eea 
in (\ref{GenBubbleM}) and setting
\bea
Y=r_1r_2,\quad x_2=\frac{1}{2}(r_1^2-r_2^2).
\eea
This leads to a very simple relation for the polar coordinates defined by (\ref{StripRed})
\bea\label{PolarA5S5}
r_1=\sqrt{2}r\cos\frac{\theta}{2},\quad r_2=\sqrt{2}r\sin\frac{\theta}{2}.
\eea
Equation for geodesics in the geometry (\ref{AdS5pp}) separates in variables $(r_1,r_2)$, which can be obtained from the elliptic coordinates (\ref{Jul20})--(\ref{XYellAdS5}) by taking the pp--wave limit: 
\bea\label{PpAdS5}
L\rightarrow \infty,\qquad \mbox{fixed}\quad r_1=Lx,\quad r_2=Ly.
\eea
We conclude that in the pp--wave limit, the elliptic coordinate degenerate into the radii of the three--spheres. 

\bigskip

The pp--wave limit of AdS$_p\times$S$^q$ in M theory,
\bea\label{PPM}
ds^2=-2dtdx_1-(r_2^2+r_5^2)dt^2+dr_2^2+r_2^2d\Omega_2^2+
dr_5^2+r_5^2d{\Omega}_5^2
\eea
is given by equation (3.14) of \cite{LLM}:
\bea
Y=\frac{r_5^2 r_2}{4},\quad 
x_2=\frac{r_5^2}{4}-\frac{r_2^2}{2},\quad 
e^D=\frac{r_5^2}{4}
\eea
This translates into
\bea
\zeta=\frac{r_5}{2},\quad \eta=\frac{r_2}{2},\quad
V=\frac{r_2r_5^2}{8}-\frac{r_2^3}{12}=\zeta^2\eta-\frac{2}{3}\eta^3
\eea
via (\ref{Ward})\footnote{Since we are dealing with translational rather than rotational symmetry, ${\tilde D}=D$ and $X=x_2$ in (\ref{Ward})}. These expressions can be obtained from (\ref{WardA})--(\ref{etaA}) or from (\ref{WardB}) by taking the large--$L$ limits, and in both cases we arrive at a counterpart of (\ref{PpAdS5}):
\bea
L\rightarrow \infty,\qquad \mbox{fixed}\quad r_2=Lx,\quad r_5=L\left(\frac{\pi}{2}-y\right).
\eea
As in the case of the type IIB pp--waves, we conclude that the elliptic coordinates degenerate into the radii of the spheres. 

\bigskip

The pp--wave limit of the AdS$_3\times$S$^3$ geometry (\ref{AdS3Elps}) is obtained by 
writing
\bea
y=\frac{\hat y}{\sqrt{Q}},\quad x=\frac{\hat x}{\sqrt{Q}},\quad
u=\frac{Q\hat u}{a},\quad t=\frac{Q}{2a}\left(
{\hat x}^++\frac{{\hat x}^-}{Q}\right),\quad
{\tilde\psi}=\frac{1}{2}\left({\hat x}^+-\frac{{\hat x}^-}{Q}\right)
\eea
and sending $Q$ to infinity, while keeping all hatted variables fixed. This results in the metric
\bea
ds^2=-d{\hat x}^+d{\hat x}^--\frac{1}{4}({\hat x}^2+{\hat y}^2)(d{\hat x}^+)^2
+d{\hat x}^2+{\hat x}^2 d{\hat u}^2+
d{\hat y}^2+{\hat y}^2 d{\tilde\phi}^2.\nonumber
\eea
Once again, elliptic coordinates degenerate to the radii of the one--spheres. 

\bigskip

To summarize, we have demonstrated that in all examples of  AdS$_p\times$S$^q$, where the HJ equation separates between the sphere and AdS in global coordinates, such separation emerges as a particular case of integrability in elliptic coordinates, and standard parameterization of AdS$_p\times$S$^q$ coincides with elliptic coordinates on the relevant flat base. In the pp--wave limits of AdS$_p\times$S$^q$, the elliptic coordinates reduce to the radii of the appropriate spheres.

\section{Discussion}

Integrability of geodesics and Klein--Gordon equation has led to numerous insights into physics of black holes. While the black hole solutions are few and far between, the large classes of supersymmetric geometries are known, and in this article we have classified such solutions with integrable geodesics. This integrability is demonstrated to imply that the HJ equation must separate in the elliptic coordinates. 
For branes with flat worldvolumes, such separation, that extends the known result for the spherical coordinates, can only occur for special distributions of sources, which are analyzed in section \ref{SecGnrGds}. For the curved supersymmetric branes, the elliptic coordinates can only be introduced in the most symmetric cases, and as demonstrated in section \ref{Bubbles}, all these situations reduce to AdS$_p\times$S$^q$ or their pp--wave limits. 

Our results rule out integrability of N=4 SYM beyond the large N limit. Specifically, we proved that the excitations of strings around heavy supersymmetric states ($\Delta\sim N^2$) are not integrable. While this is consistent with general expectations, it is somewhat surprising that none of the 1/2--BPS geometries give rise to integrable sectors. It would be interesting to extend this result to states with fewer supersymmetries. 

Our results also have unfortunate consequences for the technical progress in the fuzzball program. While a large number of geometries corresponding to microscopic states of black holes have been constructed in the last decade, the detailed calculations of the absorption/emission rates have only been performed for the simplest cases. Such calculations are based on solving the Klein--Gordon equation, and as we demonstrated in section \ref{Rotations}, this equation, as well as the Hamilton--Jacobi equation for geodesics, cannot separate beyond the known cases. Our results do not imply that a study of geodesics on a particular background is hopeless. A lot of useful information can be extracted by performing numerical integration of the equations of motion and by studying some special configurations rather than generic geodesics.

\section*{Acknowledgements}

We thank Hai Lin and Filip Petrescu for useful discussions. This work is supported in part by NSF grant PHY-1316184.

\renewcommand{\theequation}{9.\arabic{equation}}
\setcounter{equation}{0}

\appendix

\section{Examples of separable Hamilton--Jacobi equations}
\renewcommand{\theequation}{A.\arabic{equation}}
\setcounter{equation}{0}

In this appendix we provide some technical details pertaining to derivation of the results presented in sections \ref{Examples} and \ref{SecGnrGds}. In particular, we write down the explicit expressions for $S_L$ on the sphere (\ref{SLSphere}), the complete integral for $R$ in the elliptic coordinates (\ref{RActEll}) and make the connection between the holomorphic function introduced in (\ref{holomZ}) and the standard elliptic coordinates (\ref{DefEllCoord}).

\subsection{Motion on a sphere}\label{AppSphere} 

While discussing separation of variables in the Hamilton--Jacobi equation, on several occasions we have encountered equation
\bea\label{HJSphAp}
{h^{ij}}\frac{\d S^{(k)}_{L_k}}{\d y^i}
\frac{\d S^{(k)}_{L_k}}{\d y^j}=L_k^2,\qquad
\eea
on a $k$--dimensional sphere $S^k$ (see, for example, (\ref{AngHJ}), (\ref{HJSphere})). In this appendix we will write the complete integral of (\ref{HJSphere}) using induction. 

Writing the metric on $S^k$ as
\bea
h_{ij}dy^idy^j=dy_k^2+\sin^2 y_k d\Omega_{k-1}^2,
\eea
and splitting $S^{(k)}_{L}$ as
\bea
S^{(k)}_{L_k}(y_1,\dots,y_k)=F_k(y_k)+
S^{(k-1)}_{L_{k-1}}(y_1,\dots,y_{k-1}),
\eea
we can rewrite equation (\ref{HJSphAp}) as
\bea
&&(F'_k)^2+\frac{L^2_{k-1}}{\sin^2y_k}=L_k^2,\\
&&{h^{ij}}\frac{\d S^{(k-1)}_{L_{k-1}}}{\d y^i}
\frac{\d S^{(k-1)}_{L_{k-1}}}{\d y^j}=L_{k-1}^2,\qquad
\eea
Solving the first equation\footnote{Although the integral in (\ref{AnsFk}) can be performed, the result is not very illuminating.},
\bea\label{AnsFk}
F_k=\int \frac{dy_k}{\sin y_k}\sqrt{L_k^2 \sin^2y_k-L^2_{k-1}},
\eea
and applying induction, we arrive at the complete integral of (\ref{HJSphAp}) that depends on $k$ parameters $L_j$:
\bea\label{SLSphere}
S^{(k)}_{L_k}(y_1,\dots,y_k)=
\sum_{j=2}^{k}\int \frac{dy_j}{\sin y_j}\sqrt{L_j^2 \sin^2y_k-L^2_{j-1}}+
L_1y_1.
\eea
This explicit solution should be substituted into (\ref{HJact}).

\subsection{Two--center potential and elliptic coordinates}
\label{SprElliptic}

In section \ref{Examples} we reviewed separation of variables in elliptic coordinates, and 
here we present some details of that construction. 

The motion of a particle in the geometry produced by two stacks of Dp branes is governed by the Hamilton--Jacobi equation (\ref{EllHJ}) 
\bea\label{EllHJAp}
(\d_r R)^2+\frac{1}{r^2}(\d_\theta R)^2+\frac{L^2}{r^2\sin^2\theta}+ H p_\mu p^\mu=0,
\eea
where $H$ is given by (\ref{EllHarm})
\bea\label{EllHarmAp}
H=a+\frac{Q}{\rho_+^{6-p}}+\frac{Q}{\rho_-^{6-p}},\qquad
\rho_\pm=\sqrt{r^2+d^2\pm 2rd\cos\theta}
\eea
To rewrite (\ref{EllHJAp}) in terms the elliptic coordinates $(\xi,\eta)$ defined by (\ref{DefEllCoord}), we notice that 
(\ref{EllHJAp}) can be viewed as a Hamilton--Jacobi equation corresponding to an effective Lagrangian for $r$ and $\theta$:
\bea
{L}_{eff}={\dot r}^2+r^2{\dot\theta}^2-
\frac{L^2}{r^2\sin^2\theta}-H p_\mu p^\mu.
\eea
Rewriting the last expression in terms of $\xi$, $\eta$,
\bea
{L}_{eff}=d^2(\xi^2-\eta^2)\left[\frac{\dot\xi^2}{\xi^2-1}+\frac{\dot\eta^2}{1-\eta^2}\right]-
\frac{L^2}{d^2(\xi^2-1)(1-\eta^2)}-H p_\mu p^\mu
\eea
and going back to the Hamilton--Jacobi equation for $S(\xi,\eta)$, we find
\bea\label{HJell}
\frac{1}{d^2(\xi^2-\eta^2)}\left[(\xi^2-1)(\d_\xi R)^2+
(1-\eta^2)(\d_\eta R)^2+\left(\frac{1}{\xi^2-1}+\frac{1}{1-\eta^2}\right)L^2\right]+H p_\mu p^\mu=0.\nonumber\\
\eea
This equation separates in variables $(\xi,\eta)$ is and only if
\bea
(\xi^2-\eta^2)H=U_1(\xi)+U_2(\eta).
\eea
Rewriting the harmonic function (\ref{EllHarmAp}) in terms of elliptic coordinates, 
we find that the left hand side of the last expression,
\bea
(\xi^2-\eta^2)H=(\xi^2-\eta^2)\left[a+
\frac{Q}{[d(\xi+\eta)]^{6-p}}+\frac{Q}{[d(\xi-\eta)]^{6-p}}\right],
\eea
separates only for $p=5$. In this case equation (\ref{HJell}) becomes
\bea\label{SepHJell}
&&\left[(\xi^2-1)(\d_\xi R)^2+\frac{L^2}{\xi^2-1}+
p_\mu p^\mu(a\xi^2+2Q\xi)\right]
\nonumber\\
&&\qquad+
\left[(1-\eta^2)(\d_\eta R)^2+\frac{L^2}{1-\eta^2}
-p_\mu p^\mu a\eta^2\right]=0,
\eea
and its complete integral is
\bea\label{RActEll}
R&=&\int \frac{d\xi}{\sqrt{\xi^2-1}}\left[-\la-\frac{L^2}{\xi^2-1}+M^2(a\xi^2+2Q\xi)\right]^{1/2}\nonumber\\
&&+\int \frac{d\eta}{\sqrt{1-\eta^2}}\left[
\la-\frac{L^2}{1-\eta^2}
-M^2 a\eta^2
\right]^{1/2}
\eea
Here $\lambda$ is a separation constant and 
$M^2=-p_\mu p^\mu$. 

To embed the elliptic coordinates in the general framework presented in section \ref{SectPrcdr}, we have to find the holomorphic function $h(w)$ defined by (\ref{holomZ}). Comparing the kinetic terms in (\ref{HJell}) with equation (\ref{XYdiff}), we 
find the relation
\bea
\frac{1}{d^2(\xi^2-\eta^2)}\left[(\xi^2-1)(\d_\xi R)^2+
(1-\eta^2)(\d_\eta R)^2\right]=\frac{1}{A(x,y)}\left[(\d_x {\tilde R})^2+(\d_y {\tilde R})^2\right],\nonumber
\eea
which leads to expressions for $(\xi,\eta)$ in terms of $(x,y)$:
\bea\label{XiEta}
\xi=\cosh x,\qquad \eta=\cos y.
\eea
The relation between coordinates $(r,\theta)$ and $(x,y)$ looks rather complicated (see (\ref{DefEllCoord})),
\bea\label{coshOne}
&&\cosh x= \xi\equiv\frac{\rho_++\rho_-}{2d},\qquad \cos y=\eta\equiv\frac{\rho_+-\rho_-}{2d}\\
\label{coshTwo}
&&\rho_\pm=\sqrt{r^2+d^2\pm 2rd\cos\theta},
\eea
but it can be simplified by making use of the complex variables (\ref{defCmplVar}). 

First we rewrite equations (\ref{coshOne}) 
as an expression for $z$ in terms of $\xi$ and $\eta$:
\bea\label{zInterm}
&&e^x=\xi+\sqrt{\xi^2-1},\qquad e^{iy}=\eta+i\sqrt{1-\eta^2}\nonumber\\
&&z=x+iy=\ln\left[(\xi+\sqrt{\xi^2-1})(\eta+i\sqrt{1-\eta^2})\right]
\eea
Next we recall the definition (\ref{defCmplVar}) of the complex variable $w$ and use (\ref{coshTwo}) to write $\rho_\pm$ in terms of it:
\bea
\rho_+=\sqrt{(r+de^{i\theta})
(r+de^{-i\theta})}=d\sqrt{\left(\frac{l}{d}e^{\bar w}+1\right)
\left(\frac{l}{d}e^w+1\right)},\quad
\rho_-=d\sqrt{\left(\frac{l}{d}e^{\bar w}-1\right)\left(\frac{l}{d}e^w-1\right)}.\nonumber
\eea
To simplify the expressions for various ingredients appearing in (\ref{zInterm}) it is convenient to define holomorphic functions $W_\pm$:
\bea
W_+\equiv \sqrt{\frac{l}{d}e^w+1},\qquad W_-\equiv \sqrt{\frac{l}{d}e^w-1}.
\eea
Then we find
\bea
&&\xi=\frac{W_+{\overline{W}_+}+W_-{\overline{W}_-}}{2},
\qquad
\eta=\frac{W_+{\overline{W}_+}-W_-{\overline{W}_-}}{2}.\nonumber\\
&&\sqrt{\xi^2-1}=
\frac{W_-{\overline{W}_+}+W_+{\overline{W}_-}}{2},\qquad
\sqrt{1-\eta^2}=
\frac{W_-{\overline{W}_+}-W_+{\overline{W}_-}}{2i}.
\nonumber
\eea
Substitution of these results into (\ref{zInterm}) leads to the desired relation between $z$ and $w$:
\bea\label{EllEqnZ}
z&=&\ln\left[\frac{1}{4}(W_++W_-)^2((\overline{W}_+)^2-(\overline{W}_-)^2)
\right]\nonumber\\
&=&\ln\left[\frac{l}{d}e^w+
\sqrt{\frac{l^2}{d^2}e^{2w}-1}\right]
\eea
Finally, the asymptotic behavior (\ref{BoundCond}) determines $l$ in terms of $d$: $l=d/2$. 

As expected from the discussion in section \ref{SectPrcdr}, $z$ turns out to be a holomorphic function: 
\bea\label{EllEqnZpr}
z=h(w)=\ln\left[\frac{1}{2}\left\{e^w+
\sqrt{e^{2w}-4}\right\}\right].
\eea
Finally, by comparing (\ref{SepHJell}) with (\ref{XYdiff}), (\ref{XYalg}), we extract the expressions for the potentials $U_1(x)$ and $U_2(y)$:
\bea\label{EllPot}
U_1(x)&=&\frac{L^2}{\sinh^2 x}-
M^2(a\cosh^2 x+2Q\cosh x)\nonumber\\
U_2(y)&=&\frac{L^2}{\sin^2 y}+M^2 a\cos^2y.
\eea 

\bigskip

To summarize, in this section we demonstrated that the HJ equation (\ref{EllHJAp}) with $H$ given by (\ref{EllHarmAp}) separates in the elliptic coordinates (\ref{DefEllCoord}), (\ref{zInterm}), and the relevant potentials are given by (\ref{EllPot}). We also embedded the elliptic coordinates into the general discussion presented in section \ref{SectPrcdr} by deriving equation (\ref{EllEqnZpr}).

\section{Ellipsoidal coordinates}\label{AppEllips}
\renewcommand{\theequation}{B.\arabic{equation}}
\setcounter{equation}{0}

As demonstrated in section \ref{RedTo2D}, if the HJ equation (\ref{HJaa}) separates for $k>2$, such separation must occur in ellipsoidal coordinates, including degenerate cases. In this appendix we will demonstrate that a combination of the Laplace equation (\ref{LaplRk}) and the requirement (\ref{NoRotation}) rules out such separation. To avoid unnecessary complications, we will first give the detailed discussion of the $k=3$ case, and in section \ref{AppElpKG3} we will comment on minor changes which emerge from generalization to $k>3$.

\subsection{Ellipsoidal coordinates for $k=3$}

The ellipsoidal coordinates have been introduced by Jacobi \cite{Jacobi}, and there are several equivalent definitions. 
We will follow the notation of \cite{LLv8}. 

Ellipsoidal coordinates, $(x_1,x_2,x_3)$ are defined as three solutions of a cubic equation for $x$:
\bea
\frac{r_1^2}{x+a^2}+\frac{r_2^2}{x+b^2}+
\frac{r_3^2}{x+c^2}=1,
\eea
where $a>b>c>0$ and $r_1, r_2, r_3$ correspond to our $r_k$ from (\ref{BaseMetr}). The roots are arranged in the following order:
\bea\label{ElpsdRng}
x_1\ge-c^2\ge x_2\ge -b^2\ge x_3\ge -a^2
\eea
Coordinates $(r_1,r_2,r_3)$ can be expressed through $(x_1,x_2,x_3)$ by\footnote{We recall that $r_j$ must be non-negative if $d_j>0$ in (\ref{BaseMetr}), otherwise $r_j$ varies from minus infinity to infinity.}
\bea
r_1=\pm \left[\frac{(x_1+a^2)(x_2+a^2)(x_3+a^2)}{
(b^2-a^2)(c^2-a^2)}\right]^{1/2},\quad
r_2=\pm \left[\frac{(x_1+b^2)(x_2+b^2)(x_3+b^2)}{
(c^2-b^2)(a^2-b^2)}\right]^{1/2},\nonumber
\eea
\bea
r_3=\pm \left[\frac{(x_1+c^2)(x_2+c^2)(x_3+c^2)}{
(a^2-c^2)(b^2-c^2)}\right]^{1/2}.\nonumber
\eea
In terms of the elliptic coordinates, the metric of the flat three dimensional space becomes
\bea\label{IntroEpsd}
ds^2&=&dx_1^2+dr_2^2+dr_3^2=h_1^2 dx_1^2+h_2^2 dx_2^2+h_3^2 dx_3^2,\\
h_1^2&=&\frac{(x_1-x_2)(x_1-x_3)}{4R_1},\quad 
R_1=(x_1+a^2)(x_1+b^2)(x_1+c^2)\nonumber
\eea
Expressions for $h_2,h_3,R_2,R_3$ are obtained by making a cyclic permutation of indices. To simplify the formulas appearing below, it is convenient to introduce $d_{ij}\equiv x_i-x_j$. 

In ellipsoidal coordinates equation (\ref{HJaa}) becomes
\bea\label{EllisoidHJ}
\frac{4}{d_{12}d_{13}d_{23}}
\left[
d_{23}R_1(\d_1 S)^2+d_{31}R_2(\d_2 S)^2+
d_{12}R_3(\d_3 S)^2\right]+\sum_{j=1}^3 \frac{L_j^2}{r_j^2}+Hp_\mu p^\mu=0.
\eea
Before imposing the Laplace equation (\ref{LaplRk}) we will demonstrate that separation requires that $d_1=d_2=d_3=0$ in (\ref{BaseMetr}). Indeed, the separation implies that
\bea\label{AsmpSpr}
S(x_1,x_2,x_3)=S_1(x_1)+S_2(x_2)+S_3(x_3).
\eea
Then multiplying equation (\ref{EllisoidHJ}) by $d_{12}d_{13}d_{23}$ and applying 
$\d_1^2\d_2^2$ to the result, we find a relation which does not involve $S$:
\bea\label{IntCndH}
\d_1^2\d_2^2\left[d_{12}d_{13}d_{23}\left\{
\sum_{j=1}^3 \frac{L_j^2}{r_j^2}+Hp_\mu p^\mu
\right\}\right]=0.
\eea
Let us assume that $d_1>0$ in (\ref{BaseMetr}), then the last relation must hold for all values of $L_1$, thus 
\bea
\d_1^2\d_2^2\left[\frac{d_{12}d_{13}d_{23}}{r_1^2}
\right]=\d^2_1\d_2^2\left[\frac{(b^2-a^2)(c^2-a^2)d_{12}d_{13}d_{23}}{(x_1+a^2)(x_2+a^2)(x_3+a^2)}\right]=0.
\eea
This condition can only be satisfied if $a=b$, this degenerate case corresponds to oblate spheroidal coordinates:
\bea\label{ProlSpher}
r_1+ir_2=\left[\frac{(x_1+a^2)(x_2+a^2)}{
a^2-c^2}\right]^{1/2}e^{i\phi},\qquad
r_3=\pm \left[\frac{(x_1+c^2)(x_2+c^2)}{
c^2-a^2}\right]^{1/2}
\eea 
We will now demonstrate that separation of the HJ equation in spheroidal coordinates implies that $\d_\phi H=0$, i.e., violation of (\ref{NoRotation}) for $i=1,j=2$. This will falsify our assumption $d_1>0$, and similar arguments will show that separability of the HJ equation requires $d_2=d_3=0$.

To simplify notation and to connect to our discussion in section \ref{SectPrcdr} it is convenient to use an alternative form of the oblate spheroidal coordinates (\ref{ProlSpher}):
\bea
x_1=(a^2-c^2)\cosh^2 x-a^2,\quad 
x_2=(a^2-c^2)\cos^2 y-a^2
\eea 
This gives expressions for the radii:
\bea
r_1+ir_2=A\cosh x \cos y e^{i\phi},\quad r_3=\pm A\sinh x\sin y,\quad A=\sqrt{a^2-c^2},
\eea
and the metric becomes
\bea
ds^2=A^2(\sinh^2 x+\sin^2 y)(dx^2+dy^2)+A^2\cosh^2 x\cos^2 y d\phi^2
\eea
Notice that these are precisely the elliptic coordinates found in section \ref{SectPrcdr}. The HJ equation (\ref{HJaa}) in coordinates $(x,y)$ has the form
\bea\label{HJaaxy}
\frac{(\d_x S)^2+
(\d_y S)^2}{A^2(\sinh^2 x+\sin^2 y)}+
\frac{1}{A^2\cosh^2 x\cos^2 y}(\d_\phi S)^2
+\sum_j^3\frac{L_j^2}{r_j^2}+ H p_\mu p^\mu=0,
\eea
and it does not separate unless 
\bea
H\cosh^2 x\cos^2 y=X(x,y)+\Phi(\phi)
\eea
Recalling that at large values of $x$ function $H$ does not depend on $\phi$ (see (\ref{Hasymt})), we conclude that $\Phi'(\phi)=0$, then $\d_\phi H=0$ everywhere. As already mentioned, this relation violates the condition (\ref{NoRotation}), so our assumption $d_1>0$ was false. Repeating this arguments for the remaining two spheres, we conclude that $d_1=d_2=d_3=0$ in (\ref{BaseMetr}) and (\ref{LaplRk}).

\bigskip

We will now combine the integrability condition (\ref{IntCndH}),
\bea\label{IntCndHa}
\d_i^2\d_j^2\left[d_{12}d_{13}d_{23}H\right]=0,
\eea
with Laplace equation (\ref{LaplRk}) to rule out separability of the HJ equation in ellipsoidal coordinates,
\bea\label{ElpsdHJ}
\frac{4}{d_{12}d_{13}d_{23}}
\left[
R_1d_{23}(\d_1 S)^2+d_{31}R_2(\d_2 S)^2+
d_{12}R_3(\d_3 S)^2\right]+Hp_\mu p^\mu=0.
\eea
Since we have already established that $d_1=d_2=d_3=0$, equation (\ref{LaplRk}) reduces to the Laplace equation on a flat three--dimensional space formed by $(r_1,r_2,r_3)$, and it can be rewritten in terms of ellipsoidal coordinates using (\ref{IntroEpsd}):
\bea\label{ElpsdLapl}
\frac{4}{d_{12}d_{31}d_{23}}\left[
d_{23}\sqrt{R_1}\d_1(\sqrt{R_1}\d_1 H)+d_{31}\sqrt{R_2}\d_2(\sqrt{R_2}\d_2 H)+
d_{12}\sqrt{R_3}\d_3(\sqrt{R_3}\d_3 H)\right]=0.
\nonumber\\
\eea
In the remaining part of this subsection we will demonstrate that equations (\ref{IntCndHa}), (\ref{ElpsdLapl}), (\ref{NoRotation}), (\ref{Hasymt}) are inconsistent with separability of (\ref{ElpsdHJ}).

Let us assume that the HJ equation (\ref{ElpsdHJ}) separates. Although the ellipsoidal coordinates are restricted by (\ref{ElpsdRng}) the formal separation (\ref{AsmpSpr}) must persist beyond this range. In particular, it is convenient to look at the limit $x_1\rightarrow x_2$, where $x_2$ is kept as a free parameter. The Laplace equation (\ref{ElpsdLapl}) guarantees that $H$ remains finite in this limit, then equation (\ref{ElpsdHJ}) reduces to
\bea\label{x1ISx2}
\left[R_1(S_1')^2-R_2(S_2')^2\right]_{x_1=x_2}=0
\eea
This implies that
\bea
S_1(x_1)=F(x_1),\qquad S_2(x_2)=F(x_2)
\eea
with the same function $F$. Repeating this argument for $x_1=x_3$, we conclude that 
\bea
S_3(x_3)=F(x_3).
\eea
Rewriting (\ref{ElpsdHJ}) in terms of $F$, we conclude that $H$ must be symmetric under interchange of its arguments\footnote{Of course, such formal interchange takes us outside of the physical range (\ref{ElpsdRng}).}, and combination of this symmetry with integrability conditions (\ref{IntCndHa}) leads to severe restrictions on the form of $H$.

Taking $(i,j)=(1,2)$ in (\ref{IntCndHa}), and solving the resulting equation, we find
\bea
H=\frac{1}{d_{12}d_{13}d_{23}}\left[G_1(x_1,x_3)x_2+G_2(x_2,x_3)x_1
+G_{01}(x_1,x_3)+G_{02}(x_2,x_3)\right],
\eea
where $G_1,G_2,G_{01},G_{02}$ are some undetermined functions of their arguments. Applying (\ref{IntCndHa}) with different values of $(i,j)$, we find further restrictions on the form of $H$:
\bea
H=\frac{1}{d_{12}d_{13}d_{23}}\left[
\sum_{i\ne j}G_{ij}(x_i)x_j+\sum_i G_{0i}(x_i)\right]
\eea
Symmetry of $H$ under interchange of any pair of coordinates implies that $G_{0i}=0$, matrix $G_{ij}$ is anti-symmetric, 
and all $G_{ij}$ can be reduced to a single function\footnote{Specifically, $G_{12}=-G_{21}=G_{23}=-G_{32}=G_{31}=-G_{13}$ so it is convenient to introduce $G\equiv G_{12}$.} $G$:
\bea\label{HarmElpsd}
H&=&\frac{1}{d_{12}d_{13}d_{23}}\left[
G(x_1)(x_2-x_3)+G(x_2)(x_3-x_1)+G(x_3)(x_1-x_2)
\right]\nonumber\\
&=&\frac{G(x_1)}{d_{12}d_{13}}+
\frac{G(x_2)}{d_{21}d_{23}}+\frac{G(x_3)}{d_{31}d_{32}}.
\eea
Notice that constant and linear functions $G$ lead to $H=0$, and quadratic $G$ gives constant $H$. 

Substitution of (\ref{HarmElpsd}) into (\ref{ElpsdLapl}) leads to a complicated equation for function $G$:
\bea\label{ElpsdLaplA}
&&\frac{R_1 d_{23}}{d_{12}d_{13}}G_1''-2\frac{d_{23}}{d^2_{12}d^2_{13}}(d_{12}+d_{13})R_1G_1'+
\frac{d_{23}}{2d_{12}d_{13}}R_1'G_1'\nonumber\\
&&+\frac{d_{23}}{d_{12}^2d_{13}^2}\left[3(R_1-R_0)-(x_1x_2+x_1x_3+x_2x_3)\frac{R_0''}{2}-R'_0(x_1+x_2+x_3)-
3x_1x_2x_3\right]G_1\nonumber\\
&&+perm=0
\eea
Here we used a shorthand notation: $R_0=R(0)$, $R'_0=R'(0)$, $R''_0=R''(0)$, $R_1=R(x_1)$, $G_1=G(x_1)$. Equation 
(\ref{ElpsdLaplA}) should work around $x_2=x_3$ as long as $x_1$ is sufficiently large\footnote{All sources of the harmonic function are localized in some finite region of space, which cannot protrude to large $x_1$, which is analogous to radial coordinate.},  so it can be expanded in powers of $d_{23}$. Taking the leading piece proportional to $d_{23}$, multiplying by $d_{12}^3$, differentiating the result four times with respect to $x_1$, we find a closed--form equation for $F(x_1)\equiv G'''_1$:
\bea\label{FFFeqn}
10R'''F+15 R_1''F'+9R_1'F''+2R_1 F'''=0.
\eea
To analyze this equation, it is convenient to set $c=0$ by shifting $x_i$, $a^2$, $b^2$ by $-c^2$ (see definition (\ref{IntroEpsd})), and
set $a=1$ by rescaling $x_i$. The resulting equation (\ref{FFFeqn}) has a general solution 
\bea\label{ElpsdFb}
F(x)=\frac{(b^2+2x+x^2)(b^2-x^2)(b^2+2xb^2+x^2)}{x^{5/2}(b^2+x)^{5/2}(1+x)^{5/2}}
\left(C_1+C_2 I_2+C_3 I_3\right),
\eea
where
\bea
I_2&=&\int_{q}^x \frac{x^{3/2}(1+x)^{3/2}(b^2+x)^{3/2}}{(x^2+b^2+2xb^2)^2(b^2+2x+x^2)^2}dx,
\nonumber\\
I_3&=&\int_{q}^x
\frac{x^{3/2}(1+x)^{3/2}(b^2+x)^{3/2}}{(x^2+b^2+2xb^2)^2(b^2+2x+x^2)^2}\frac{x(1+x)(b^2+x)}{(b^2-x^2)^2}dx
\eea
The low limit of integration, $q$, will be defined in a moment. 
Function $H$ should remain finite and smooth as long as $x_1$ is sufficiently large, this implies that functions $G(x_2)$ and $G(x_3)$ must be finite for all 
$0\ge x_2\ge-b^2\ge x_3\ge -1$ (recall the region (\ref{ElpsdRng}) and our normalization $c=0$, $a=1$), so function $G(x)$ must be well--defined for all $0>x>-1$. This gives restrictions on $C_2$ and $C_3$. 

We will now demonstrate that function $G(x)$ cannot remain finite for $x<-b$ and $x>-b$ unless $C_3=0$. 
Let us choose $q=-b-\eps$ and assume that function $G(p)$ is
finite. Recalling the definition of $F$ ($F(x)=G'''(x)$), we find
\bea
G(x)=\int_q^x dy_1\int_q^{y_1} dy_2
\int_q^{y_2} dy_3 F(y_3)+G(q)+(x-q)G'(q)+
\frac{1}{2}(x-q)^2G''(q)
\eea
As $x$ changes from $-b-\eps$ to $-b+\delta$, the last three terms as well as contributions proportional to $C_1$ and $C_2$ remain finite, so we focus on the term containing $C_3$:
\bea\label{G3Elpsd}
G_3(x)&=&C_3\int_q^x dy_1\int_q^{y_1} dy_2
\int_q^{y_2} dy_3
(b+y_3)f(y_3)I_3(y_3)\nonumber\\
&=&C_3\int_q^x dy_1\int_q^{y_1} dy_2
\int_q^{y_2} dy_3
(b+y_3)f(y_3)\int_q^{y_3}\frac{g(y)}{(b+y)^2}
\eea
Here we introduced two functions, 
\bea
f(x)&=&\frac{(b^2+2x+x^2)(b-x)(b^2+2xb^2+x^2)}{x^{5/2}(b^2+x)^{5/2}(1+x)^{5/2}},\nonumber\\
g(x)&=&\frac{x^{3/2}(1+x)^{3/2}(b^2+x)^{3/2}}{(x^2+b^2+2xb^2)^2(b^2+2x+x^2)^2}\frac{x(1+x)(b^2+x)}{(b+x)^2},
\eea
which remain finite and non--zero in some vicinity of $x=-b$. Changing the order of integration in (\ref{G3Elpsd}), we find 
\bea
G_3(x)&=&C_3\int_q^{x}\frac{g(y)}{(b+y)^2}dy
\int_y^{x} dx_3
(b+y_3)f(y_3)\int_{y_3}^{x} dy_2\int_{y_2}^{x} dy_1
\nonumber\\
&=&C_3\int_q^{x}\frac{g(y)}{(b+y)^2}dy
\int_y^{x} dy_3
(b+y_3)f(y_3)\frac{(x-y_3)^2}{2}
\eea
The integral in the right--hand side does not make sense at $x>-b$ unless 
\bea\label{IntEqlZero}
\int_{-b}^{x} dy_3
(b+y_3)f(y_3)\frac{(x-y_3)^2}{2}=0,
\eea
which is clearly not the case. We conclude that $G_3(x)$ (and thus $G(x)$) is ill--defined at $x>-b$ unless $C_3=0.$\footnote{ 
Although a simpler analysis shows that $I_3$ is ill--defined for $x>-b$, this condition by itself is not sufficient to rule out  $C_3$: in particular, had function $f$ satisfied (\ref{IntEqlZero}), $G_3(x)$ would have existed for $x>-b$, even though $I_3$ would have not.} Once $C_3$ is set to zero, a similar analysis in a vicinity of $x=\sqrt{1-b^2}-1$ leads to $C_2=0$, and integrating the resulting $F(x)$, we find 
\bea
G(x)=C_1\sqrt{x(x+b^2)(x+1)}+C_3x^2+C_4 x+C_5
\eea
Putting back $a$ and $c$, we  can rewrite the last expression as
\bea\label{ElpsdSlnG}
G(x)=C_1\sqrt{(x+a^2)(x+b^2)(x+c^2)}+C_3x^2+C_4 x+C_5,
\eea
This function gives the most general solution of (\ref{FFFeqn}) consistent with physical requirements imposed on $G$,
and direct substitution of (\ref{ElpsdSlnG}) into (\ref{ElpsdLaplA}) shows that the corresponding function $H$ (see (\ref{HarmElpsd})) is harmonic. 

Although solution (\ref{ElpsdSlnG}) is well-defined everywhere (unlike contributions proportional to $C_2$ and $C_3$), the corresponding harmonic function (\ref{HarmElpsd}) has singular points at arbitrarily large $x_1$ unless $C_1=0$. Indeed, consider (\ref{HarmElpsd}) near $x_2=-b^2,x_3=-b^2$ for large $x_1$:
\bea
H
&=&\frac{G(x_1)}{d_{12}d_{13}}+
\frac{G(x_2)}{d_{21}d_{23}}+\frac{G(x_3)}{d_{31}d_{32}}
\nonumber\\
&=&
C_1\frac{\sqrt{(b^2-c^2)(b^2-a^2)}}{d_{21}d_{23}}\left[\sqrt{x_2+b^2}-\sqrt{x_3+b^2}\right]+
finite\nonumber\\
&=&C_1\frac{\sqrt{(b^2-c^2)(b^2-a^2)}}{d_{21}d_{23}}
\frac{d_{32}}{\sqrt{x_2+b^2}+\sqrt{x_3+b^2}}+finite\\
&=&C_1\frac{\sqrt{(b^2-c^2)(b^2-a^2)}}{d_{21}(\sqrt{x_2+b^2}+\sqrt{x_3+b^2})}+finite\nonumber
\eea
This harmonic function diverges as $x_2$ and $x_3$ approach $-b^2$ unless $C_1=0$. 

\bigskip

To summarize, we have demonstrated that separation of the HJ equation in ellipsoidal coordinates with $k=3$ implies that the transverse space is three--dimensional, and the harmonic function is given by (\ref{HarmElpsd}) with 
\bea
G(x)=C_3x^2+C_4 x+C_5. 
\eea
Such harmonic function, $H=C_3$, does not satisfy the required boundary condition (\ref{Hasymt}), so ellipsoidal coordinates for $k=3$ do not no separate the HJ equation for geodesics in D--brane backgrounds. In the next subsection we will outline the extension of this result to $k>3$. 

\subsection{Ellipsoidal coordinates for $k>3$}
\label{AppElpKG3}

After giving a detailed description of ellipsoidal coordinates for $k=3$, we will briefly comment on the extension of the results obtained in the last subsection to $k>3$. 

The ellipsoidal coordinates in higher dimensions were introduced in \cite{Jacobi}\footnote{We use a slightly modified notation to connect with discussion in the last subsection.}:
\bea
r_i^2=-\left[\prod_j(a_i^2+x_j)\right]\left[\prod_{j\ne i}\frac{1}{(a_j^2-a_i^2)}\right].
\eea
Here $r_i$ are the $k$ radii introduced in (\ref{BaseMetr}), and $x_i$ are ellipsoidal coordinates corresponding to $k$ root of an algebraic equation
\bea\label{EqnElpsd}
\sum_{i=1}^k\frac{r_k^2}{x+a_k^2}=1,
\eea
The ranges of ellipsoidal coordinates are analogous to (\ref{ElpsdRng}) in the three--dimensional case:
\bea
x_1\ge -a_1^2\ge x_2\ge -a_2^2\ge \dots x_k\ge -a_k^2.
\eea 
In terms of the ellipsoidal coordinates, the radial part of the metric (\ref{BaseMetr}),
\bea
ds^2_r=dr_1^2+\dots+dr_k^2,
\eea
becomes
\bea
ds_r^2=\sum_{i=1}^k h_i^2 (dx_i)^2,\qquad h_i^2\equiv\frac{1}{4}\left[\prod_{j\ne i}(x_i-x_j)\right]\left[\prod_j\frac{1}{a_j^2+x_i}\right]
\eea
To simplify some formulas appearing below, it is convenient to 
define
\bea
D_i=\prod_{j\ne i}(x_i-x_j),\qquad 
R_i=\prod_j(a_j^2+x_i),\qquad D=\prod_{i< j}(x_i-x_j).
\eea

A counterpart of the HJ equation (\ref{EllisoidHJ}) for $k>3$ is 
\bea\label{ElpsNHJ}
4\sum_{i=1}^k\frac{R_i}{D_i}(\d_i S)^2+\sum_{j=1}^k \frac{L_j^2}{r_j^2}+Hp_\mu p^\mu=0.
\eea
Assuming that this equation separates in ellipsoidal coordinates,
\bea
S(x_1,\dots,x_k)=S_1(x_1)+\dots+S_k(x_k),
\eea
and applying derivatives to (\ref{ElpsNHJ}), we find one of the integrability conditions (cf. (\ref{IntCndH})):
\bea\label{ElpsNInt}
\d_i^{n-1}\d_j^{n-1}\left[D\left\{\sum_{l=1}^k \frac{L_l^2}{r_l^2}+Hp_\mu p^\mu\right\}\right]=0
\eea

As before, this relation can be used to show that $d_1=d_2=\dots=d_k=0$ in (\ref{BaseMetr}). Indeed, for  $d_1>0$, equation (\ref{ElpsNInt}) must be satisfied for all values of $L_1$, then
\bea
\d_1^{n-1}\d_2^{n-1}\left[\frac{D}{r_1^2}\right]=
-\left[\prod_{j>1}(a_j^2-a_1^2)\right]
\d_1^{n-1}\d_2^{n-1}\left[\frac{D}{R_1}\right]=0
\eea
The last relation is false, and the easiest way to see this is to notice that the leading contribution in the vicinity of 
$x_1=-a_1^2$ comes when all $\d_1$ derivatives hit $(x_1+a_1^2)^2$, giving $n$--order pole, however, the remaining $\d_2$ derivatives do not kill $D$. As before the degenerate case (e.g., $a_2=a_1$) requires a separate consideration, and it can be eliminated using the arguments that followed equation (\ref{ProlSpher}). 

Next, we demonstrate that functions $S$ and $H$ must have a formal symmetry under interchange of their arguments. Indeed, taking the limit $x_1\rightarrow x_2$ in (\ref{ElpsNHJ}), we find a counterpart of (\ref{x1ISx2}):
\bea
\left[\frac{R_1d_{12}}{D_1}(S_1')^2-\frac{R_2d_{21}}{D_2}(S_2')^2\right]_{x_1=x_2}=0
\eea
This and other similar limits imply that
\bea
S_1(x_1)=F(x_1),\quad S_2(x_2)=F(x_2),\quad\dots\quad S_k(x_k)=F(x_k),
\eea
then equation (\ref{ElpsNHJ}) ensures the symmetry of $H$.\footnote{Recall that we already established that $d_1=d_2=\dots=d_k=0$, so angular momenta disappear from 
(\ref{ElpsNHJ}).}

Modifying the arguments that led to (\ref{HarmElpsd}), we conclude that
\bea\label{ElpsNHarm}
H=\sum \frac{G(x_i)}{D_i}.
\eea
Indeed, integrability condition (\ref{ElpsNInt}) for $(i,j)=(1,2)$ gives
\bea
H=\frac{1}{D}\left[
\sum_{l=0}^{k-2}C_l(x_2,x_3\dots x_k)x_1^l
-\sum_{l=0}^{k-2}{\tilde C}_l(x_1,x_3\dots x_k)x_2^l
\right].
\eea
Symmetry of $H$ implies that $C_l$ and ${\tilde C}_l$ have the same functional form. Repeating this argument for other pairs $(i,j)$, we find
\bea
H=\frac{1}{D}\sum_j\sum_{l=0}^{(k-2)(k-1)} 
{G}_l(x_j)P_l\left[x_1\dots x_{j-1},x_{j+1}\dots x_k\right],
\eea 
where $P_l$ is a polynomial of $l$--th degree, which is anti-symmetric under interchange of any pair of its arguments. Any such polynomial is proportional to
\bea
{\tilde P}\left[y_1\dots \dots y_{k-1}\right]\equiv\prod_{i<j}
(y_i-y_j),
\eea
which already has degree $(k-1)(k-2)$, so 
$P_{(k-1)(k-2)}={\tilde P}$ and using the relation 
\bea
{\tilde P}\left[x_1\dots x_{j-1},x_{j+1}\dots x_k\right]D_j=
(-1)^{j+1}D,\nonumber
\eea
we find
\bea
H=\frac{1}{D}\sum_j {G}_{(k-2)(k-1)}(x_j)\left\{c{\tilde P}\left[x_1\dots x_{j-1},x_{j+1}\dots x_n\right]\right\}=
\sum_j \frac{G(x_j)}{D_j}
\eea
This proves (\ref{ElpsNHarm}).

Since we have already established that $d_1=d_2=\dots=d_k=0$, it becomes easy to write the Laplace equation (\ref{LaplRk}) in ellipsoidal coordinates (cf. (\ref{ElpsdLapl})):
\bea\label{ElpsNLpl}
\sum_{i=1}^k\frac{\sqrt{R_i}}{D_i}\d_i\left[
\sqrt{R}_i\d_i H\right]=0.
\eea
Substitution of (\ref{ElpsNHarm}) into (\ref{ElpsNLpl}) leads to a counterpart of equation (\ref{ElpsdLaplA}):
\bea\label{ElpsdLaplAN}
&&\frac{R_1}{D_1^2}G_1''-2\frac{R_1}{D_1^2}\left[
\frac{1}{d_{12}}+\frac{1}{d_{13}}+\dots+
\frac{1}{d_{1k}}\right]G_1'+
\frac{1}{2D_1^2}R_1'G_1'\nonumber\\
&&+\left[\frac{R_1}{D_1}\d_1^2\left(\frac{1}{D_1}\right)+
\frac{R_1'}{2D_1}\d_1\left(\frac{1}{D_1}\right)+
\sum_{j=2}^k\left\{\frac{R_j}{D_j}~\frac{2}{D_1d_{1j}^2}+
\frac{R_j'}{2D_j}~\frac{1}{D_1d_{1j}}\right\}
\right]G_1\nonumber\\
&&+perm=0,
\eea
Repeating the analysis which led from equation (\ref{ElpsdLaplA}) to (\ref{ElpsdSlnG}) for $k=3$, we find the most general solution 
of (\ref{ElpsdLaplAN}) that exists for all $x\in (-a_j^2,-a_{j+1}^2)$:
\bea
G(x)=C \sqrt{\prod_j(a_j^2+x)}+\sum_{j=0}^{k-1}C_j x^j.
\eea
Requiring $H$ to remain finite at sufficiently large $x_1$, we conclude that most integration constants in the last relation must vanish, and $H$ must be a constant, just as in the $k=3$ case. 

\section{Equation for geodesics in D--brane backgrounds}
\label{AppPrcdr}
\renewcommand{\theequation}{C.\arabic{equation}}
\setcounter{equation}{0}

In this appendix we implement the program outlined in section \ref{SectPrcdr}. Specifically, we introduce the perturbative expansions (\ref{AppPert}) for functions $h(w)$, $U_1(x)$ and $U_2(y)$, substitute the results into (\ref{HarmRTheta}), and find the restrictions imposed by the Laplace equation (\ref{LaplEqn}). 

\subsection{Particles without angular momentum}

First we analyze the harmonic function (\ref{LaplEqn}) for $L_1=L_2=0$, and angular momenta will be added in the next subsection. To stress the fact that we are dealing with this special case, the potentials will be denoted as ${\tilde U}_1(x)$, ${\tilde U}_2(y)$.
Far away from the sources, the harmonic function is given by (\ref{LeadHarm}), and Laplace equation separates in spherical coordinates, so we find the leading asymptotics\footnote{Due to linearity of the Laplace equation (\ref{LaplEqn}), a constant term can always be added to $H$, so the boundary condition (\ref{AsympApp}) can be easily extended to the asymptotically--flat case $H\approx a+\frac{Q}{r^{7-p}}$.}:
\bea\label{AsympApp}
x+iy\approx\ln\frac{r}{l}+i\theta,\qquad H\approx \frac{Q}{r^{7-p}}.
\eea  
In this appendix we will construct the most general function $H$ and coordinates $(x,y)$ which satisfy four conditions:
\label{PageCondAD}
\begin{enumerate}[(a)]
\item{The Hamilton--Jacobi equation (\ref{HJone}) separates in variables $(x,y)$.}
\item{Away from the sources, function $H(r,\theta)$ satisfies the Laplace equation (\ref{LaplEqn}).}
\item{At large values of $r$, function $H$ and coordinates $(x,y)$ approach the asymptotic expressions given by (\ref{AsympApp}).}
\item{The sources are localized at finite values of $r$, in particular, $H$ is regular for $r>R$.}
\end{enumerate}

To derive the expressions for $(x,y)$ and $H$ consistent with (\ref{AsympApp}), it is convenient to introduce expansions in powers of 
$l/r\sim e^{-w}\sim e^{-x}$:
\bea\label{AppPert}
h(w)=w+\sum_{k>0}a_k e^{-kw},\qquad
{\tilde U}_1(x)=e^{(2-n-m)x}
\left[1+\sum_{k>0}b_k e^{-kx}\right].
\eea
Substituting this into the Laplace equation (\ref{LaplEqn}), matching the results for all powers of $l/r$, and resumming the series using Mathematica, we find that the most general solution for $h(w)$ is parameterized by one constant $a_2$:
\bea\label{TheSoln}
h(w)&=&\ln\left[\frac{1}{2}\left\{e^w+\sqrt{4a_2+e^{2w}}\right\}\right],
\eea
and corresponding potentials are given by\footnote{For $p\ge 5$ we find additional solutions for ${\tilde U}_1$ and ${\tilde U}_2$, which will be discussed below.}
\bea\label{PertPotential}
{\tilde U}_1(x)=\frac{C_1}{(e^x+a_2 e^{-x})^{n-1}
(e^x-a_2 e^{-x})^{m-1}},\quad {\tilde U}_2(y)=0.
\eea
We now start with coordinates defined by (\ref{TheSoln}) and find the most general separable solution of the Laplace equation (\ref{LaplEqn}).

First we observe that starting with an arbitrary solution (\ref{TheSoln}) and adjusting $l$, we can set $a_2$ to one of three values $(-1,0,1)$. Indeed, solution (\ref{TheSoln}), definitions (\ref{defCmplVar}), and  boundary conditions (\ref{AsympApp}) remain invariant under the transformation
\bea\label{GaugeTr}
l\rightarrow e^{\la} l,\quad w\rightarrow w-\la,\quad
x\rightarrow x-\la,\quad a_2\rightarrow a_2 e^{-2\la}
\eea
for any real $\la$. Since $a_2$ is a real number\footnote{This follows from reality of potential  ${\tilde U}_1(x)$ in (\ref{PertPotential}).}, it can be set to zero or to $\pm 1$ by choosing an appropriate $\la$ in (\ref{GaugeTr}). It is convenient to analyze these cases separately.
~\\
~\\
\noindent
{\bf Case I:} $a_2=0$, $z=w$.\\
Variables $(x,y)$ correspond to spherical coordinates, and equation (\ref{HarmRTheta}) becomes
\bea\label{HUtilde}
H(r,\theta)=\frac{1}{M^2}\left[-\frac{1}{r^2}\left\{
{\tilde U}_1\left(\ln\frac{r}{l}\right)+
{\tilde U}_2(\theta)\right\}\right]
\eea
Definig a new function
\bea
{\hat U}_1(r)\equiv {\tilde U}_1\left(\ln\frac{r}{l}\right),
\eea
we find that the Laplace equation (\ref{LaplEqn}) reduces to two relations:
\bea
&&\frac{1}{r^{8-p}}\frac{d}{dr}\left[r^{8-p}\frac{d}{dr}\frac{{\hat U}_1}{r^2}\right]+\frac{\la}{r^4}=0,
\nonumber\\
&&\frac{1}{r^{8-p}}\frac{d}{dr}\left[
r^{8-p}\frac{d}{dr}\frac{1}{r^2}\right]{\tilde U}_2+
\frac{1}{r^4\sin^{n}\theta\cos^{m}\theta}
\d_\theta(\sin^{n}\theta\cos^{m}\theta\d_\theta 
{\tilde U}_2)=\frac{\la}{r^4}\nonumber
\eea
with separation constant $\la$. Solving these equations, we find the harmonic function 
\bea
H(r,\theta)&=&-\frac{1}{M^2r^2}
[{\tilde U}_1+{\tilde U}_2]=-\frac{1}{M^2r^2}\left[\frac{C_1}{r^{5-p}}+C_2r^2\right.\nonumber\\
&&\left.+\frac{C_3}{\sin^{n-1}\theta\cos^{m-1}\theta}
+\frac{C_4}{\sin^{n-1}\theta} F\left(\frac{m-1}{2},
\frac{3-n}{2},\frac{m+1}{2};\cos^2\theta\right)\right]\nonumber
\eea
Notice that $\la$ gives constant contributions to ${\tilde U}_1$ and ${\tilde U}_2$, which cancel in the sum. Condition (c) implies that $C_2=0$, and condition (d) gives $C_3=C_4=0$. We conclude that in this case the harmonic function is sourced by a single stack of Dp branes:
\bea
H(r,\theta)=\frac{Q}{r^{7-p}}.
\eea
\noindent
{\bf Case II:} $a_2=-1$, 
$
\displaystyle z=\ln\left[\frac{1}{2}(e^w+\sqrt{e^{2w}-4})\right]
$.\\
This change of variables has been analyzed in Appendix \ref{SprElliptic} (see equation (\ref{EllEqnZ})). Reversing the steps which led from (\ref{coshOne})--(\ref{coshTwo}) to (\ref{EllEqnZ}), we can rewrite (\ref{TheSoln}) as
\bea\label{CaseTwoRep}
\cosh x=\frac{\rho_++\rho_-}{2d},\quad \cos y=\frac{\rho_+-\rho_-}{2d},\quad
\rho_\pm=\sqrt{r^2+d^2\pm 2rd\cos\theta},\quad
d=2l.
\eea

~\\
\noindent
Let us now implement the requirements (a)--(d) listed in page \pageref{PageCondAD}.

(a) As discussed in section \ref{SectPrcdr}, the Hamilton--Jacobi equation is separable if $z=h(w)$ is a holomorphic function, and $H$ has the form (\ref{HarmRTheta}):
\bea\label{SeparHS}
H&=&-
\frac{1}{M^2}\left[
\frac{|h'|^2}{r^2}\left[
{\tilde U}_1\left(x\right)+
{\tilde U}_2\left(y\right)\right]
\right].
\eea
We recall that in this section we are focusing on the special case 
$L_1=L_2=0$.

(b) We will now find the most general function  
${\tilde U}(r,\theta)={\tilde U}_1(x)+{\tilde U}_2(y)$ which  
satisfies the Laplace equation (\ref{LaplEqn}) in the metric
\bea\label{metrBB}
ds^2_{9-p}=
dr^2+r^2d\theta^2+
r^2\sin^{2}\theta d\Omega_{n}^2+
r^2\cos^{2}\theta d\Omega_{m}^2.
\eea
Using a relation
\bea
\left|\frac{h'}{r}\right|^2=\left|\frac{e^w}{\sqrt{e^{2w}-4}}
\frac{1}{le^{(w+{\bar w})/2}}\right|^2=
\frac{1}{l^2|e^{2w}-4|}=\frac{1}{d^2}
\left|\frac{4e^{2z}}{(e^{2z}-1)^2}\right|=
\frac{1}{d^2(\cosh^2x-\cos^2y)},\nonumber
\eea
we can rewrite (\ref{metrBB}) in terms of $x$ and $y$:
\bea\label{metrAA}
ds^2_{9-p}=
d^2\left[(\cosh^2x-\cos^2y)(dx^2+dy^2)+
\sinh^2 x\sin^2 y d\Omega_{n}^2+
\cosh^2 x\cos^2 y d\Omega_{m}^2\right]
\eea
and to rewrite the Laplace equation (\ref{LaplEqn}) in these coordinates:
\bea\label{LaplInXY}
\frac{1}{\sinh^{n}x\cosh^{m}x}\frac{\d}{\d x}\left[
\sinh^{n}x\cosh^{m}x\frac{\d H}{\d x}
\right]+
\frac{1}{\sin^{n}y\cos^{m}y}\frac{\d}{\d y}\left[
\sin^{n}y\cos^{m}y\frac{\d H}{\d y}
\right]=0
\eea
We are looking for a separable solution of equation (\ref{LaplInXY}) which has the form (\ref{SeparHS}):
\bea\label{SeparH}
H
&=&-\frac{1}{(Md)^2(\cosh^2 x-\cos^2 y)}
\left[
{\tilde U}_1\left(x\right)+
{\tilde U}_2\left(y\right)\right].
\eea
Straightforward algebraic manipulations with (\ref{LaplInXY}) lead to equation for 
${\tilde U}={\tilde U}_1\left(x\right)+{\tilde U}_2\left(y\right)$
\bea\label{EqnForTilde}
0&=&\frac{1}{\sinh^{n}x\cosh^{m}x}\frac{\d}{\d x}\left[
\sinh^{n}x\cosh^{m}x\frac{\d \tilde U}{\d x}
\right]+
\frac{1}{\sin^{n}y\cos^{m}y}\frac{\d}{\d y}\left[
\sin^{n}y\cos^{m}y\frac{\d \tilde U}{\d y}
\right]\nonumber\\
&&-2\frac{\left[\{(m+n-2)(\cosh 2x+\cos 2y)-2(m-n)\}+2(\sinh 2x\d_x+\sin 2y\d_y)\right]{\tilde U}}{\cosh 2x-\cos2y}.\nonumber\\
\eea
Since the first line of this equation is the sum of $x$-- and $y$--dependent terms, the second line must have the same structure, so it can be rewritten as 
\bea\label{SplitEqn}
F(x)+G(y)&=&2(m+n-2)({\tilde U}_1(x)-{\tilde U}_2(y))-4\frac{K(x)+L(y)}{\cosh 2x-\cos2y}\\
\label{SplitEqnK}
K(x)&\equiv&\left[(m+n-2)\cosh 2x-(m-n)+\sinh 2x\d_x\right]{\tilde U}_1\\
\label{SplitEqnL}
L(y)&\equiv&\left[(m+n-2)\cos 2y-(m-n)+\sin 2y\d_y\right]{\tilde U}_2
\eea
Consistency of equation (\ref{SplitEqn}) requires that
\bea\label{KandL}
K(x)=\la_1\cosh 2x+\la_2\cosh^2 2x+\la_0,\quad L(y)=-\la_1\cos 2y-\la_2\cos^2 2y-\la_0
\eea
with constant $\la_0$, $\la_1$ and $\la_2$. Substituting this $K(x)$, $L(y)$ in (\ref{SplitEqnK})--(\ref{SplitEqnL}) and solving the resulting equations for 
${\tilde U}_1(x)$ and ${\tilde U}_2(y)$, we find
\bea\label{SolnUtilde}
{\tilde U}_1&=&\frac{1}{\cosh^{m-1}x\sinh^{n-1}x}\left[C_1+\frac{1}{2}\int dx K(x)\cosh^{m-2}x\sinh^{n-2}x\right]\nonumber\\
{\tilde U}_2&=&\frac{1}{\cos^{m-1}y\sin^{n-1}y}\left[C_2+\frac{1}{2}\int dy L(y)\cos^{m-2}y\sin^{n-2}y\right]
\eea
The last two equations give necessary, but not sufficient condition for ${\tilde U}={\tilde U}_1\left(x\right)+{\tilde U}_2\left(y\right)$ to solve 
(\ref{EqnForTilde}). Going back to equation (\ref{EqnForTilde}), we can now rewrite it as
\bea
&&\frac{1}{\sinh^{n}x\cosh^{m}x}\frac{\d}{\d x}\left[
\sinh^{n}x\cosh^{m}x\frac{\d \tilde U_1}{\d x}
\right]+
\frac{1}{\sin^{n}y\cos^{m}y}\frac{\d}{\d y}\left[
\sin^{n}y\cos^{m}y\frac{\d \tilde U_2}{\d y}
\right]\nonumber\\
&&\qquad\qquad+2(m+n-2)({\tilde U}_1(x)-{\tilde U}_2(y))-4\la_1-4\la_2(\cosh 2x+\cos 2y)=0.\nonumber
\eea
This relation is satisfied identically for all $\la_0$, $\la_1$ and $\la_2$.

(c) Although expressions (\ref{SolnUtilde}) produce solutions 
of the Laplace equation (\ref{EqnForTilde}) for all $K(x)$ and $L(x)$ given by 
(\ref{SplitEqnK})--(\ref{SplitEqnL}), the resulting harmonic function (\ref{SeparH}) may not satisfy the boundary conditions (\ref{AsympApp}). 
To analyze these boundary conditions, we find the leading behavior of $H$ at large values of $r$ by substituting 
the asymptotic expressions for various coordinates,
\bea\label{BBeqn}
h(w)\approx w,\quad x\approx \ln\frac{r}{l},\quad y\approx\theta,
\eea
in equations (\ref{SeparH}), (\ref{SplitEqnK})--(\ref{SplitEqnL}), (\ref{SolnUtilde}). 
We conclude that the leading contribution to $H$ satisfies the boundary conditions (\ref{AsympApp})\footnote{We recall that $7-p=m+n$ due 
to (\ref{MNasPQ}).} for the following values of $(C_2,\la_1,\la_2)$:
\bea\label{FromBC}
m=n=0:&&\la_2=0;\nonumber\\
m+n=1:&&\la_2=0;\nonumber\\
m=n=1:&&\la_0=\la_1=\la_2=0;\\
(m,n)=(2,0)\mbox{ or }(0,2):&&\la_1=\la_2=0,\quad C_2=0;\nonumber\\\
m+n>2:&&\la_0=\la_1=\la_2=0,\quad C_2=0.\nonumber
\eea

(d) Function $H$ given by (\ref{SeparH}), (\ref{SolnUtilde}), (\ref{KandL}) satisfies the Laplace equation and the boundary conditions (\ref{AsympApp}) for all values of $(C_2,\la_1,\la_2)$ listed in (\ref{FromBC}). However, regularity of $H$ at sufficiently large values of $r$ imposes some additional requirements. First we notice that regularity in the asymptotic region (\ref{BBeqn}) implies that ${\tilde U}_2(y)$ must remain finite for all values of $y$. This gives additional restrictions for the first three cases listed in (\ref{FromBC})
\bea\label{FromReg}
m=n=0\mbox{ or }m+n=1\mbox{ or }m=n=1:&&\la_1=0,\quad\la_2=-\la_0.
\eea
and to additional restriction $\la_0=0$ for $(m,n)=(2,0)\mbox{ or }(0,2)$. Since for $m=n=1$ the constant $C_2$ can be absorbed into $C_1$, restrictions (\ref{FromBC}), (\ref{FromReg}) can be summarized as
\bea\label{FromAll}
m+n\le1:&&\la_0=\la_1=\la_2=0;\nonumber\\
m+n>1:&&\la_0=\la_1=\la_2=0,\quad C_2=0.
\eea
For $m+n>1$ we recover the solution (\ref{PertPotential}):
\bea\label{SolnUtildeGen}
p<6:\quad {\tilde U}_1&=&\frac{C_1}{\cosh^{m-1}x\sinh^{n-1}x},\qquad
{\tilde U}_2=0,
\eea
and for $m+n\le1$ we find three special cases:
\bea\label{SolnUSpec}
(p,m,n)=(7,0,0):&&
{\tilde U}_1=\frac{C_1}{2}\sinh 2x,\quad
{\tilde U}_2=\frac{C_2}{2}\sin 2y\nonumber\\
(p,m,n)=(6,1,0):&&
{\tilde U}_1=C_1\sinh x,\quad
{\tilde U}_2=C_2\sin y\\
(p,m,n)=(6,0,1):&&
{\tilde U}_1=C_1\cosh x,\quad
{\tilde U}_2=C_2\cos y\nonumber
\eea
The corresponding harmonic function is given by (\ref{HarmTilde}). 
~\\
~\\
~\\
\noindent
{\bf Case III:} $a_2=1$, 
$z=\ln\left[e^w+\sqrt{e^{2w}+4}\right]
$.\\
This change of variables can be obtained from case 2 by making replacements\footnote{Notice that this replacement also interchanges $S^n$ and $S^m$ in (\ref{SSmetr}).}
\bea\label{ShiftTheta}
w\rightarrow w-\frac{i\pi}{2},\quad 
z\rightarrow z-\frac{i\pi}{2}\quad\Rightarrow\quad
\theta\rightarrow \theta-\frac{\pi}{2},\quad
y\rightarrow y-\frac{\pi}{2},
\eea
which lead to
\bea\label{RepCase3}
\cosh x=\frac{\rho_++\rho_-}{2d},\quad 
\sin y=\frac{\rho_+-\rho_-}{2d},\quad
\rho_\pm=\sqrt{r^2+d^2\pm 2rd\sin\theta},
\eea
Repeating the steps which led to (\ref{SolnUtildeGen}), we arrive at 
\bea\label{SolnUtildeGenII}
p<6:\quad {\tilde U}_1&=&\frac{C_1}{\cosh^{n-1}x\sinh^{m-1}x},\qquad
{\tilde U}_2=0.
\eea
while the counterpart of (\ref{SolnUSpec}) becomes
\bea\label{SolnUSpecII}
(p,m,n)=(7,0,0):&&
{\tilde U}_1=\frac{C_1}{2}\sinh 2x,\quad
{\tilde U}_2=\frac{C_2}{2}\sin 2y\nonumber\\
(p,m,n)=(6,0,1):&&
{\tilde U}_1=C_1\sinh x,\quad
{\tilde U}_2=C_2\cos y\\
(p,m,n)=(6,1,0):&&
{\tilde U}_1=C_1\cosh x,\quad
{\tilde U}_2=C_2\sin y\nonumber
\eea
The corresponding harmonic function is given by (\ref{HarmTildeAlt}).

\subsection{General case}\label{Momenta}

As shown in the last subsection, the harmonic function (\ref{HarmRTheta}),
\bea\label{Harmbb}
H(r,\theta)=
\frac{1}{M^2}\left[
\frac{L_2^2}{r^2\sin^2\theta}+
\frac{L_1^2}{r^2\cos^2\theta}-
\frac{|h'|^2}{r^2}\left[
U_1\left(x\right)+
U_2\left(y\right)\right]
\right].
\eea
solves the Laplace equation (\ref{LaplEqn}) as long as $L_1=L_2=0$ and $U_1(x)$, $U_2(y)$ are given by (\ref{SolnUtildeGen}) or (\ref{SolnUSpec}). We will now find the potentials $U_1(x)$ and $U_2(y)$ for non-zero values of $L_1$ and $L_2$. 

Recalling the change of coordinates (\ref{CaseTwoRep}), we can write the first two terms in (\ref{Harmbb}) as
\bea
&&\frac{L_2^2}{r^2\sin^2\theta}+
\frac{L_1^2}{r^2\cos^2\theta}=
\frac{L_2^2}{d^2\sinh^2 x\sin^2 y}+
\frac{L_1^2}{d^2\cosh^2 x\cos^2 y}\nonumber\\
&&\qquad=\frac{1}{d^2(\cosh^2 x-\cos^2 y)}
\left[\left(\frac{L_2^2}{\sinh^2 x}-
\frac{L_1^2}{\cosh^2 x}\right)+\left(
\frac{L_2^2}{\sin^2 y}+\frac{L_1^2}{\cos^2 y}\right)\right]
\eea
Then the harmonic function (\ref{Harmbb}) can be rewritten as
\bea\label{HarmTilde}
H(r,\theta)&=&-\frac{|h'|^2}{M^2r^2}\left[{\tilde U}_1(x)+
{\tilde U}_2(y)\right]\nonumber\\
&=&-\frac{1}{(Md)^2(\cosh^2 x-\cos^2 y)}\left[{\tilde U}_1(x)+
{\tilde U}_2(y)\right],\\
{U}_1(x)&=&{\tilde U}_1(x)+\frac{L_2^2}{\sinh^2 x}-\frac{L_1^2}{\cosh^2 x},\qquad
{U}_2(y)={\tilde U}_2(y)+
\frac{L_1^2}{\cos^2 y}+\frac{L_2^2}{\sin^2 y},\nonumber
\eea
where ${\tilde U}_1$ and ${\tilde U}_2$ are given by (\ref{SolnUtildeGen}) or (\ref{SolnUSpec}).

Reparametrization (\ref{RepCase3}) can be obtained from (\ref{CaseTwoRep}) by making the replacement (\ref{ShiftTheta}), then instead of (\ref{HarmTilde}) we 
find\footnote{Notice that the interchange $\sin\theta\leftrightarrow\cos\theta$ is accompanied by the interchange of angular momenta $L_1\leftrightarrow L_2$}
\bea\label{HarmTildeAlt}
H(r,\theta)&=&-\frac{|h'|^2}{M^2r^2}\left[{\tilde U}_1(x)+
{\tilde U}_2(y)\right]\nonumber\\
&=&-\frac{1}{(Md)^2(\cosh^2 x-\sin^2 y)}\left[{\tilde U}_1(x)+
{\tilde U}_2(y)\right],\\
{U}_1(x)&=&{\tilde U}_1(x)+\frac{L_1^2}{\sinh^2 x}-\frac{L_2^2}{\cosh^2 x},\qquad
{U}_2(y)={\tilde U}_2(y)+\frac{L_1^2}{\cos^2 y}+\frac{L_2^2}{\sin^2 y}.\nonumber
\eea
In this case, functions ${\tilde U}_1$ and ${\tilde U}_2$ are given by (\ref{SolnUtildeGenII}) or (\ref{SolnUSpecII}).

\section{Equations for Killing tensors}
\label{AppKill}
\renewcommand{\theequation}{D.\arabic{equation}}
\setcounter{equation}{0}

In section \ref{SecKill} we discussed the Killing tensors associated with separation of the HJ equation in elliptic coordinates. The detailed calculations are presented in this appendix.

\subsection{Killing tensors for the metric produced by D--branes}

In this appendix we will solve the equations for Killing tensors in the geometry (\ref{EinstFrame})--(\ref{WWbase}) and derive the expression (\ref{Jul25}). To simplify some formulas appearing below, we modify the parameterization of 
(\ref{EinstFrame})--(\ref{WWbase}):
\bea\label{SmplPrmMtr}
ds^2=FG dx^\mu dx_\mu+F\left[A(dx^2+dy^2)+e^B d\Omega_m^2+e^C d\Omega_n^2\right]
\eea
and introduce four types of indices:
\bea
R^{1,p}:\, \mu,\nu,\dots;\qquad
(x,y):\, i,j,\dots;\qquad S^m:\, a,b,\dots;\qquad 
S^n:\, {\dot a},{\dot b},\dots
\eea
Separation of the wave equation implies an existence of rank-two conformal Killing tensors which satisfy equation 
(\ref{CKTdef})
\bea\label{CnfKillEqn}
\nabla_{(M}K_{NL)}=\frac{1}{2}W_{(M}g_{NL)}.
\eea
We begin with analyzing $(xxx)$ and $(yyy)$ components of this relation:
\bea\label{Jul30f}
&&\nabla_xK_{xx}=\frac{1}{2}W_x g_{xx}\,\Rightarrow\, W^x=2\nabla_x K^{xx},\nonumber\\
&&\nabla_yK_{yy}=\frac{1}{2}W_y g_{yy}\,\Rightarrow\, W^y=2\nabla_y K^{yy}.
\eea
Substituting this into $(xxy)$ component and recalling that $g_{xx}=g_{yy}$, we find
\bea\label{Jul30}
\nabla_yK_{xx}+2\nabla_x K_{xy}=\frac{1}{2}W_y g_{xx}=\nabla_y K_{yy}\,\Rightarrow\,
\nabla_y(K^{yy}-K^{xx})=2\nabla_x K^{xy}.
\eea
Using the explicit expressions for the Christoffel symbols,
\bea
\Gamma_{xx}^x=\Gamma_{xy}^y=-\Gamma_{yy}^x=\frac{1}{2}\d_x\ln g_{xx},\quad
\Gamma_{yy}^y=\Gamma_{xy}^x=-\Gamma_{xx}^y=\frac{1}{2}\d_y\ln g_{yy},
\eea
equation (\ref{Jul30}) can be rewritten as
\bea
\d_y(K^{yy}-K^{xx})=2\d_x K^{xy}.
\eea
Combining this with a similar relation coming from $(xyy)$ component of (\ref{CnfKillEqn}),
\bea
\d_x(K^{xx}-K^{yy})=2\d_y K^{xy},
\eea
we conclude that
\bea\label{KillXY}
K^{yy}=K^{xx}+N,\qquad 2d K^{xy}=\star_2 dN.
\eea
so $K^{xy}$ and $N\equiv K^{xx}-K^{yy}$ are dual harmonic functions. We also quote the expressions for $W_i$, which can be obtained by evaluating the covariant derivatives in  (\ref{Jul30f}):
\bea\label{KillVecXY}
W_x=2\left(\d_x\left[g_{xx}K^{xx}\right]+K^{xy}\d_y g_{yy}
\right),
\quad
W_y=2\left(\d_y\left[g_{yy}K^{yy}\right]+K^{xy}\d_x g_{xx}\right).
\eea

Next we look at the Killing equation (\ref{CnfKillEqn}) with three legs on $R^{1,p}$:
\bea\label{Jul30a}
{\d}_{(\mu} K_{\nu\la)}-2\Gamma_{(\mu\nu}^i K_{\la)i}=\frac{1}{2}FGW_{(\la}\eta_{\mu\nu)}.
\eea
Using the expression for the relevant Christoffel symbol,
\bea
\Gamma_{\mu\nu}^i=-\frac{1}{2}g^{ij}\d_jg_{\mu\nu}=-\frac{1}{2FA}\d_i(FG)\eta_{\mu\nu},
\eea
we can rewrite (\ref{Jul30a}) as an equation for a conformal Killing tensor on $R^{1,p}$:
\bea\label{FlatCnfKill}
\d_{(\la} K_{\mu\nu)}=\frac{1}{2}\eta_{(\mu\nu}Z_{\la)},\qquad
Z_\la\equiv\left[GFW_{\la}-
\frac{2}{FA}\d_i(FG)K_{\la i}
\right].
\eea
Tensors satisfying this equation have already been encountered in section \ref{RedTo2D}, and the general solution of 
(\ref{FlatCnfKill}) has been found in \cite{Weir}:
\bea\label{WeirKill}
K_{\mu\nu}&=&A\eta_{\mu\nu}+B_{ab}K^{(a)}_\mu K^{(b)}_\nu,\\
Z_\mu&=&4B_{ab}f^{(a)}K^{(b)}_\mu.\nonumber
\eea
Here $K^{(a)}_\mu$ are conformal Killing vectors on $R^{1,p}$,
\bea\label{FlatKillEqn}
\nabla_{(\mu}K^{(a)}_{\nu)}=f^{(a)}\eta_{\mu\nu},
\eea
and coefficients $A$, $B_{ab}$ can depend on the directions transverse to $R^{1,p}$. Solutions of equation (\ref{SmplPrmMtr}) are reviewed in Appendix \ref{AppFltKil}. The conformal Killing vectors $K^{(a)}_\mu$ are responsible for separation on $R^{1,p}$ (which is already taken into account in the ansatz 
(\ref{GenAction})), and to study the separation in $(x,y)$ coordinates\footnote{In contrast to the explicit symmetries related to Killing vectors, the origin of separation in $(x,y)$ is usually called a ``hidden symmetry".}, it is sufficient to focus on the first term in (\ref{WeirKill}), which is invariant under $ISO(p,1)$ transformations. Repeating this arguments for equation (\ref{CnfKillEqn}) with three legs on the spheres, we conclude that the Killing tensor responsible for separation of $x$ and $y$ is invariant under $SO(m+1)\times SO(n+1)\times ISO(p,1)$, in particular, this implies that 
\bea\label{SymmKT}
K^{MN}p_Mp_N=K_1(x,y)\eta^{\mu\nu}p_\mu p_\nu+
K_2(x,y)h^{ab}p_ap_b+
K_3h^{\dot a\dot b}p_{\dot a}p_{\dot b}+K^{ij}p_ip_j,
\eea 
\bea
W_\mu=0,\qquad W_a=0,\qquad W_{\dot a}=0.\nonumber
\eea
Here $h_{ab}$ and $h_{\dot a\dot b}$ are metrics on $S^m$ and $S^n$. Ansatz (\ref{SymmKT}) ensures that all components of equation (\ref{CnfKillEqn}) which do not contain legs along $x$ or $y$ are satisfied.

Next we look at $(x,\mu,\nu)$ and $(y,\mu,\nu)$ components:
\bea\label{Jul30b}
\d_i K_{\mu\nu}-4\Gamma^\la_{i(\mu}K_{\nu)\la}-
2\Gamma^j_{\mu\nu}K_{ij}=\frac{1}{2}W_i g_{\mu\nu}
\eea
Taking into account the expressions for the Christoffel symbols,
\bea
\Gamma^\la_{i\mu}=\frac{1}{2}\delta^\la_\mu \d_i \ln(FG),
\quad \Gamma^j_{\mu\nu}=
-\frac{1}{2FA}\eta_{\mu\nu}\d_j(FG),
\eea
we can rewrite equation (\ref{Jul30b}) as
\bea\label{Jul30mnk}
\d_i(K^{\mu\nu})-
K_{ij}\eta^{\mu\nu}\frac{1}{FA}\d_j\frac{1}{FG}=
\frac{1}{2}W_i g^{\mu\nu}\,\Rightarrow\,
\d_i K_1+\frac{K_{ij}}{FA}\d_j\frac{1}{FG}=
\frac{1}{2FG}W_i
\eea
Similar analysis of equations along sphere directions gives:
\bea\label{Jul30sph}
\d_i K_2-\frac{K_{ij}}{FA}\d_j\frac{e^{-B}}{F}=
\frac{e^{-B}}{2F}W_i ,\qquad
\d_i K_3-\frac{K_{ij}}{FA}\d_j\frac{e^{-C}}{F}=\frac{e^{-C}}{2F}W_i 
\eea
Equations (\ref{KillXY}), (\ref{KillVecXY}), (\ref{Jul30mnk}), 
(\ref{Jul30sph}) give a complete system which is equivalent to 
(\ref{CnfKillEqn}) for the ansatz (\ref{SymmKT}). 

In the special case $K^{xx}=K^{xy}=0$ (which corresponds to the Killing tensor (\ref{Jul25})), we find
\bea\label{Oct23a}
K^{yy}=c,\quad W_x=0,\quad W_y=2c\d_y(AF),\quad
\d_x K_1=\d_x K_2=\d_x K_3=0
\eea
\bea\label{Oct23b}
\d_y K_1=c\d_y\frac{A}{G},\quad
\d_y K_2=c\d_y\frac{A}{e^B},\quad
\d_y K_2=c\d_y\frac{A}{e^C}
\eea
Integrability conditions require that
\bea\label{Oct23}
\d_x\d_y\frac{A}{G}=\d_x\d_y\frac{A}{e^B}=\d_x\d_y\frac{A}{e^C}
\eea
and they are satisfied by our solution (\ref{EinstFrame}), (\ref{WWbase}), (\ref{KGharm}). As expected, we did not get any restrictions on function $F$ from (\ref{SmplPrmMtr}) since conformal rescaling of the metric does not affect equations for the null geodesics. However, as demonstrated in section \ref{SecWave}, separability of the HJ equation does not persist for the wave equation unless function $F$ has a special form. In particular, we found that the wave equation on the D--brane background is only separable in the Einstein frame. 

To summarize, we have demonstrated that metric (\ref{SmplPrmMtr}) possesses a conformal Killing tensor of the form 
(\ref{SymmKT}) if and only if equations (\ref{KillXY}), (\ref{KillVecXY}), (\ref{Jul30mnk}), (\ref{Jul30sph}) are satisfied. In the special case $K^{xx}=K^{xy}=0$, the system reduces to equations (\ref{Oct23a})--(\ref{Oct23b}), and their integrability conditions are given by (\ref{Oct23}). Application of this construction to the geometry (\ref{EinstFrame})--(\ref{WWbase}) 
gives (\ref{Jul25}).

\subsection{Review of Killing tensors for flat space}
\label{AppFltKil}

In the last subsection we have encountered Killing equations on $R^{1,p}$ space and on a sphere,
\bea\label{KillFlat}
\d_{(\la} K_{\mu\nu)}=g_{(\mu\nu}W_{\la)}
\eea
and we chose the simplest solution,
\bea
K_{\mu\nu}=K g_{\mu\nu},\qquad W_\mu=0,\quad
\d_\mu K=0.
\eea
In this section we review the most general solution of (\ref{KillFlat}).

Killing tensors on symmetric spaces were studied in \cite{Weir}, where it was shown that the most general solution of (\ref{KillFlat}) on a sphere can be written as 
\bea\label{WeirKillA}
K_{\mu\nu}&=&A\eta_{\mu\nu}+B_{ab}K^{(a)}_\mu K^{(b)}_\nu,\\
W_\mu&=&4B_{ab}f^{(a)}K^{(b)}_\mu,\nonumber
\eea
where $K^{(a)}_\mu$ are conformal Killing vectors satisfying equation
\bea\label{FlatKillEqnA}
\nabla_{(\mu}K^{(a)}_{\nu)}=f^{(a)}\eta_{\mu\nu}. 
\eea
Although an explicit solution of equation (\ref{FlatKillEqnA}) was not given in \cite{Weir}, it can be easily derived, and in this section we will present such derivation for flat space since this case played an important role in the construction presented in section \ref{RedTo2D}\footnote{Specifically, an explicit construction of Killing tensors (\ref{WeirKillA}), (\ref{FltCKV}) can be used to show that any coordinate system leading to separation of the Laplace equation in flat space with $d>2$ is a special case of the ellipsoidal coordinates (\ref{ref317}).}. Out result can be easily extended to the conformal Killing tensors on a sphere.

Consider the equation for the Conformal Killing Vector (CKV) on flat space:
\bea\label{CKVeqn}
\d_i K_j+\d_j K_i=2f \delta_{ij}
\eea
From equations with $i=j$ we conclude that
\bea\label{CKVfK}
\d_1 K_1=\d_2 K_2=\dots=f
\eea
Then applying $\d_1\d_2$ to the $(1,2)$ component of equation (\ref{CKVeqn}),
\bea\label{KV12}
\d_1 K_2+\d_2 K_1=0,
\eea
we find a restriction on $f$:
\bea\label{Aug17}
(\d_1^2+\d_2^2)f=0\quad \Rightarrow\quad
f=g(x_1+ix_2,x_3,x_4\dots)+\overline{g(x_1+ix_2,x_3,x_4\dots)}.
\eea
In two dimensions this exhausts all equations for the CKV, thus the most general solution is parameterized by one holomorphic function $F$:
\bea
d=2:\quad K_1=F(x_1+ix_2)+cc,\quad K_2=iF(x_1+ix_2)+cc,\quad
f=\d_1 K_1.
\eea

For $d\ge 3$ we can combine (\ref{KV12}) with its counterparts for the $(1,3)$ and $(2,3)$ component of (\ref{CKVeqn}) to further restrict the form of $f$. Differentiating 
(\ref{KV12}) with respect to $x_3$ and combining the result with
\bea
\d_2\d_1 K_3+\d_2\d_3 K_1=0,\quad 
\d_1\d_2 K_3+\d_1\d_3 K_2=0,
\eea
we conclude that
\bea\label{Aub14b}
\d_2\d_3 K_1=0\quad\Rightarrow\quad \d_2\d_3 f=0.
\eea
Repetition of this argument for all $i\ne j$ gives
\bea
\d_i\d_j f=0\quad\Rightarrow\quad f=\sum h_j(x_j).
\eea
Substituting this into (\ref{Aug17}), we find a relation between $D_1$ and $D_2$:
\bea
h_1''(x_1)+h_2''(x_2)=0\quad\Rightarrow\quad
\begin{array}{ll}
h_1=D_1 x_1^2+A_1 x_1+B_1\\
h_2=D_2 x_2^2+A_2 x_2+B_2
\end{array},
\quad D_2=-D_1.
\eea
Similar arguments show that 
\bea
D_3=-D_2,\quad D_3=-D_1 \quad\Rightarrow\quad
D_i=0.
\eea
To summarize, we have demonstrated that $f$ is a linear function,
\bea\label{Aug14a}
f=A_i x_i+B,
\eea
and (\ref{CKVfK}) leads to the following expressions for $K_1$, $K_2$:
\bea
K_1&=&
x_1A_j x_j-\frac{A_1x_1^2}{2}+Bx_1+h_1(x_2,x_3,\dots),
\nonumber\\
K_2&=&
x_2B_j x_j-\frac{B_2x_2^2}{2}+Bx_2+h_2(x_1,x_3,\dots).
\nonumber
\eea
Substitution of these relations into (\ref{KV12}) gives an equation for $h_1$ and $h_2$: 
\bea
&&A_2x_1+\d_2h_1(x_2,x_3,\dots)
+A_1 x_2+\d_1 h_2(x_1,x_3,\dots)=0,
\eea
and the solution reads
\bea
h_1(x_2,x_3,\dots)&=&-\frac{A_1 x_2^2}{2}+x_2h_{12}(x_3,\dots)+
{\tilde h}_1(x_3,\dots)\nonumber\\
h_2(x_1,x_3,\dots)&=&-\frac{A_2 x_1^2}{2}-x_1h_{12}(x_3,\dots)+
{\tilde h}_2(x_3,\dots)\nonumber
\eea
Equation (\ref{Aub14b}) implies that $\d_3 h_{12}=0$.  Analogs of (\ref{Aub14b}) with different indices ensure that $h_{12}$ does not depend on $(x_4,\dots,x_d)$, this leads to the complete determination of $h_1$ as a function of $x_2$:
\bea
h_1(x_2,x_3,\dots)&=&-\frac{A_1 x_2^2}{2}+C_{12}x_2+
{\tilde h}_1(x_3,\dots)\nonumber\\
\eea
Repeating this argument for other pairs, we find the most general expression 
for $K_i$:
\bea\label{FltCKV}
K_i&=&
(A_j x_j)x_i-\frac{A_i r^2}{2}+Bx_i+C_{ij}x_j+D_i,\qquad
C_{ij}=-C_{ji}
\eea
To summarize, we have demonstrated that any conformal Killing vector in flat space has form (\ref{FltCKV}) with constant parameters $A_i$, $B$, $C_{ij}$. The arguments of \cite{Weir} imply that the general Killing tensor on such space can be written as combination of such Killing vectors (\ref{WeirKillA})
with additional constants $A$, $B_{ab}$.

\section{Non--integrability of strings on D--brane backgrounds}
\renewcommand{\theequation}{E.\arabic{equation}}
\setcounter{equation}{0}
\label{AppNVE}
In section \ref{SecGnrGds} we have classified Dp--brane backgrounds leading to integrable geodesics, and in section 
\ref{NonintStr} we outlined the procedure for studying dynamics of strings on such geometries. This appendix contains 
technical details supporting the analysis of section \ref{NonintStr}. Appendix \ref{NVEReview} reviews one of the 
approaches to integrable systems, and section \ref{NonIntStrsDBr} applies this general construction to specific configurations of strings on Dp--brane backgrounds. In particular, in appendix \ref{AppNintGen} we focus on the geometries leading to integrable geodesics\footnote{As discussed in the Introduction, integrability of geodesics is a necessary condition for integrability of strings.} and rule out integrability of strings for the majority of such backgrounds.
 
\subsection{Review of analytical non--integrability}
\label{NVEReview}

A dynamical system is called integrable if the number of integrals of motion is equal to the number of degrees of freedom. To demonstrate that a system is {\it not integrable}, it is sufficient to look at linear perturbations around a particular solution and to demonstrate that the resulting {\it linear equations} do not have a sufficient number of integrals of motion. The equation for linear perturbations is known as the Normal Variational Equation (NVE), and the Kovacic algorithm \cite{Kovacic_alg} gives a powerful analytical tool for studying this equation \cite{ziglinMor}. Let us outline the procedure developed in \cite{ziglinMor}.

Consider a Hamiltonian system parameterized by canonical variables as $(x_i, p_i)$, ($i$ ranges from $1$ to $N$). To rule 
out integrability using NVE, one has to perform the following steps.
\begin{enumerate}
\item Start with a Hamiltonian and write down the equations of motion
\bea
H=\frac{1}{2}\sum_{i=1}^N p_i+V(x_1,..., x_N),
\eea
\bea\label{eom}
\dot{x}_i&=&\frac{\partial H}{\partial p_i}=f_i(x_1,...,x_N,p_1,...,p_N),\nonumber\\
\dot{p}_i&=&-\frac{\partial H}{\partial x_i}=g_i(x_1,...,x_N,p_1,...,p_N),\\ i&=&1,...,N.\nonumber
\eea
\item Write the full variational equations of (\ref{eom}):
\bea\label{var_eqs}
\delta\dot{x}_i&=&\sum_{j=1}^N\frac{\partial f_j}{\partial x_j} \delta x_j+\sum_{i=j}^N\frac{\partial f_j}{\partial p_j} \delta p_j,\nonumber\\
\delta\dot{p}_i&=&\sum_{j=1}^N\frac{\partial g_j}{\partial x_j} \delta x_j+\sum_{j=1}^N\frac{\partial g_j}{\partial p_j} \delta p_j,\\ i&=&1,...,N.\nonumber
\eea
\item Choose a particular solution $x_i=\hat{x}_i,p_i=\hat{p}_i, i=1,...,N$ and one pair of the canonical coordinates, 
$(x_k, p_k)$. The Normal Variational Equation (NVE) is a subsystem of (\ref{var_eqs}) with $i=k$:
\bea\label{Nnv01}
\delta\dot{x}_k=\sum_{j=1}^N\frac{\partial f_j}{\partial x_j} \delta x_j\Big|_{\hat{x}_i,\hat{p}_i}+\sum_{i=j}^N\frac{\partial f_j}{\partial p_j} \delta p_j\Big|_{\hat{x}_i,\hat{p}_i},\\
\delta\dot{p}_k=\sum_{j=1}^N\frac{\partial g_j}{\partial x_j} \delta x_j\Big|_{\hat{x}_i,\hat{p}_i}+\sum_{j=1}^N\frac{\partial g_j}{\partial p_j} \delta p_j\Big|_{\hat{x}_i,\hat{p}_i}.\nonumber
\eea
\item Rewrite the system (\ref{Nnv01}) as a second-order differential equation for $\delta x_k$:
\bea
\delta \ddot{x}_k+q(t)\delta \dot{x}_k+r(t)\delta x_k=0,
\eea
or in the standard notation
\bea\label{nve}
\ddot{\eta}+q(t)\dot{\eta}+r(t)\eta=0,\quad \delta x_k=\eta.
\eea
Equation (\ref{nve}) is known as NVE \cite{ziglinMor}.
\item Make the NVE suitable for using the Kovacic algorithm, namely, algebrize the NVE (i.e. rewrite it as differential equation with rational coefficients) by using a change of variables $t\to x(t)$:
\bea\label{alg_nve}
\eta^{''}+\left(\frac{\ddot{x}}{\dot{x}^2}+\frac{q(t(x))}{\dot{x}}\right)\eta'+\frac{r(t(x))}{\dot{x}^2}\eta=0, \quad F'\equiv\frac{dF}{dx}.
\eea
The Kovacic algorithm requires the coefficients in this equation to be rational functions of $x$. The next step is to convert (\ref{alg_nve}) to the normal form by redefining function $\eta$:
\begin{eqnarray}\label{norm_nve}
{\tilde \eta}{''}(x)+U(x){\tilde\eta}(x)=0.
\end{eqnarray}

\item Apply the Kovacic algorithm to (\ref{norm_nve}), and if it fails then the system is not integrable. 
\end{enumerate}
A more detailed explanation of these steps, containing several examples, can be found in \cite{StepTs}. 

\subsection{Application to strings in Dp--brane backgrounds}\label{NonIntStrsDBr}

Let us now apply the NVE method described in Appendix \ref{NVEReview} to equations for strings in the D$p$--brane background given by (\ref{SSmetr}):
\begin{eqnarray}\label{p-brane_ansatz_1}
ds^2&=&\frac{1}{\sqrt{H}} \eta_{\mu\nu} dx^{\mu} dx^{\nu}+\sqrt{H}(dr^2+r^2d\theta^2+r^2\cos^2\theta d\Omega_{m}^2+r^2\sin^2\theta d\Omega_{n}^2).
\end{eqnarray}
To compare with \cite{StepTs} we introduce a notation $f=H^{1/4}$. The Polyakov action for the string,
\begin{equation}\label{Polyakov}
S=-\frac{1}{4 \pi \alpha^{'}} \int d\sigma d\tau G_{MN}(X) \partial_a X^{M} \partial^{a} X^{N},
\end{equation}
leads to the equations of motion and Virasoro constraints,
\bea\label{Virasoro}\label{Virasoro1}
G_{MN}\dot{X^M} X^{'N}=0,\\
\label{Virasoro2}
G_{MN}(\dot{X^M} \dot{X^N}+X^{'M} X^{'N})=0.
\eea

To demonstrate that system is not integrable, it is sufficient to look only at a specific sector and show that the number of conserved quantities does not match the number of variables. Specifically, we consider the following ansatz:
\begin{equation}\label{string_ansatz}
x^{0}=t(\tau),\quad r=r(\tau), \quad \phi=\phi(\sigma), \quad \theta=\theta(\tau).
\end{equation}
All other coordinates are considered to be constants. Substitution (\ref{string_ansatz}) into (\ref{Polyakov}) leads to the Lagrangian
\begin{equation}\label{OrLag}
L=-f^{-2}\dot{t}^2+f^2\dot{r}^2+f^2r^2(-\sin^2\theta\phi^{'2}+\dot{\theta}^2).
\end{equation}
Solving the equations of motion for cyclic variables $t,\phi$ 
\begin{eqnarray}
\dot{t}&=&Ef^2,\nonumber\\
\phi'&=&\nu=const,
\end{eqnarray}
and substituting the Virasoro constraint (\ref{Virasoro2}), 
\bea
E^2&=&\dot{r}^2+r^2\dot{\theta}^2+\nu^2r^2\sin\theta^2,
\eea
back to the original Lagrangian (\ref{OrLag}) we obtain the effective Lagrangian
\begin{equation}
L=f^2(\dot{r}^2+r^2\dot{\theta}^2-r^2\sin^2\theta\nu^2-E^2),
\end{equation}
and the corresponding Hamiltonian
\begin{equation}
H=\frac{p_r^2}{4f^2}+\frac{p_{\theta}^2}{4f^2r^2}+f^2\nu^2r^2\sin^2\theta+E^2.
\end{equation}
Hamiltonian equations for $\theta$ and $p_{\theta}$ are
\bea\label{Heqptheta}
\dot{\theta}&=&\frac{p_{\theta}}{2f^2r^2},\nonumber\\
\\
\dot{p_{\theta}}&=&\frac{f'_{\theta}}{2f^3}\left(p_r^2+\frac{p_{\theta}^2}{r^2}\right)-2ff'_{\theta}\nu^2r^2\sin^2\theta-2f^2\nu^2r^2\sin\theta\cos\theta.\nonumber
\eea

Let us choose a particular solution
\bea\label{Oct23s}
\theta=\frac{\pi}{2},\qquad p_{\theta}=0
\eea
corresponding to a string wrapped on the equator of $S^2$ and moving only in $r$. Then by shifting coordinate $\tau$ we can set $r(0)=0$, and the Virasoro constraint (\ref{Virasoro2}) gives
\begin{equation}\label{Oct23s1}
r=\hat{r}(\tau)=\frac{E}{\nu}\sin{\nu\tau}.
\end{equation}
Expanding (\ref{Heqptheta}) around solution (\ref{Oct23s})--(\ref{Oct23s1}), we get equations for variations:
\bea\label{Jj23}
\delta\dot{\theta}&=&\frac{\delta p_{\theta}}{2f^2r^2},\nonumber\\
\\
\delta\dot{p_{\theta}}&=&
\frac{f^{''}_{\theta\theta}\delta\theta}{2f^3}p_r^2-\frac{3}{2}\frac{f'_{\theta}}{f^4}f'_{\theta}\delta\theta p_r^2-2f'_{\theta}\delta\theta-2ff^{''}_{\theta\theta}\delta\theta\nu^2r^2+2f^2\nu^2r^2\delta\theta.\nonumber
\nonumber
\eea
Substituting $p_r=2f^2\dot{r}$ into the last equation, we find equation for $\delta\dot{p_{\theta}}$,
\bea\label{Jj23a}
\delta\dot{p_{\theta}}&=&
\left[ 2f^{''}_{\theta\theta}\dot{r}^2 - 6(f'_{\theta})^2\dot{r}^2-2(f'_{\theta})^2\nu^2r^2 -2ff^{''}_{\theta\theta}\nu^2r^2+2f^2\nu^2r^2 \right]\delta\theta,\nonumber
\eea
and substitution of this result into (\ref{Jj23}) leads to the NVE for $\delta\theta\equiv\eta$:
\bea
\ddot\eta+2\dot{r}\left( \frac{f'_r}{f}+\frac{1}{r} \right) \dot\eta - \left[ \frac{f^{''}_{\theta\theta}}{f} \left(\frac{\dot{r}}{r}\right)^2 -3\left(\frac{f'_{\theta}}{f}\right)^2\left(\frac{\dot{r}}{r}\right)^2-\left(\frac{f'_{\theta}}{f}\right)^2\nu^2-\frac{f^{''}_{\theta\theta}}{f}\nu^2+\nu^2 \right]\eta=0.\nonumber
\eea
Finally by changing variable $\displaystyle r=\frac{E}{\nu}\sin{\nu\tau}$ we obtain
\begin{equation}\label{NVE_frtheta}
\eta^{''}+\left[ \frac{\ddot{r}}{\dot{r}^2}+2\left(\frac{f'_r}{f}+\frac{1}{r}\right) \right] \eta' - \left[ \frac{f^{''}_{\theta\theta}}{f} \frac{1}{r^2} -3\left(\frac{f'_{\theta}}{f}\right)^2\frac{1}{r^2}-\left(\frac{f'_{\theta}}{f}\right)^2\frac{\nu^2}{\dot{r}^2}-\frac{f^{''}_{\theta\theta}}{f}\frac{\nu^2}{\dot{r}^2}+\frac{\nu^2}{\dot{r}^2} \right]\eta=0.
\end{equation}

To summarize, in this Appendix we have derived the NVE for the metric (\ref{p-brane_ansatz_1}) and the ``pulsating'' string ansatz (\ref{string_ansatz}). To analyze the resulting equation (\ref{NVE_frtheta}) the function $f$ should be specified. In the next Appendix \ref{AppNintGen} we consider a particular function $f$ corresponding to the most general geodesics-integrable harmonic function derived in section \ref{SecGnrGds}.

\subsection{Application to geometries with integrable geodesics}
\label{AppNintGen}

In this Appendix we consider the most general harmonic function $H=f^4$ leading to integrable geodesics (see (\ref{TheSolnMone}) and (\ref{TheSoln67})). First we express this particular $f$ in terms of $(r,\theta)$, then we use NVE (\ref{NVE_frtheta})  derived in the previous Appendix to check its integrability. The first step leads to
\bea
f^4&=&\frac{d^2}{\rho_+\rho_-}\Bigg\{ \tilde{Q} \left[ \left(\frac{\rho_++\rho_-}{2d}\right)^2-1 \right]^{\frac{1-n}{2}}\left[ \frac{\rho_++\rho_-}{2d} \right]^{1-m}  \nonumber\\ && + P\left[  1-\left( \frac{\rho_+-\rho_-}{2d} \right)^2 \right]^{\frac{1-m}{2}} \left[  \frac{\rho_+-\rho_-}{2d} \right]^{1-n} \Bigg\}, \quad n+m<2 \label{fnmll2}\\
f^4&=&\frac{d^2\tilde{Q}}{\rho_+\rho_-}\left[ \frac{\rho_++\rho_-}{2d} \right]^{1-m}\left[ \left(\frac{\rho_++\rho_-}{2d}\right)^2-1\right]^{\frac{1-n}{2}},\quad n+m\ge2\label{fnmge2}
\eea
where we used the mapping between $(x,y)$ and $(\rho_+,\rho_-)$, namely 
\bea
\displaystyle \cosh x=\frac{\rho_++\rho_-}{2d}, \qquad
\cos y=\frac{\rho_+-\rho_-}{2d}.
\eea
The NVEs for $P=0$ are the same in both cases (\ref{fnmll2}), (\ref{fnmge2}):
\bea\label{NVEmnge2}
\eta^{''}&+&\frac{U}{D}\eta=0,\nonumber\\
U&=&E^4 \left[-d^4(n-1)^2-2 d^2 r^2 \left((m-3) n-5 m+n^2+2\right)-r^4
   (m+n-4) (m+n)\right] \nonumber\\
   &&+2 E^2 r^2 \nu^2 \Big[d^4 ((n-2) n+5)+2 d^2 r^2 (m (n-5)+(n-4) n+9)\nonumber\\
   &&+r^4 \left(m^2+2 m (n-3)+(n-6) n+4\right)\Big]-r^4 \nu^4 \Big[d^4 (n-3) (n+1)\\&&+2 d^2
   r^2 \left((m-5) n-5 m+n^2+4\right)+r^4 \left(m^2+2 m (n-4)+(n-8) n-4\right)\Big],\nonumber\\
D&=&16 r^2 \left(d^2+r^2\right)^2 [E^2-(r \nu)^2]^2.\nonumber
\eea
Application of the Kovacic algorithm shows non--integrability of the system unless $d=0$ and $m+n=4$. The integrable case corresponds to strings on $AdS_5\times S^5$, then  
equation (\ref{NVEmnge2}),
\begin{equation}\label{NVEAdS5S5}
\eta^{''}+\frac{5x^2-2}{4(x^2-1)^2}\eta=0,\qquad x\equiv \frac{r\nu}{E},
\end{equation}
coincides with (3.26) from \cite{StepTs}. 

Non--vanishing $P$--term in (\ref{fnmll2}) for $n=0,m=0$ and $n=0,m=1$ gives rise to divergences at $\theta=\pi/2$ ($f(r,\pi/2)=0$), so the NVE around solution (\ref{Oct23s}) is not well defined. The last remaining case of (\ref{fnmll2}) is $n=1,m=0$. 
Setting $\tilde Q=0$ gives the following NVE
\bea\label{Nov5}
\eta''+\frac{E^4(3d^2+2r^2)+E^2\nu^2(2d^4-d^2r^2-5r^4)+r^2\nu^4(d^4+4d^2r^2+6r^4)}{4(d^2+r^2)^2(E^2-r^2\nu^2)^2}\eta=0.
\eea
Using the Kovacic algorithm we see that this term corresponds to an integrable system. Unfortunately the corresponding geometry is unphysical since it has singularities at arbitrarily large values of $r$. To see this we recall that metric 
(\ref{p-brane_ansatz_1}) becomes singular when $f=0$, and function $f$ vanishes at $\rho_+-\rho_-=2d$, which corresponds to 
$\theta=0$ and arbitrary value of $r$ (recall (\ref{EllHarm})). 

To summarize, in this appendix we have demonstrated that supergravity backgrounds with integrable geodesics do not lead to integrable string theories, with the exception of $AdS_5\times S^5$.

\section{Geodesics in bubbling geometries.}\label{AppBbl}

\renewcommand{\theequation}{F.\arabic{equation}}
\setcounter{equation}{0}

In this appendix we will present the analysis of geodesics in the 1/2--BPS geometries constructed in \cite{LLM}. Our conclusions are summarized in section \ref{Bubbles}. 

The BPS geometries constructed in \cite{LLM} are supported by the five--form field strength, and the metric is given by 
\bea\label{GenBubble}
ds^2&=&-h^{-2}(dt+V_i dx^i)^2+h^2(dY^2+dx_1^2+dx_2^2)+Ye^G d\Omega_3^2+Ye^{-G} d{\tilde\Omega}_3^2,\\
h^{-2}&=&2Y\cosh G,\qquad 
YdV=\star_3 dz,\qquad
z=\frac{1}{2}\tanh G,\nonumber
\eea
where function $z(x_1,x_2,Y)$, which satisfies the Laplace equation,
\bea\label{bbLpl}
\d_i\d_i z+Y\d_Y\left(Y^{-1}\d_Y z\right)=0,
\eea
obeys the boundary conditions 
\bea\label{Droplet}
z(Y=0)=\pm \frac{1}{2}.
\eea
As discussed in section \ref{Bubbles}, only three classes of configurations can potentially lead to integrable geodesics, and we will discuss these classes in three separate subsections.  

\subsection{Geometries with AdS$_5\times$S$^5$ asymptotics}
\label{BubblesCircles}

The boundary conditions depicted in figure \ref{Fig:Bubbles}(a) lead to geometries (\ref{GenBubble}) which are invariant under rotations in $(x_1,x_2)$ plane, and such solutions are conveniently formulated in terms of coordinates introduced in 
(\ref{RingRed}):
\bea\label{RngBubble}
ds^2=-h^{-2}(dt+V_\phi d\phi)^2+h^2(dr^2+r^2d\theta^2+r^2\cos^2\theta d\phi^2)+
Ye^G d\Omega_3^2+Ye^{-G} d{\tilde\Omega}_3^2.
\nonumber\\
\eea
The complete solution of the Laplace equation (\ref{bbLpl}) and expression for $V_\phi$ for this case were found in \cite{LLM}:
\bea\label{zV}
{z}=\frac{1}{2}+\frac{1}{2}\sum_{i=1}^n(-1)^{i+1}\left[\frac{r^2-R_i^2}{\sqrt{(r^2+R_i^2)^2-4R_i^2r^2\cos^2\theta}}-1\right],
\nonumber \\
V_\phi=-\frac{1}{2}\sum_{i=1}^n(-1)^{i+1}\left[ \frac{r^2+r_i^2}{\sqrt{(r^2+R_i^2)^2-4R_i^2r^2\cos^2\theta}} - 1 \right],
\eea 
Summation in (\ref{zV}) is performed over $n$ circles with radii $R_i$, and following conventions of \cite{LLM}, we take $R_1$ to be the radius of the largest circle. For example, a disk corresponds to one circle, a ring to two circles, and so on. 

To write the HJ equation (\ref{HJone}) in the metric 
(\ref{RngBubble}), we notice that coordinates $(t,\phi)$, 
as well as two spheres, separate in a trivial way, this gives
\bea\label{HJbblSep}
S=-Et+J\phi+S^{(3)}_{L_1}(y)+
{\tilde S}^{(3)}_{L_2}({\tilde y})+T(r,\theta),
\eea
where $S^{(3)}_{L_1}(y)$ and 
${\tilde S}^{(3)}_{L_2}({\tilde y})$ satisfy equations (\ref{HJSphere}). Then equation (\ref{HJone}) becomes
\bea\label{HJ_rings}
0&=&-\left(h^4-\frac{V_{\phi}^2}{r^2\cos^2\theta}
\right)E^2+
\left(\frac{\partial T}{\partial r}\right)^2+
\frac{1}{r^2}\left(
\frac{\partial T}{\partial \theta}\right)^2\nonumber\\
&&+2\frac{V_{\phi}}{r^2\cos^2\theta}JE+
\frac{J^2}{r^2\cos^2\theta}+
\frac{L_1^2(\frac{1}{2}-z)}{Y^2}+\frac{L_2^2(z+\frac{1}{2})}{Y^2}.
\eea
We begin with analyzing equations for geodesics with $L_1=L_2=J=0$:
\bea\label{RingsA}
(\d_r T)^2+\frac{1}{r^2}(\d_\theta T)^2
-\left(h^4-\frac{V_{\phi}^2}{r^2\cos^2\theta}\right)E^2=0,
\eea
which is similar to (\ref{TheHJ}). To evaluate the last term in (\ref{RingsA}), we need expressions (\ref{zV}) as well as relation
\bea
h^4=-\frac{(z+\frac{1}{2})(z-\frac{1}{2})}{Y^2}=
-\frac{(z+\frac{1}{2})(z-\frac{1}{2})}{r^2\sin^2\theta}.\nonumber
\eea

Applying the techniques developed in section \ref{SecGnrGds} to (\ref{RingsA}), we conclude that this equation is separable if and 
only if there exists a holomorphic function $g(w)=x+iy$, such that
\bea\label{RingsCompl}
\frac{r^2}{|g'(w)|^2}\left(h^4-\frac{V_{\phi}^2}{r^2\cos^2\theta}\right)=U_1(x)+U_2(y).
\eea
Complex variable $w$ is defined in terms of $(r,\theta)$ by (\ref{defCmplVar}). To find $g(w)$, we employ the same perturbative technique that was used to derive (\ref{TheSolnMthree})--(\ref{TheSolnMtwo}): starting with a counterpart of (\ref{AppPert}), 
\bea\label{AppPertRing}
g(w)=w+\sum_{k>0}a_k e^{-kw},\qquad
{U}_1(x)=e^{-2x}
\left[1+\sum_{k>0}b_k e^{-kx}\right],
\eea
and solving (\ref{RingsCompl}) order--by--order in $e^{-x}$, we find the necessary condition for the existence of a solution:
\bea\label{ReqnRing}
(D_2)^{k-1}D_{2(k+1)}=(D_4)^k.
\eea
Here we defined combinations of $R_j$ appearing in (\ref{zV}):
\bea\label{RiRing}
D_{p}\equiv\sum_{j=1}^n (-1)^{j+1} (R_{j})^p.
\eea
Relations (\ref{ReqnRing}) are clearly satisfied for $n=0$ (flat space) and for $n=1$ (AdS$_5\times$S$^5$), and we will now demonstrate that they fail for $n\ge 2$.

If $n\ge 2$ in (\ref{RiRing}), we can define 
$\eps\equiv R_2/R_1< 1$ and 
\bea\label{defDtld}
{\tilde D}_{p}\equiv\sum_{j=2}^n (-1)^{j+1} (R_{j})^p.
\eea
Notice that
\bea\label{IneqDtld}
|{\tilde D}_{p}|\le \sum_{j=2}^n (R_{j})^p\le
(n-1)(R_{2})^p<(n-1)(\eps R_1)^p.
\eea
Rewriting an infinite set of relations (\ref{ReqnRing}) as
\bea\label{RdfnDD}
D_2D_{2(k+1)}=D_4 D_{2k},
\eea
and extracting explicit powers of $R_1$ in the last relation we find an equation
\bea
{\tilde D}_2-R_1^{-2}{\tilde D}_4=
R_1^{2(1-k)}{\tilde D}_{2k}-R_1^{-2k}{\tilde D}_{2k+2}.
\eea
Then inequality (\ref{IneqDtld}) implies that
\bea\label{DDineq}
|{\tilde D}_2-R_1^{-2}{\tilde D}_4|\le R_1^{2(1-k)}|{\tilde D}_{2k}|+R_1^{-2k}|{\tilde D}_{2k+2}|<
(n-1)\eps^{2k}R_1^2(1+\eps^2).
\eea
The left--hand side of this inequality does not depend on $k$, and the right--hand side goes to zero as $k$ goes to infinity ($n$ is an arbitrary but fixed number), then we conclude that
\bea
{\tilde D}_2=R_1^{-2}{\tilde D}_4\quad\Rightarrow\quad
D_2=R_1^{-2}D_4.
\eea
Substituting this into (\ref{RdfnDD}), we express all 
${\tilde D}_k$ through ${\tilde D}_2$ and $R_1$:
\bea
{\tilde D}_{2k+2}=R_1^2{\tilde D}_{2k}=\dots=R_1^{2k}{\tilde D}_2,
\eea
then inequality (\ref{IneqDtld}) ensures that ${\tilde D}_2=0$:
\bea
|{\tilde D}_2|=R_1^{-2k}|{\tilde D}_{2k+2}|=
(n-1)R_1^2\eps^{2k+2}\rightarrow 0 \quad \mbox{as}\ k\rightarrow\infty.
\eea
On the other hand, the definition (\ref{defDtld}) implies that ${\tilde D}_2$ is strictly negative\footnote{Recall that $(R_2)^2>(R_3)^2>\dots (R_n)^2$.}, as long as $n\ge 2$. We conclude that relation (\ref{ReqnRing}) holds only for $n=0,1$. To show this for all values of $n$ we used an infinite set of relations (\ref{ReqnRing}), however, for any given $n$ is it sufficient to use (\ref{ReqnRing}) with $k=0,\dots,n$. Note that the equations with $k=0,1$ are trivial. 

Going back to the solution with $n=1$, we can extract the relevant holomorphic function and potential $U_1(x)$ (see (\ref{RingsCompl}))
\bea\label{hDisk}
g(w)&=&\ln\left[ \frac{1}{2}\left( e^w+\sqrt{e^{2w}-(R_1/l)^2} \right) \right],\nonumber\\
U_1(x)&=&\frac{e^{2x}E^2R_1^2}{l^2(e^{2x}+(R_1/2l)^2)^2},\quad U_2(y)=0.
\eea
We recall the $w$ is defined by (\ref{defCmplVar}), and this relation has a free dimensionful parameter $l$, which can be chosen in a convenient way. Setting $l=R_1/2$, we recover the standard elliptic coordinates (\ref{TheSolnMone}), which can also be rewritten as (\ref{ElliptCase}):
\bea\label{BblElptCrdA}
\cosh x=\frac{\rho_++\rho_-}{2d},\quad \cos y=\frac{\rho_+-\rho_-}{2d},\quad
\rho_\pm=\sqrt{r^2+R_1^2\pm 2rR_1\cos\theta}.
\eea
The relation between $(x,y)$ and standard coordinates on AdS$_5\times$S$^5$ will be discussed in section \ref{StdPar}.

We now go back to the general HJ equation (\ref{HJ_rings}) and show that presence of (angular) momenta does not spoil separation of the HJ equation for $n=1$ (\ref{hDisk}). To prove this we express the contribution from momenta as sums of two functions $\tilde{U}_1(x)+\tilde{U}_2(y)$ (analogously to (\ref{RingsCompl})). A series of transformations is resulted in
\bea
&&\frac{r^2}{|g'(w)|^2}\left(2\frac{V_{\phi}}{r^2\cos^2\theta}JE+\frac{J^2}{r^2\cos^2\theta}\right)\nonumber\\
&&\qquad=-\frac{32e^{2x}(R_1/l)^2JE}{(4e^{2x}+(R/l)^2)^2}-\left[\frac{16e^{2x}(R_1/l)^2}{(4e^{2x}+(R_1/l)^2)^2} +\frac{1}{\cos^2y} \right]J^2,\nonumber\\
&&\frac{r^2}{|g'(w)|^2}\frac{L_1^2(\frac{1}{2}-z)}{Y^2}=\frac{16L_1^2e^{2x}(R_1/l)^2}{(-4e^{2x}+(R_1/l)^2)^2},\\
&&\frac{r^2}{|g'(w)|^2}\frac{L_2^2(z+\frac{1}{2})}{Y^2}=\frac{L_2^2}{\sin^2y}.\nonumber
\eea
Clearly the right hand sides of these relations are separable.

\subsection{Geometries with pp--wave asymptotics}
\label{StripsPlane}

We will now discuss the geometries with translational $U(1)$ symmetry (\ref{StripRed}), which correspond to parallel strips in the $(x_1,x_2)$ plane (see figure \ref{Fig:Bubbles}(b,c)). It is convenient to distinguish two possibilities: $z$ can either approach different values $x_2\rightarrow \pm\infty$ (as in figure \ref{Fig:Bubbles}(b)) or approach the same value on both sides (as in figure \ref{Fig:Bubbles}(c)). In this subsection we will focus on the first option, which corresponds to geometries with plane wave asymptotics, and the second case will be discussed in subsection \ref{SecStrips}.

Pp--wave can be obtained as a limit of $AdS_5\times S^5$ geometry by taking the five--form flux to infinity \cite{BMN}. This limit has a clear representation in terms of boundary conditions in $(x_1,x_2)$ plane: taking the radius of a disk to infinity, we recover a half-filled plane corresponding to the pp--wave (see figure \ref{Fig:BubblesInt}(c)). Taking a similar limit for a system of concentric circles, we find excitations of pp--wave geometry by 
a system of parallel strips (see figure \ref{Fig:Bubbles}(b)). Since strings are integrable on the pp--wave geometry \cite{BMN}, it is natural to ask whether such integrability persists for the deformations represented in figure \ref{Fig:Bubbles}(b). In this subsection we will rule out integrability on the deformed backgrounds by demonstrating that even equations for massless geodesics are not integrable. 

Solutions of the Laplace equation (\ref{bbLpl}) corresponding to the boundary conditions depicted in figure \ref{Fig:Bubbles}(b) are 
given by \cite{LLM}
\bea\label{zVpwave}
z&=&-\frac{x_2-d_0}{2\sqrt{(x_2-d_0)^2+Y^2}}-\frac{1}{2} \sum_{i=1}^n \left[ \frac{x_2-d_{2i}}{\sqrt{(x_2-d_{2i})^2+Y^2}}-\frac{x_2-d_{2i-1}}{\sqrt{(x_2-d_{2i-1})^2+Y^2}} \right],\nonumber \\
&&\\
V_1&=&-\frac{1}{2\sqrt{(x_2-d_0)^2+Y^2}}-\frac{1}{2} \sum_{i=1}^n \left[ \frac{1}{\sqrt{(x_2-d_{2i})^2+Y^2}}-\frac{1}{\sqrt{(x_2-d_{2i-1})^2+Y^2}} \right],\nonumber\\
V_2&=&0.\nonumber
\eea
Here $n$ is the number of black strips (strip number $i$ is located at $d_{2i-1}<x_2<d_{2i}$), and $x_2=d_0$ is the boundary of the half--filled plane. We will assume that the set of $d_j$ is ordered:
\bea
d_{2N}>d_{2N-1}>\dots>d_1>d_0.
\eea
Although one can shift $x_2$ to set $d_0=0$, we will keep this value free and use the shift symmetry later to simplify some equations. 

Repeating the steps performed in the last subsection, we write the counterpart of (\ref{HJbblSep})
\bea\label{F26}
S=-Et+px_1+S^{(3)}_{L_1}(y)+
{\tilde S}^{(3)}_{L_2}({\tilde y})+T(r,\theta),
\eea
where coordinates $(r,\theta)$ are defined by (\ref{StripRed}). The HJ equation is given by the counterpart of (\ref{HJ_rings}):
\bea\label{HJppWave}
0&=&-\left(h^4-V_1^2\right)E^2+
\left(\frac{\partial T}{\partial r}\right)^2+
\frac{1}{r^2}
\left(\frac{\partial T}{\d\theta}\right)^2\nonumber\\
&&+2pV_1E+p^2+
\frac{L_1^2(\frac{1}{2}-z)}{Y^2}+\frac{L_2^2(z+\frac{1}{2})}{Y^2}.
\eea
As before, we begin with analyzing this equation for  $L_1=L_2=p=0$:
\bea\label{PwaveA}
(\d_r T)^2+\frac{1}{r^2}(\d_\theta T)^2
-\left(h^4-V_1^2\right)E^2=0.
\eea
Results of section \ref{SecGnrGds} ensure that equation (\ref{PwaveA}) is separable if and only if there exists a holomorphic function $g(w)$, such that
\bea\label{PwaveCompl}
\frac{r^2}{|g'(w)|^2}\left(h^4-V_1^2\right)=U_1\left(x\right)+
U_2\left(y\right),
\eea
where complex variable $w$ is defined by (\ref{defCmplVar}) and $x+iy=g(w)$.
Solving equation (\ref{PwaveCompl}) in perturbation theory, we find relations for $d_i$, which are much more complicated that (\ref{ReqnRing}). Introducing the convenient notation $D_{p}=\displaystyle\sum_{j=1}^{2n} (-1)^{j} (d_{j})^p$ as in the previous subsection we find two options for the relations between $D_i$:
\bea\label{CondStripsPlane1}
&&D_4=\frac{D_2^3+D_3^2}{D_2},\quad
D_5=2D_2D_3+\frac{D_3^3}{D_2^2},\quad
D_6=D_2^3+3D_3^3+\frac{D_3^4}{D_2^3}, \\
&&D_7=\frac{D_3(D_2^3+D_3^2)(3D_2^3+D_3^2)}{D_2^4},\quad
D_8=D_2^4+6D_2D_3^2+\frac{5D_3^4}{D_2^2}+\frac{D_3^6}{D_2^5},\quad\dots\nonumber 
\eea
or
\bea\label{CondStripsPlane2}
&&D_4=\frac{D_2^2}{2}+\frac{8D_3^2}{9D_2},\quad
D_5=\frac{5D_2D_3}{6}+\frac{20D_3^3}{27D_2^2},\quad
D_6=\frac{D_2^3}{4}+D_3^2+\frac{16D_3^4}{27D_2^3},\\
&&D_7=\frac{7D_3(9D_2^3+8D_3^2)^2}{972D_2^4},\quad
D_8=\frac{D_2^4}{8}+\frac{8D_2D_3^2}{9}+\frac{80D_3^4}{81D_2^2}+\frac{256D_3^6}{729D_2^5},\quad\dots\nonumber
\eea
The systems (\ref{CondStripsPlane1}) and (\ref{CondStripsPlane2}) are complicated, but they can be analyzed using Mathematica, and 
here we just quote the result: equation (\ref{PwaveCompl}) has no solutions if $n>1$ in (\ref{zVpwave}). 

One special case can be studied analytically: setting $D_1=D_3=0$ in (\ref{CondStripsPlane1}) and (\ref{CondStripsPlane2}) we find simple sets of relations:
\bea\label{Oct14}
D_{2i+1}=0,\quad D_{2i}=D_2^i,\quad\mbox{or}\quad
D_{2i+1}=0,\quad D_{2i}=\frac{D_2^i}{2^{i-1}}, \quad i\ge 1.
\eea
Interestingly, the same system of equations (\ref{CondStrips}) will be encountered in the next subsection\footnote{Solutions (\ref{Oct14}) correspond to $D_1=1$ and to $D_1=2$ in (\ref{CondStrips}).}, where we will prove that 
$n\le 1$.

For $n=0$ equation (\ref{PwaveCompl}) gives no restrictions on $g(w)$ since $h^4-V_1^2=0$ for the pp--wave. Going back to (\ref{HJppWave}) and requiring it to separate for all $(p,L_1,L_2)$, we find the standard pp--wave coordinates:
\bea
ds^2&=&-2dt dx_1-(r_1^2+r_2^2)dt^2+dr_1^2+r_1^2d\Omega^2+dr_2^2+r_2^2d{\tilde\Omega}^2,\\
&&r_1=x,\quad r_2=y.\nonumber
\eea
To write the solution for $n=1$, it is convenient to choose $d_1=0$ by shifting $x_2$. This gives for (\ref{PwaveCompl})
\bea\label{hOneStripHalfPlane}
g(w)&=&w+\ln\left[\frac{1}{2}\left(\sqrt{1-\frac{d_0}{l}e^{-w}}+1\right)\right] +\ln\left[\frac{1}{2}\left(\sqrt{1-\frac{d_2}{l}e^{-w}}+1\right)\right]\\
U_1(x)&=&-\frac{128d_0 d_2 l^2 e^{2x} E^2}{(d_0 d_2-16l^2e^{2x})^2},\quad U_2(y)=0.\nonumber
\eea 
Analysis of the HJ equation (\ref{HJppWave}) with non-vanishing momenta $p,L_1,L_2$ is much more complicated than in case of geometries with $AdS_5\times S^5$ asymptotics, so we performed perturbative analysis. Specifically we compared the function ${\tilde g}(w)$ separating momenta terms with $g(w)$ from (\ref{hOneStripHalfPlane}) separating the rest of the HJ equation (\ref{HJppWave}) and saw that $g(w)$ and ${\tilde g}(w)$ are not compatible for $n=1$.

We conclude that solution with $n=0$ (the standard pp--wave) always gives separable geodesics, solution with $n=1$ leads to integrable geodesics only for vanishing momenta, and solutions with $n\ge 2$ never gives integrable geodesics. 

\subsection{Geometries dual to SYM on a circle}
\label{SecStrips}

Finally we consider configuration depicted in figure \ref{Fig:Bubbles}(c). As discussed in \cite{LLM}, these configurations are dual to Yang--Mills theory on $S^3\times S^1\times R$, and since we are only keeping zero modes on the sphere, the solutions (\ref{GenBubble}) correspond to BPS states in two--dimensional gauge theory on a circle.   

Following the discussion presented in the last subsection, we arrive at equations (\ref{HJppWave}) and (\ref{PwaveA}), however, instead of (\ref{zVpwave}) we find
\bea\label{zVstrip}
z&=&-\frac{1}{2} \sum_{i=1}^n \left[ \frac{x_2-d_{2i}}{\sqrt{(x_2-d_{2i})^2+Y^2}}-\frac{x_2-d_{2i-1}}{\sqrt{(x_2-d_{2i-1})^2+Y^2}} \right],\nonumber \\
V_1&=&-\frac{1}{2} \sum_{i=1}^n \left[ \frac{1}{\sqrt{(x_2-d_{2i})^2+Y^2}}-\frac{1}{\sqrt{(x_2-d_{2i-1})^2+Y^2}} \right],\\
V_2&=&0.\nonumber
\eea
where $n$ is the number of black strips (strip number $i$ is located at $d_{2i-1}<x_2<d_{2i}$), and we assume that the set 
of $d_j$ is ordered:
\bea\label{OrderStrip}
d_{2n}>d_{2n-1}>\dots>d_1.
\eea
Solving equation (\ref{PwaveCompl}) in perturbation theory (here we again start with (\ref{HJppWave}) in absence of momenta), we find a sequence of restrictions on the set of $d_i$. Introducing a convenient notation 
\bea\label{IneqD}
D_{p}=\sum_{j=1}^{2n} (-1)^{j} (d_{j})^p,
\eea
we can write the first three equations as
\bea\label{EqnForDst}
D_1^2D_4-2D_1D_3D_2+D_2^3=0,\nonumber\\
D_1^3D_5-D_1^2D_3^2-D_1D_3D_2^2+D_2^4=0,\\
D_1^4 D_6 - 3 D_1^2 D_2 D_3^2 + D_3 D_2^3 D_1^2=0.
\nonumber
\eea
Notice that inequality (\ref{OrderStrip}) ensures that $D_1>0$, but we can set $D_2=0$ by introducing a shift
\bea
d_n\rightarrow d_n+\alpha.
\eea 
Indeed, $D_1$ remains invariant under this shift, and $D_2$ transforms as 
\bea
D_2\rightarrow D_2+2\alpha D_1,
\eea
so by choosing an appropriate $\alpha$, we can set $D_2=0$. This choice leads to great simplifications in (\ref{EqnForDst}), in particular, we find that $D_4=0$. We will now show that $D_{2p}=0$ for all values of $p$.

First we observe that the structure of our perturbative expansion guarantees that restriction at order $p-1$ has the form
\bea\label{Dpolyn}
D_1^{p-2}D_{p}+P[D_p,D_{p-1},\dots,D_1]=0.
\eea
where $P$ is some polynomial of $p$ arguments. Next we notice that if any configuration of strips leads to a separable Hamilton--Jacobi equation, the configuration which is obtained by the reflection about $x_2$ axis must have the same property. 
The reflection corresponds to the transformation
\bea\label{MirrorD}
d'_{i}= -d_{2N+1-i}\ \Rightarrow\ 
D'_{2k}= -D_{2k},\quad D'_{2k+1}= D_{2k+1},
\eea
so if equations (\ref{Dpolyn}) are solved by the set 
$\{D_k\}$, they should also be solved by the set $\{D'_k\}$. 
We can now use induction to demonstrate that any solution of  (\ref{Dpolyn}) with $D_2=0$ must have
\bea\label{DkZero}
D_{2k}=0\quad\mbox{for all}\quad k.
\eea
The statement is trivial for $k=1$, and we assume that it holds for all $k<K$. Then equation (\ref{Dpolyn}) for $p=2K$ becomes
\bea
D_1^{2K-2}D_{2K}+P[D_{2K-1},0,D_{2K-3},0\dots,D_1]=0
\eea
This equation is symmetric under (\ref{MirrorD}) if and only if $D_{2K}$ is equal to zero\footnote{We also used that $D_1$ is always positive.}. This completes the proof of (\ref{DkZero}) by induction.

For configurations satisfying (\ref{DkZero}) perturbative expansion becomes very simple, and restrictions of $d_i$ can be formulated as
\bea\label{CondStrips}
(D_1)^{p-1}D_{2p+1}=(D_3)^p, \quad p\ge 1.
\eea
Moreover, restrictions (\ref{DkZero}) can be used to simplify the expressions for $D_{2p+1}$. Let us demonstrate that two sets,
\bea\label{StripBBa}
\{d_{2n-1}^2,d_{2n-3}^2,\dots,d_1^2\}\quad\mbox{and}\quad
\{d_{2n}^2,d_{2n-4}^2,\dots,d_2^2\},
\eea 
contain the same elements, although these elements may appear in different order. Indeed, consider 
\bea
d_+=\max\{d_{2n-1}^2,d_{2n-3}^2,\dots,d_1^2\},\quad
d_-=\max\{d_{2n}^2,d_{2n-4}^2,\dots,d_2^2\}
\eea
and assume that $d_+>d_-$. Then
\bea
D_{2p}=\sum_{j=1}^{n}
\left[(d_{2j-1})^{2p}-(d_{2j})^{2p}\right]\ge
(d_+)^{2p}-(n-1)(d_-)^{2p}
\eea
becomes positive for sufficiently large $p$, thus relations (\ref{DkZero}) imply that $d_+\le d_-$. Similar argument 
shows that $d_+\ge d_-$, so $d_+=d_-$ and using (\ref{OrderStrip}) we conclude that $d_{2n}=-d_1$. Repeating this argument for the sets 
\bea\label{StripBB}
\{d_{2n-3}^2,\dots,d_1^2\}\quad\mbox{and}\quad
\{d_{2n-2}^2,\dots,d_2^2\},
\eea 
we find $d_{2n-2}=-d_3$, and continuing this procedure we conclude that\footnote{Notice that symmetry (\ref{MirrorD}) and conditions (\ref{DkZero}) are not sufficient for this conclusion: equation (\ref{OrderStrip}) played a crucial role in or derivation.}
\bea\label{ddOne}
d_{j}= -d_{2n+1-j}\,.
\eea
For such distribution we find an alternative expression for $D_{2p+1}$:
\bea\label{IneqDprm}
D_{2p+1}=2\sum_{j=n+1}^{2n} (-1)^{j+1} (d_{j})^{2p+1}
\eea
Combining (\ref{IneqD}) and (\ref{ddOne}), we conclude that 
\bea\label{OrderDDsr}
d_{2n}>d_{2n-1}>\dots>d_{n+1}>0,
\eea
so equations (\ref{CondStrips}), (\ref{IneqDprm}) are very similar to equations (\ref{RiRing}), (\ref{ReqnRing}), and we can used the same logic\footnote{The proof would not work without inequality (\ref{OrderDDsr}).}  to conclude that $n=1$.  

To separate equation (\ref{PwaveA}) for one strip, it is convenient to set $d_1=0$ by shifting the origin of $x_2$. Then we find
\bea\label{h_one_strip}
g(w)&=&2\ln\left[ \frac{1}{2}
\left(\sqrt{e^w-(d_2/l)}+e^{w/2}\right)\right],\nonumber\\
U_1(x)&=&\frac{8d_2le^xE^2}{(d_2+4le^x)^2},\quad U_2(y)=0.
\eea

Now we go back to the original HJ equation (\ref{HJppWave}) containing non-vanishing momenta $L_1,L_2,p$ and demonstrate that momenta do not spoil separability. Here we consider the separable case of one strip.

Based on the logic used through the entire paper separability of each momentum $p,L_1,L_2$ requires 
\bea\label{ppWaveMomenta}
\frac{r^2}{|g'(w)|^2}\left[
2pV_1E+p^2+\frac{L_1^2(\frac{1}{2}-z)}{Y^2}+\frac{L_2^2(z+\frac{1}{2})}{Y^2}
\right] =\tilde{U}_1(x)+\tilde{U}_2(y),
\eea
Expressing the left--hand side of (\ref{ppWaveMomenta}) in terms of $(x,y)$ and using the holomorphic function $g(w)$ from (\ref{h_one_strip}), we find
\bea
{\tilde U}_1(x)&=&p^2\left(\frac{d_2^4}{256l^4}e^{-2x}+e^{2x}\right)+\frac{8L_1^2d_2e^x[(d_2/l)^2+16e^{2x}]}{l[(d_2/l)^2-16e^{2x}]^2}+
\frac{8d_2L_2^2e^x}{l(d_2/l+4e^x)^2},\nonumber\\
{\tilde U}_2(y)&=&-2pE\frac{d_2}{l}\cos y-\frac{p^2}{8l^2}d_2^2\cos2y+\frac{L_2^2}{\sin^2y}.
\eea
Thus addition of momenta does not spoil separability.

\bigskip

All results obtained in this appendix can be summarized by listing all bubbling geometries leading to separable geodesics (here we use the language introduced in \cite{LLM} to describe the solutions):
\begin{enumerate}
\item $AdS_5\times S^5$: the boundary conditions are given by the disk depicted in figure \ref{Fig:BubblesInt}(a).
\item Pp--wave: the boundary conditions are depicted in figure \ref{Fig:BubblesInt}(c).
\item Single M2 brane: the boundary conditions (one strip) are depicted in figure \ref{Fig:BubblesInt}(b).
\item Pp--wave with an additional strip (see figure \ref{Fig:BubblesInt}(d)): geodesics are only separable in all momenta and angular momenta vanish.
\end{enumerate}

\end{document}